\documentclass[10pt,b5paper,twoside,onecolumn]{scrartcl}
\usepackage[cp1250]{inputenc}
\usepackage{color,fancyvrb}
\usepackage{graphicx}
\usepackage{hyperref}
\usepackage{fancyhdr}
\usepackage{setspace}
\usepackage{float}
\usepackage{pdfpages}
\usepackage{verbatim}
\usepackage{lmodern} 
\usepackage{amsthm}
\usepackage{amssymb}
\usepackage{multicol}
\usepackage{mathpazo}
\usepackage{epsfig}
\usepackage{booktabs}
\usepackage[T1]{fontenc}
\usepackage[titles]{tocloft}
\begin{document}
	
\pagestyle{fancyplain}
\fancyhf{}
\fancyhead[LE]{\textit{Instance Scale, Numerical Properties and Design of Metaheuristics}}
\fancyhead[RO]{\textit{D. Chalupa, P. Nielsen}}
\fancyfoot[C]{\thepage}
\fancypagestyle{plain}
{
	\fancyhf{} 
	\renewcommand{\headrulewidth}{0pt} 
	\renewcommand{\footrulewidth}{0pt}
}

\thispagestyle{empty}
	
\begin{center}\textbf{\LARGE\sffamily\noindent
Instance Scale, Numerical Properties and Design of Metaheuristics: A Study for the Facility Location Problem
}\end{center}

\begin{center}{\large\sffamily\noindent David Chalupa, Peter Nielsen}\end{center}
\begin{center}
{
\noindent
Operations Research Group\\
Department of Materials and Production\\
Aalborg University\\
Fibigerstr\ae de 16, Aalborg 9220, Denmark\\
Email: \texttt{\{dc,peter\}@m-tech.aau.dk}
} \vspace{15pt}
\end{center}

\vspace{15pt}
	
\paragraph{Abstract.} 
Metaheuristics are known to be strong in solving large-scale instances of computationally hard problems. However, their efficiency still needs exploration in the context of instance structure, scale and numerical properties for many of these problems. In this paper, we present an in-depth computational study of two local search metaheuristics for the classical uncapacitated facility location problem. We investigate four problem instance models, studied for the same problem size, for which the two metaheuristics exhibit intriguing and contrasting behaviours. The metaheuristics explored include a local search (LS) algorithm that chooses the best moves in the current neighbourhood, while a randomised local search (RLS) algorithm chooses the first move that does not lead to a worsening. The experimental results indicate that the right choice between these two algorithms depends heavily on the distribution of coefficients within the problem instance. This is also put further into context by finding optimal or near-optimal solutions using a mixed-integer linear programming problem solver. Since the facility location problem is a relatively simple example of a choice-and-assignment problem, similar phenomena are likely to be discovered in a number of other, possibly more complex computational problems in science and engineering.


\section{Introduction}

Many problems in science and engineering are widely regarded as computationally hard. Within operations research, these involve a number of planning, scheduling or production optimisation problems. Such problems include a variety of facility location \cite{gao1992dual} and supply chain optimisation problems \cite{melo2009facility}, as well as shop scheduling problems \cite{blum2004ant}, including job shop scheduling \cite{jobshopscheduling} and flow shop scheduling \cite{flowshopscheduling}. Typical application areas for solving this type of problems include product demand analysis \cite{nielsen2010analyzing}, warehouse location \cite{michel2004simple}, planning and scheduling technology design \cite{steger2011advanced}, portfolio selection for product development \cite{Relich2015102}, container pick-up operations \cite{Do2016285}, or applications in preventive health care \cite{Haase2015}.

A variety of optimisation algorithms have been developed and experimentally verified over the last decades. This holds not only for real-world variants of these problems \cite{canel2001algorithm}, but also for a wider range of NP-hard optimisation problems such as knapsack \cite{garcia2014tabu}, resource allocation \cite{lee2005hybrid} or the $k$-reachability problem \cite{CHALUPA20171}.

The \textit{No free lunch theorems} for optimisation have had a vital impact on design of efficient algorithms to solve combinatorial optimisation problems \cite{wolpert1997no}. The theorems have been interpreted from an algorithmic point of view \cite{culberson1998futility}, as well as in a more application-oriented context \cite{ho2002simple}. One of the implications is that approximation capabilities and runtime of algorithms for combinatorial optimisation problems are now increasingly interpreted in their relation to specific problem instances. This is particularly pronounced in theoretical literature on evolutionary computation \cite{eacomplexity}. 

In this paper, we present a computational study of local search strategies for the classical uncapacitated facility location problem \cite{aikens1985facility,cornuejols1983uncapacitated}. The problem is very well suited for this type of study not only for its NP-hardness \cite{MEGIDDO1982194}, but also because of a wide range of its multiplicative, capacitated and dynamic variants \cite{farahani2009dynamic,pirkul1998multi,Wollenweber2008}. In addition, the problem has very good scaling properties, as well as relatively less constrained search space that does not seem to change its structural properties heavily by scaling. This makes it highly suitable for this computational investigation. Facility location also has important applications in production systems, operations design and management, as well as supply chain design \cite{daskin2005facility}.

\paragraph{Contributions.} In this research we use two local search algorithms to solve the facility location problem and obtain that their performance comparison indicators depend vastly on the coefficients within the instance. To further investigate this, we extend on a previous study of local search algorithms for very large instances of the problem, modelling large-scale applications such as customer service centre location \cite{OR-027}.

The first algorithm is steepest descent local search (LS), choosing the move to open or close a facility at each time step such that the best objective value is obtained. Another algorithm studied will be randomised local search (RLS), which attempts to open or close a facility and accepts the move whenever it does not lead to a worsening. These algorithms are well-studied in evolutionary computation theory \cite{eacomplexity} and have been used to solve other combinatorial optimisation problems, e.g. the minimum conductance graph partitioning problem \cite{CSI-0076}. For each set of instances with up to $90$ potential facility sites, we also use a mixed-integer linear programming solver to compare the results of the local search algorithms to distribution of the actual optima (or near-optimal solutions, in case of very hard instances).

The experimental results are presented for four sets of problem instances, with each set containing instances with $1000$ customers and numbers of facility sites ranging from $50$ to $140$ (for instances with up to $90$ facility sites we also used the ILP-based solver). These four sets do not differ in size but only in the distribution of values within the instances themselves. Intriguingly we observe that the behaviour of LS and RLS depend heavily on the particular facility cost and distance model. Some of the observations contrast with each other, highlighting the tight coupling of numerical properties of an instance and the choice of a suitable algorithm to solve it.

Unsurprisingly, the choice of the right local search strategy is indeed closely tied to the numerical properties of a particular instance. If these properties are known for a particular real-world application, they can be taken into account in an efficient algorithm design. Otherwise, our results provide evidence that simple and scalable strategies are suitable to explore unknown search landscapes. As the facility location problem belongs to a class of classical assignment and cost optimisation problems, we find it reasonable that the results can be generalised to other popular real-world combinatorial optimisation problems.

The paper is structured as follows. In Section 2, we introduce and review the uncapacitated facility location problem. In Section 3, we propose our four cost and distance models for instances of the problem, as well as the local search algorithms explored. Section 4 presents the experimental results and provides a brief discussion. Our conclusions are presented in Section 5.

\section{The Uncapacitated Facility Location Problem}

We will firstly formalise the problem as an integer linear program and provide an overview of related results on heuristics and metaheuristics to solve the problem in large-scale. Next, we will point to the results underpinning the aim of this study. An illustration of the interpretation of the problem and the concepts used is depicted in Figure 1.

\begin{figure*}
\begin{center}
\includegraphics[scale=0.9]{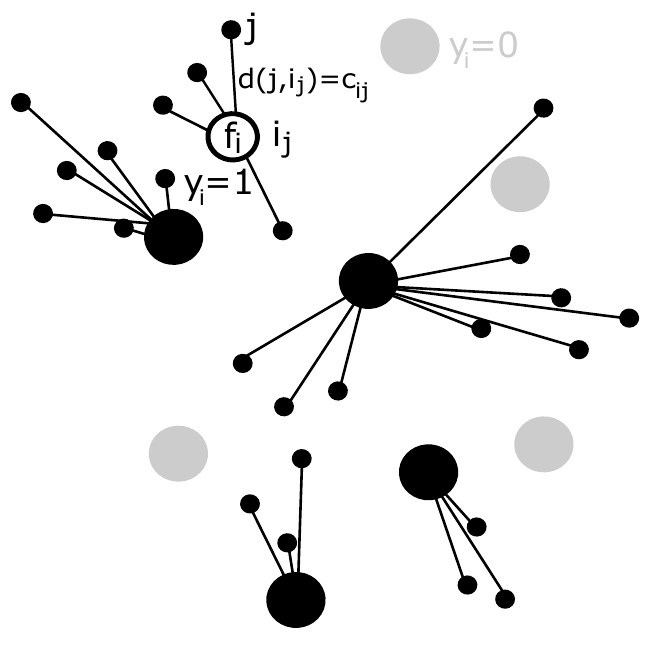}
\end{center}
\caption{Illustration of the concepts and interpretation of the variables used in the formulation of the facility location problem in formulas (1) and (2).}
\end{figure*}

Let $\mathcal{F'} \subseteq  \mathcal{F}$ be a subset of selected potential facility locations. Let $f_i$ be the cost of facility $i \in \mathcal{F}$ and let $d(j,i_j)$ be the distance from a customer $j \in \mathcal{C}$ to the nearest facility $i_j \in \mathcal{F'}$. Then, the objective in the uncapacitated facility location problem will be to minimise the following objective function:

\begin{equation}
\min \sum_{i \in \mathcal{F'}} f_i + \sum_{j \in \mathcal{C}} d(j,i_j).
\end{equation}

\noindent
It is possible to transform this formulation into an integer linear programming (ILP) formulation of the problem \cite{al1999tabu}. Let $n$ be the number of facilities and let $m$ be the number of customers. Then, alternatively, the objective is solve the following ILP formulation of the problem:

\begin{equation}
\min \sum_{i=1}^{n} f_iy_i + \sum_{i=1}^{n}\sum_{j=1}^{m} c_{ij}x_{ij},
\end{equation}

\noindent
s.t.

\begin{equation}
\sum_{i=1}^{n} x_{ij} = 1, ~~~ j=1,...,m
\end{equation}

\begin{equation}
x_{ij} \leq y_{i}, ~~~ i=1,...,n, ~~~ j=1,...,m,
\end{equation}

\noindent
where $c_{ij}$ is the cost of meeting the demand of customer $j$ from facility $i$, and $x_{ij}$ and $y_{i}$ are binary decision variables determining if facility $i$ is used to serve customer $j$ and if a facility is established at position $i$.

With this exact formulation, an out-of-the-box mixed-integer linear programming solver can be used to solve the problem up to a certain scale. We have previously conducted a brief study of such an approach with a high number of customers \cite{OR-027}. As an alternative, specific branch-and-bound algorithms have also been proposed for the problem \cite{tcha1984branch}. For larger instances, as with most NP-hard problems, one needs to use approximation algorithms or heuristics.

The approximation algorithms for the problem use quite specific features of the problem to find solutions of good quality. The best general approximation algorithm for the problem achieves approximation ratio $\mathcal{O}(\log n)$ \cite{Hochbaum1982}. The metric facility location problem is approximable within a factor of 1.488 \cite{Li2011}. However, it is NP-hard to approximate it within a factor better than 1.463 \cite{GUHA1999228}. This documents that both for general and special cases of the problem, there is a gap in approximation results that is well-suited for exploration using heuristics.

Several optimisation algorithms have been applied to solve large scale problem instances. These range from local search algorithms to hybrid population-based approaches \cite{aarts1997local,blum2003metaheuristics}. However, little is still known about the choice of operators to explore the landscape and solve the problem efficiently. This seems to be the case even though the problem is relatively simple and easy to test computationally. An empirical comparison has previously been conducted, comparing genetic algorithms, tabu search and simulated annealing in solving the problem \cite{arostegui2006empirical}. Interestingly but perhaps not surprisingly, this study finds that tabu search works best for most instances, while genetic algorithms or simulated annealing may perform very well in specific cases.

The ideas of local search for the problem have been analysed theoretically \cite{arya2004local,korupolu2000analysis}. Several tabu search approaches have been proposed to solve the problem \cite{al1999tabu,arostegui2006empirical,sun2006solving}, generally differing in more general versus more specific design and parameterisation. Genetic algorithms have also been used to solve the problem \cite{jaramillo2002use}, as well as evolutionary simulated annealing \cite{yigit2006solving}, or particle swarm optimisation \cite{guner2008discrete}.

Many of these algorithms combine several ideas to solve the problem efficiently and usually incorporate a set of parameters with tuned values. However, the interplay of different components of these algorithms is often still not fully understood. This is the case not only for the facility location problem but holds for a number of combinatorial optimisation problems. We will investigate the performance of two local search algorithms for a set of carefully chosen instances. These will highlight the contrast between search space structures and their influence on the actual behaviour of different optimisation techiniques. The aim is to establish how the behaviours of metaheuristics can vastly differ with only a slight variation in values within the problem instance. This will not only highlight the need for hybrid metaheuristics in these problems but will also strengthen the case for hybrid algorithms with a compact parameter suite.

\section{Our Facility Cost and Distance Models and Local Search Algorithms}

In this section, we introduce our four facility cost and distance models, as well as the algorithms we use to solve them.

\subsection{Facility Cost and Distance Models}

Each of the four problem models have specific quantitative properties, leading to contrasting search landscapes. In Model 1, all facilities have the same cost, while the distances follow a moderately varied uniform distribution. Model 2 works with binary facility costs and a bimodal distribution of distances, with many distant connections but a few close ones. Model 3 assumes both binary facility costs and binary distances. Last but not least, Model 4 works with facility costs and distances following a Poissonian distribution. For illustration, the probability distributions for occurrences of specific distance values in all four models are plotted in Figure 2. Similar plots are omitted for the distributions of facility costs, as these were very simple in all four models. These models are chosen, as they lead to a number of different landscapes. For example, while Model 1 represents a moderately rugged landscape, Model 2 is expected to feature ``spikes'' and Model 3 is expected to lead to relatively large plateaus in the search space.

\subsubsection{Model 1: Flat Facility Cost, Moderately Varied Random Distances}
The first model will assign the same unit cost to all facilities. The distances will be represented by integers taken uniformly at random from a limited interval between $1$ and $10$. This way plateaus will be generated in the search space. The corresponding values of $f_i$ and $c_{ij}$ will be:

\begin{equation}
f_i = 1, ~~~ i = 1,...n,
\end{equation}

\begin{equation}
c_{ij} = random(1,10), ~~~ i=1,...,n, ~~~ j=1,...,m.
\end{equation}

\begin{figure*}
\begin{center}
\includegraphics[scale=0.17]{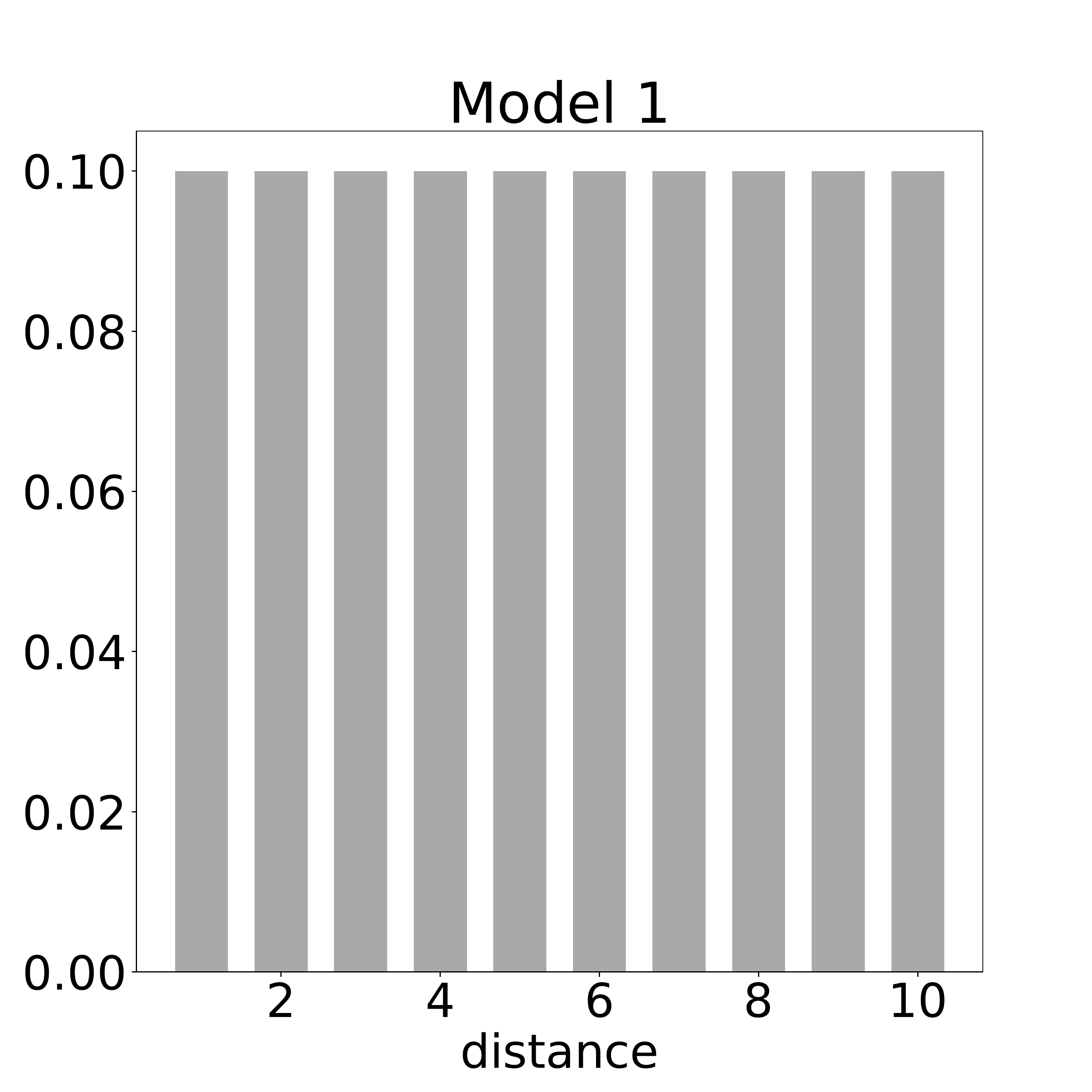}
\includegraphics[scale=0.17]{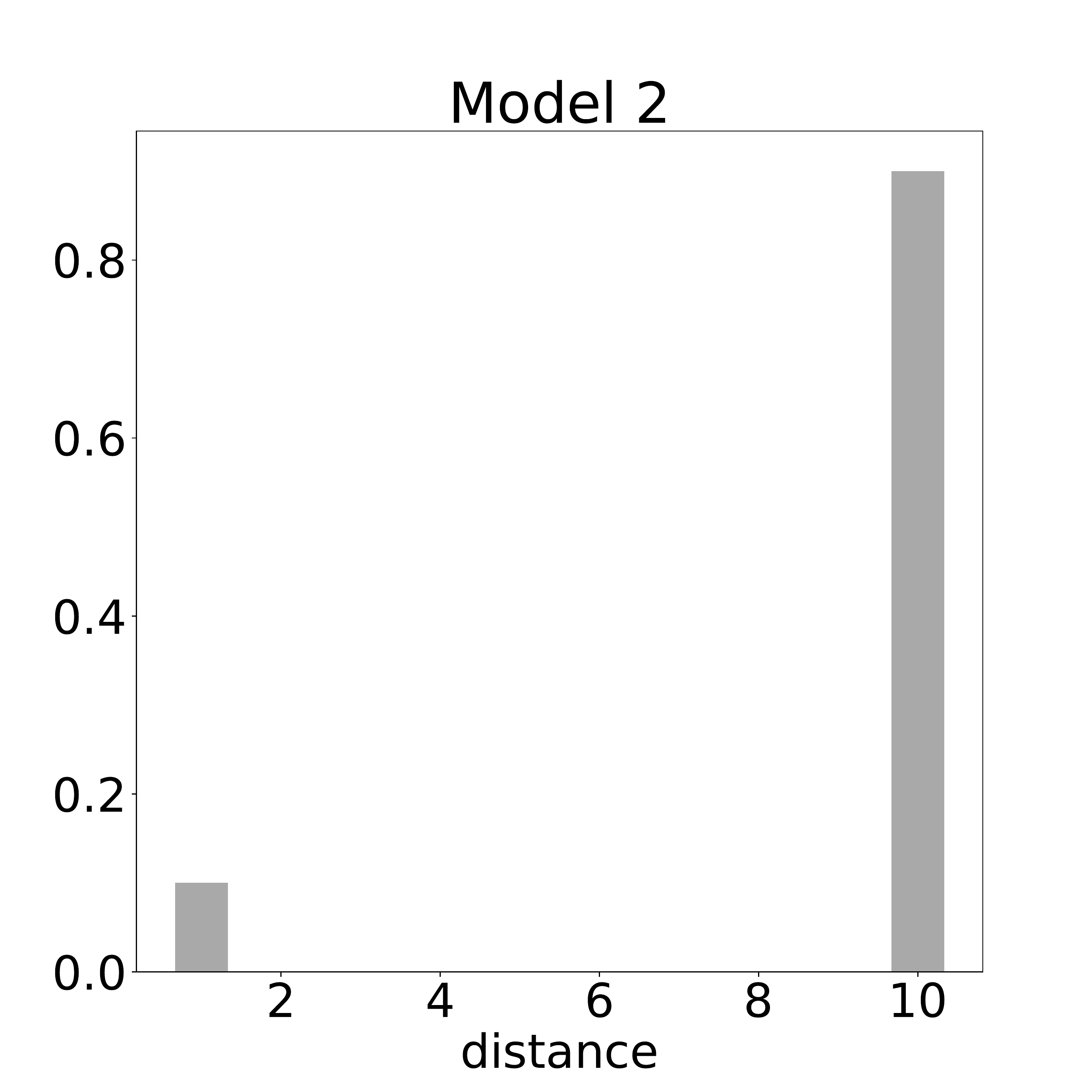}
\includegraphics[scale=0.17]{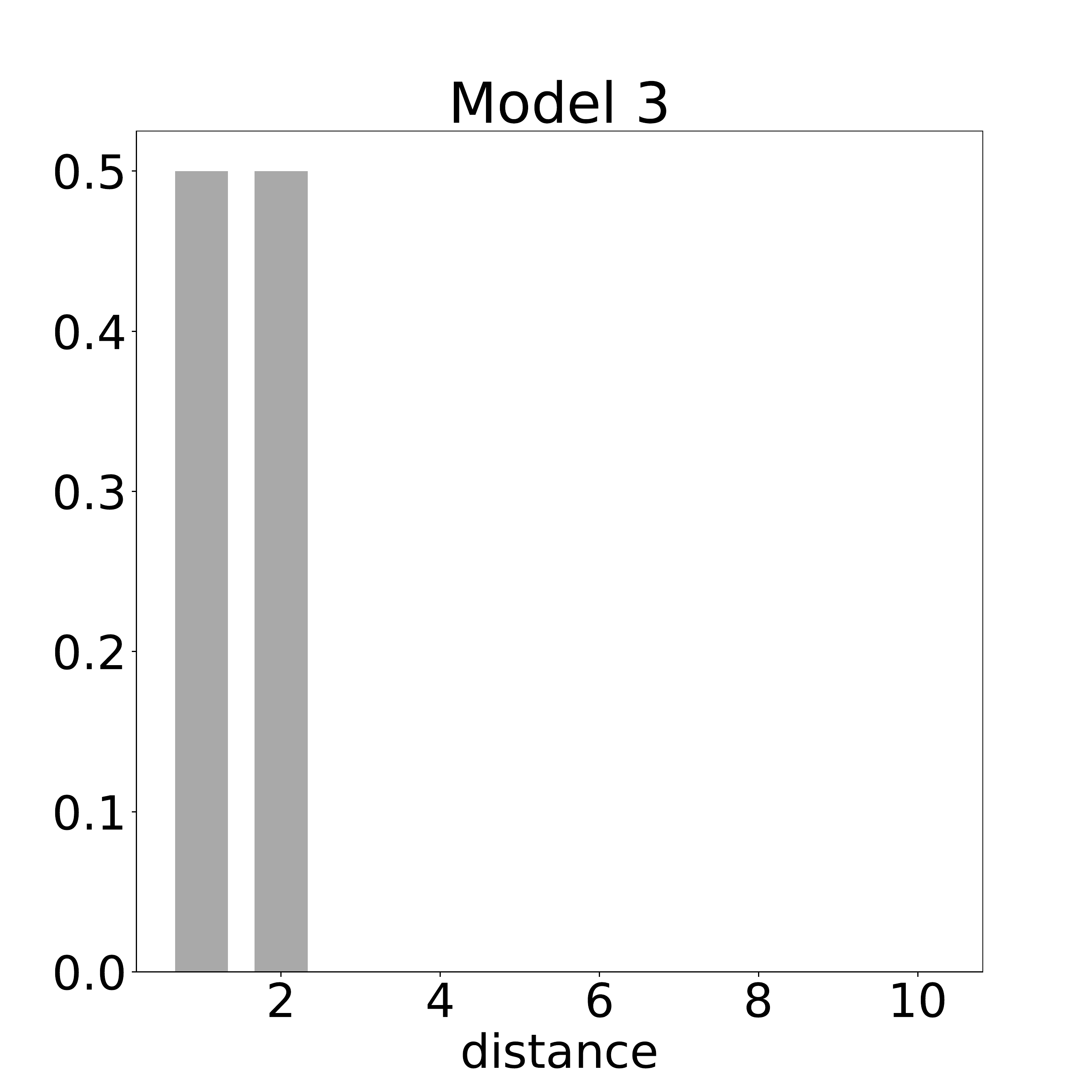}
\includegraphics[scale=0.17]{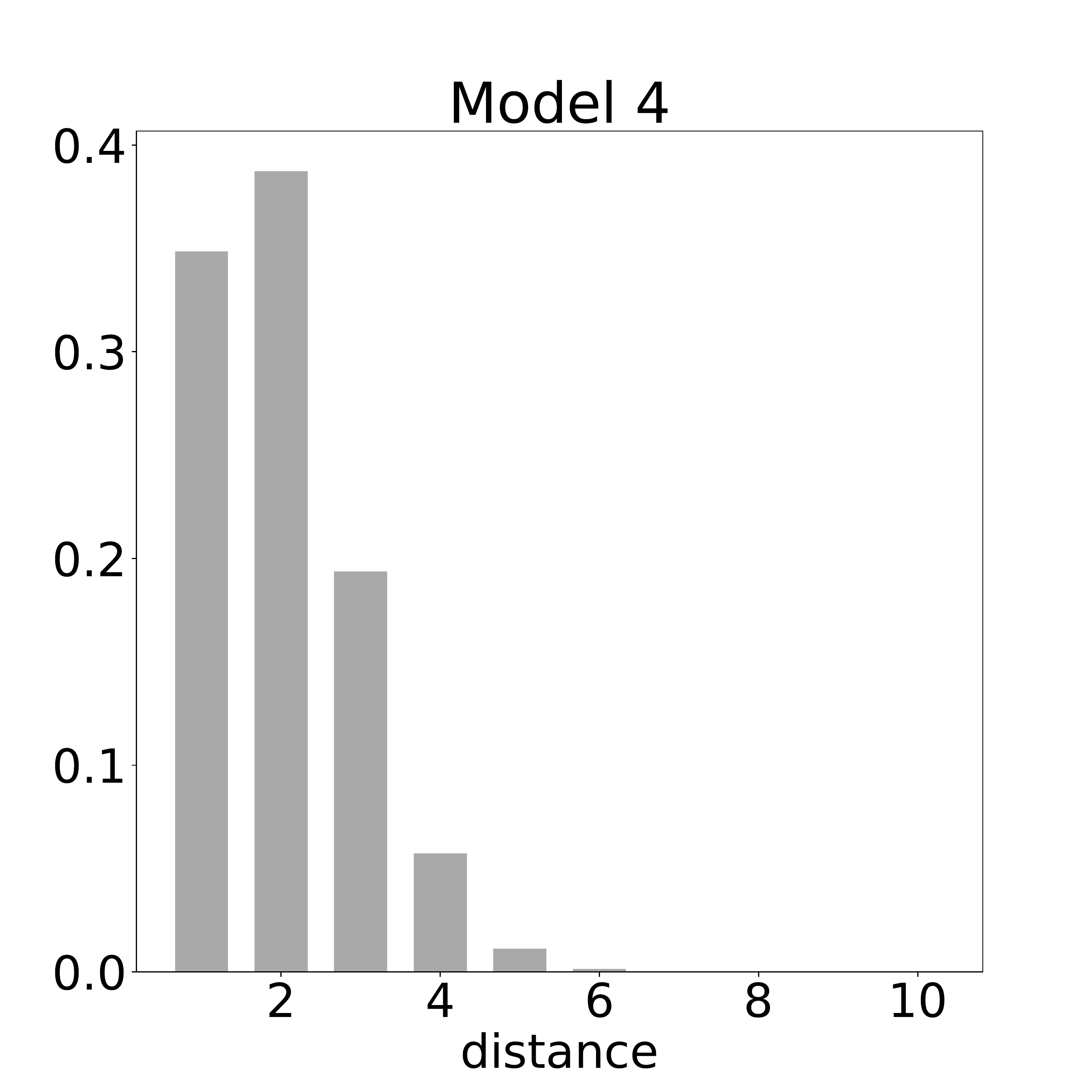}
\end{center}
\caption{Illustrations of the probability distributions for customer-facility distance values in all four of our models (from Model 1 in the top left corner, to Model 4 in the bottom right corner).}
\end{figure*}

\subsubsection{Model 2: Binary Facility Cost, Bimodal Distribution of Distances}

In the second model, we use a binary choice facility cost. Each facility costs either $1$ or $2$ units. The distances will follow a bimodal distribution. Similarly to the previous model, we will generate uniformly random integer values between $1$ and $10$. However, each value higher than $1$ will be overwritten by the maximum possible value. This generates instances with many high distances, but also several low distances between customers and facilities. The values within an instance will be the following:

\begin{equation}
f_i = random(1,2), ~~~ i = 1,...n,
\end{equation}

\begin{equation}
c_{ij} = \left\{ \matrix{ 1 & with~~probability~~0.1 \cr 10 & otherwise } \right., ~~~ i=1,...,n, ~~~ j=1,...,m.
\end{equation}

\subsubsection{Model 3: Binary Facility Cost, Binary Distances}

The facility cost structure in this model is the same as in the previous model. However, the distances will also be generated as $1$ or $2$. This will lead to a relative flat problem landscape. The aim will be to effectively search for a solution with as many facility cost and distance values equal to $1$ as possible.

\begin{equation}
f_i = random(1,2), ~~~ i = 1,...n,
\end{equation}

\begin{equation}
c_{ij} = random(1,2), ~~~ i=1,...,n, ~~~ j=1,...,m.
\end{equation}

\subsubsection{Model 4: Flat Facility Cost, Poissonian Distribution of Distances}

The last model is a modification of Model 1. The uniform facility cost will be used, with each facility costing a single unit. The distances will be taken from the Poissonian distribution $Po_{k}(\lambda)$, with $k$ trials and a probability of success $\lambda / k$ per trial. The distance will be equal to the number of successful trials, incremented by $1$, to avoid zero distance in case that all trials fail. This leads to a highly skewed distribution of distances. In our implementation, we choose $k=10$ and $\lambda=1$. The average distance will be $2$, while there will also be many distances equal to $1$ and also a few distance values of $3$ or more. The parameters of the model are the following:

\begin{equation}
f_i = 1, ~~~ i = 1,...n,
\end{equation}

\begin{equation}
c_{ij} = 1 + Po_{k=10}(\lambda = 1), ~~~ i=1,...,n, ~~~ j=1,...,m.
\end{equation}

\subsection{Local Search Algorithms}

We use two local search algorithms that have been explored in our previous study on a similar large-scale model of customer service centre applications \cite{OR-027}. The algorithms have been chosen for their simplicity and scalability, as well as the fact that neither of them requires extensive explicit or implicit parameter tuning. This makes them highly suitable for our investigation of the implications of the No Free Lunch theorem in real-world applications. The algorithms also have somewhat complementary qualities, as supported by our further experiments. In these experiments, the comparative performance of LS and RLS surprisingly depends on the coefficients within the problem instance.

The first one is a systematic local search (LS) algorithm that, at each time step, chooses the move that leads to the largest drop in the objective value. The second algorithm is a randomised local search (RLS), which attempts opening or closing a single randomly chosen facility in each time step and accepts the move if it does not lead to a worsening. It is worth noting that this type of algorithms is well-studied in evolutionary computation theory \cite{eacomplexity} and have been used to solve other optimisation problems, such as searching for bottlenecks in complex networks \cite{CSI-0076}.

\paragraph{Local search (LS)} The algorithm will search in space of bit strings, i.e. in $\{0,1\}^n$. Each bit $y_i$ determines whether facility $i$ is open or closed, similarly to the ILP formulation of the problem. The algorithm starts with all facilities open, i.e. with a bit string consisting solely of $1$-bits. Each customer is then assigned to the closest facility. Let $y$ be the current bit string and let $y'(i)$ be the bit string obtained by flipping bit $y_i$ in $y$. Then, LS chooses $i$ such that the objective value of $y'(i)$ is minimised within the neighbourhood. This leads to a steepest descent behaviour of the algorithm. In our investigations, LS will be terminated after $n$ iterations, where $n$ is the number of candidate facilities.

\paragraph{Randomised local search (RLS)} This algorithm also starts with all facilities open. Let $y$ be the current bit string and let $y'(i)$ be the bit string obtained by flipping bit $y_i$ in $y$. Then, RLS chooses $1 \leq i \leq n$ at random in each time step. The new bit string $y'(i)$ is accepted if the objective value of $y'(i)$ is not higher than the objective value of $y$. In terms of the number of bit flips attempted, each iteration of LS corresponds to $n$ iterations of RLS. RLS will therefore be terminated after $n^2$ iterations in our further experiments.

The simplicity of both LS and RLS will enable us to investigate the intricate interplay between the algorithm design and numerical properties of problem instance. This impacts both the choice of efficient local search and design of hybrid algorithms for a particular cost structure. One can intuitively expect this to be inherently tied to the properties of a particular real-world application. It is also worth noting that both algorithms can be implemented efficiently using lookups in precomputed data structures that can be used to recalculate the objective value for a potential bit flip. This requires storing the lists of customers assigned to each facility, as well as the mapping of customers to particular facilities. Such implementation options allow both algorithms to scale well to quite large problem instances \cite{OR-027}.

\section{Experimental Results and Discussion}

In this section, we present the experimental results obtained. We first describe the experimental settings and then present the results for the four problem instance models. This is followed by a brief discussion and the potential impact of these findings on future investigations and designs of algorithms for the facility location problem.

\subsection{Experimental Settings}

For each of the facility cost and distance models, we generated 10 instances and ran LS and RLS 1000 times for each instance. We analyse the results from the perspectives of both the ability to estimate the distribution of optima, as well as the ability to sample high-quality solutions in ``lucky runs''.

To determine actual optima or near-optimal solution we use the CBC mixed-integer programming branch-and-cut solver from the COIN-OR package \cite{BonamiAlgorithmicMixedIntegerPrograms,LinderothMilp}. We use a precompiled 64-bit Windows binary of CBC compiled by the Intel 11.1 compiler. The time limit for CBC solution search was set to 24 hours per instance.

The experiments were performed on a machine with an Intel Core i7-6820 CPU @ 2.70 GHz, 32 GB RAM and Windows 10 operating system. LS and RLS were implemented in C++ using Qt, compiled by the MinGW 32-bit compiler.

Each run of LS was stopped after $n$ iterations, as in each iteration, a flip of each bit separately is attempted. For RLS, $n^2$ local search iterations were allowed, since a single bit is always flipped in each iteration.

\subsection{Results for the Four Facility Cost and Distance Models}

The experimental results are presented for each model separately using box-whisker plots to illustrate the distribution of solutions found by LS and RLS. For smaller instances, the distribution of optimal or near-optimal solutions and lower bounds will also be presented. This will allow us to estimate the actual approximation capabilities of LS and RLS in these costing and distance models. In addition, we will investigate the growth of computational needs to solve the problem exactly in all of the models studied. The results will then be discussed and all the techniques investigated will be compared.

\subsubsection{Model 1: Flat Facility Cost, Moderately Varied Random Distances}

Figure 3 presents the box-whisker plots obtained for the instances generated by Model 1. For these instances, RLS outperforms LS. This can be explained by the possibility that closing or opening a facility, which leads to a higher objective value at the moment, could not necessarily lead to the best moves in the future. Such a possibility is linked to the randomised distance structure and the presence of potential local optima. CBC was able to find optima for instances with up to $60$ customers, as well as for most instances with $70$ customers. One can observe that the gap between the typical performances of LS and RLS tends to widen with growing instance size. However, this seems to hold also for the gap between the performance of RLS and the actual optima. This suggests that more advanced techniques could offer some improvement in this model, including tabu search \cite{sun2006solving}, population-based local search or evolutionary algorithms \cite{jaramillo2002use}.

\begin{figure*}
\begin{center}
\includegraphics[scale=0.13]{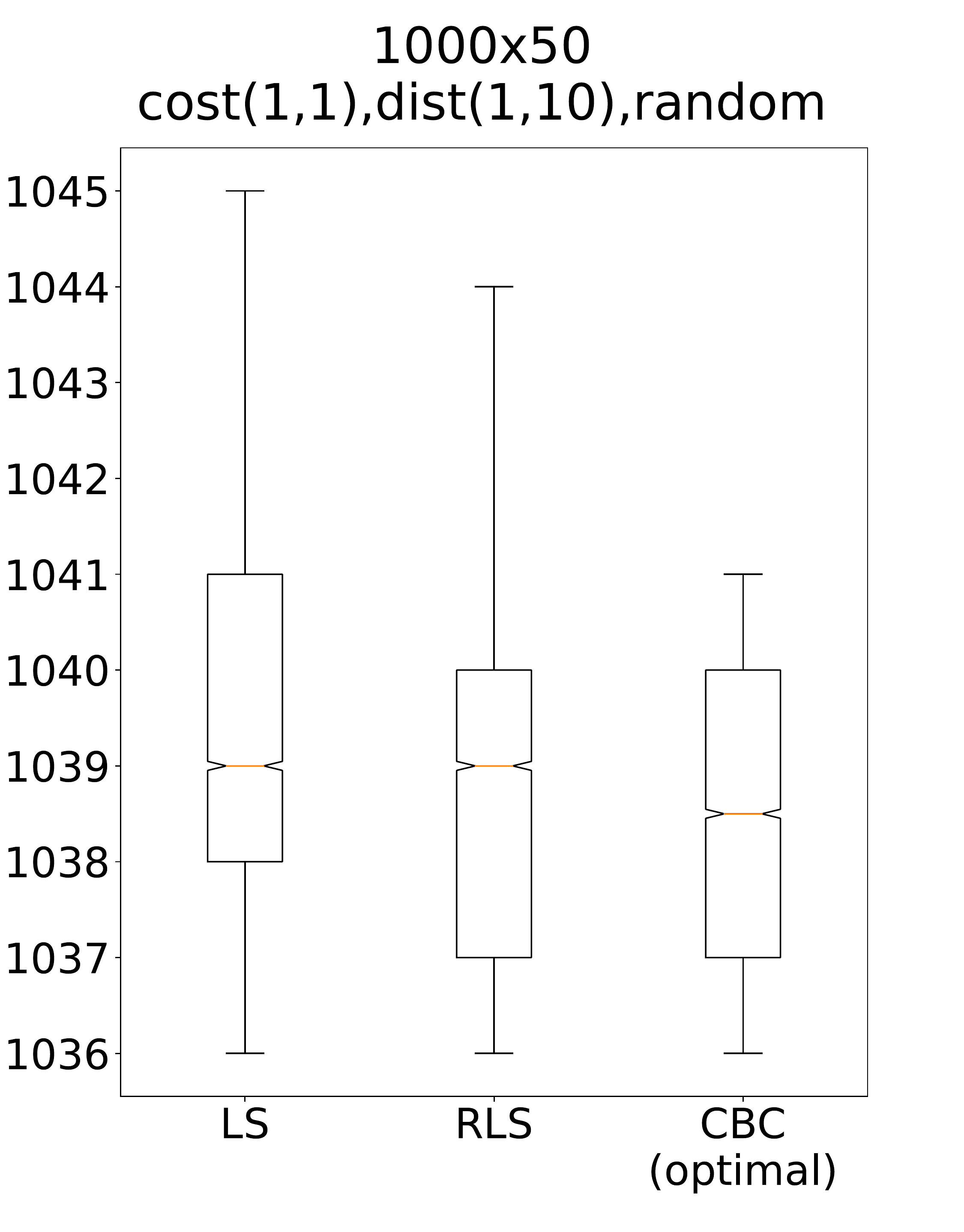}
\includegraphics[scale=0.13]{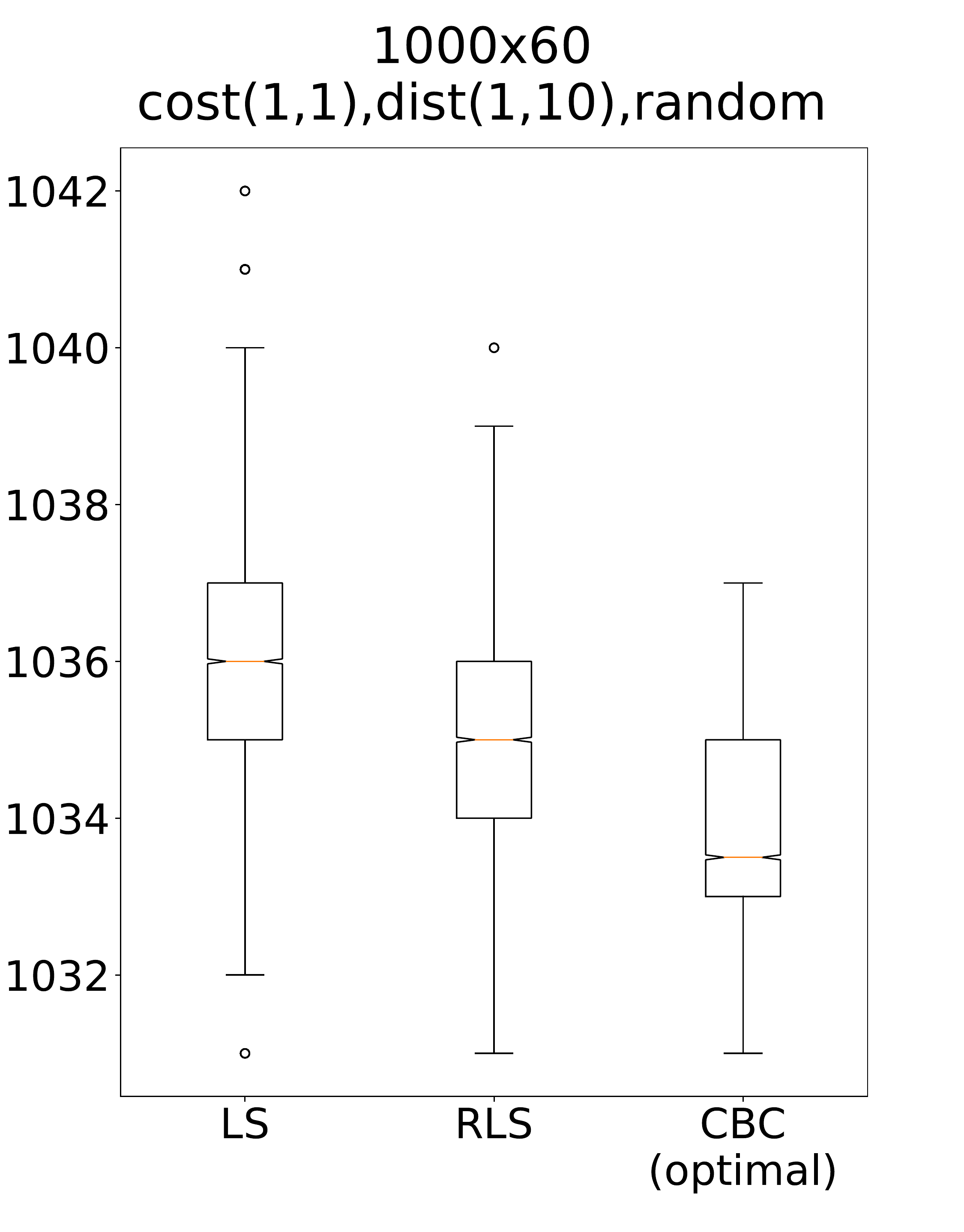}
\includegraphics[scale=0.13]{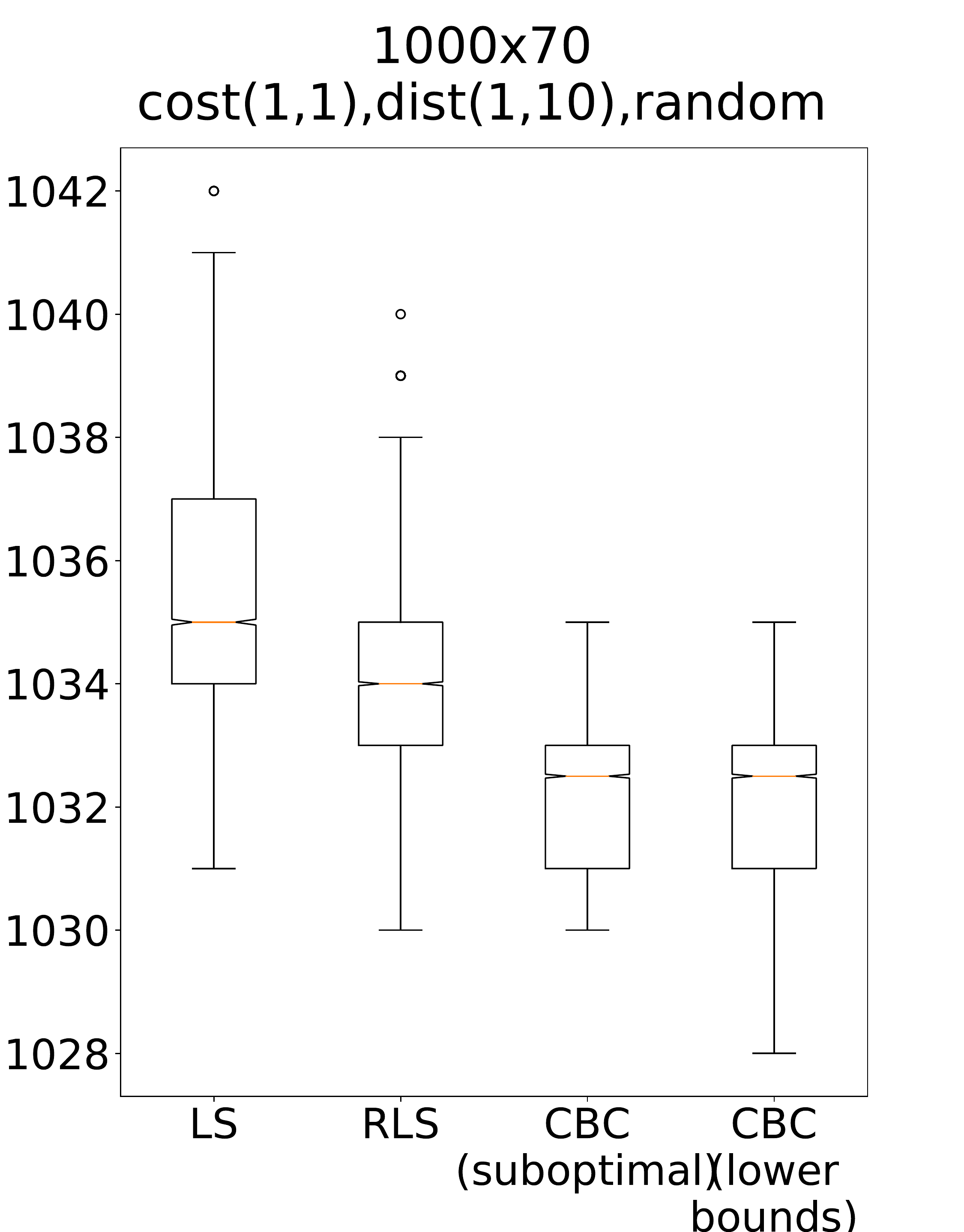}
\includegraphics[scale=0.13]{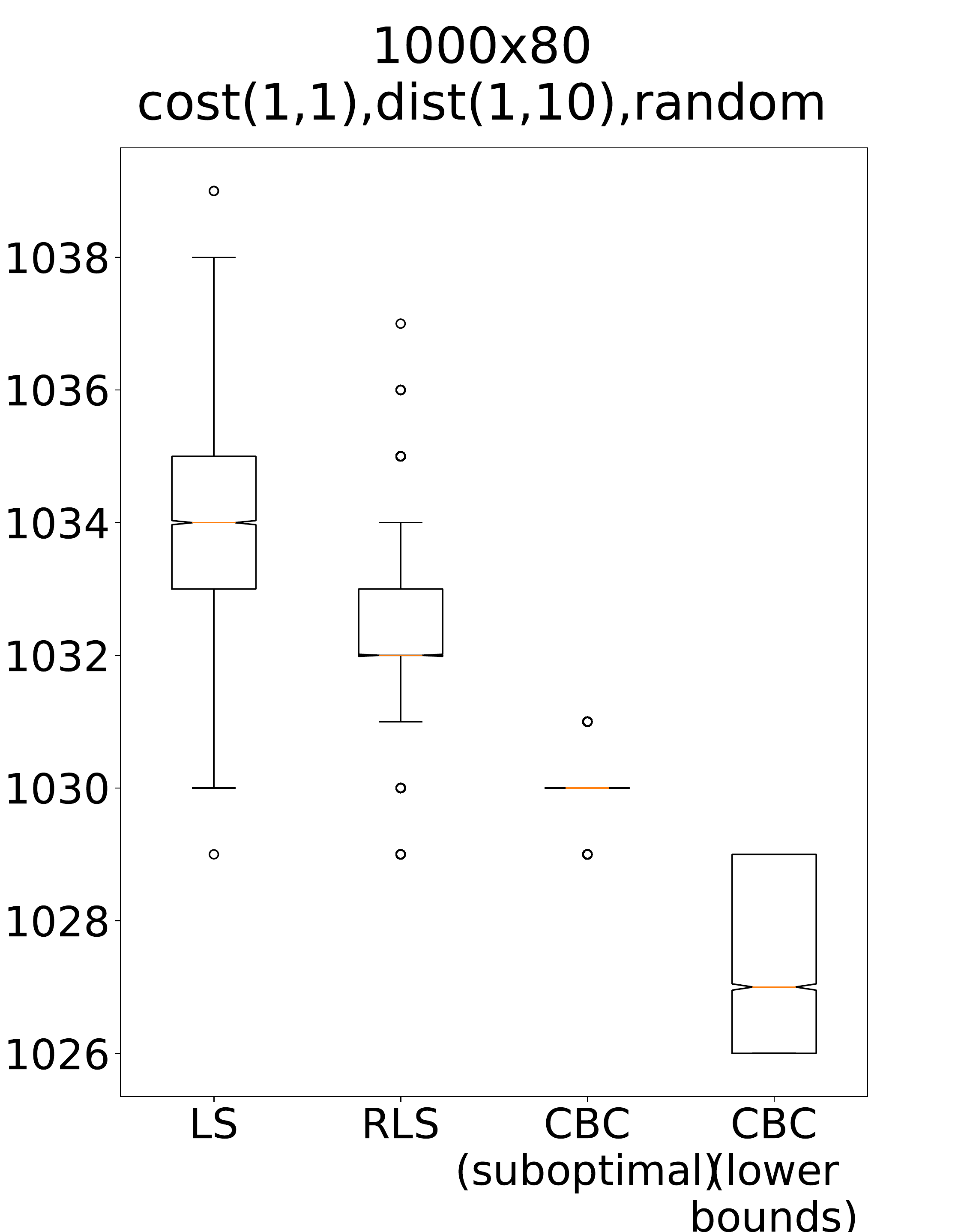}
\includegraphics[scale=0.13]{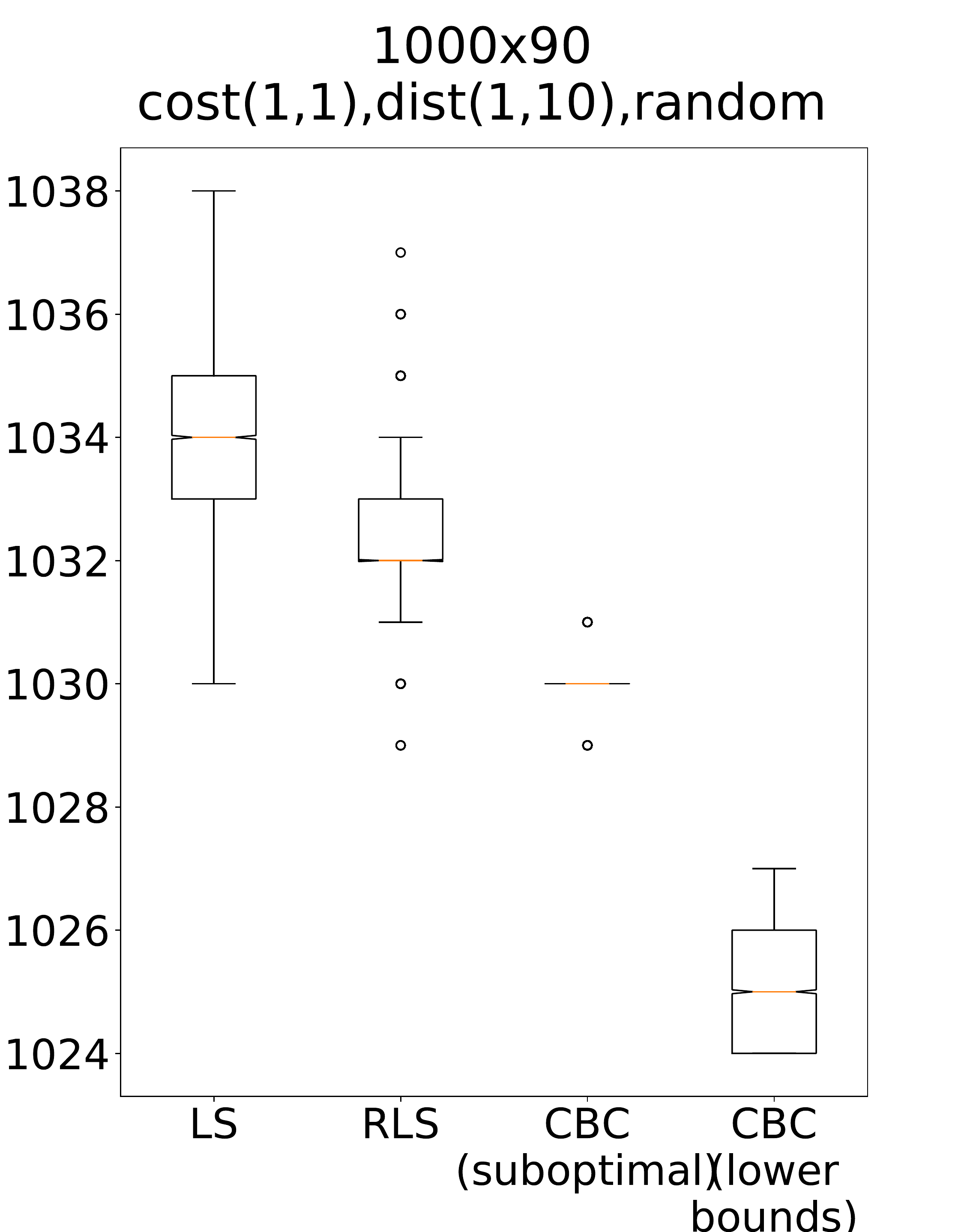}\\
\includegraphics[scale=0.13]{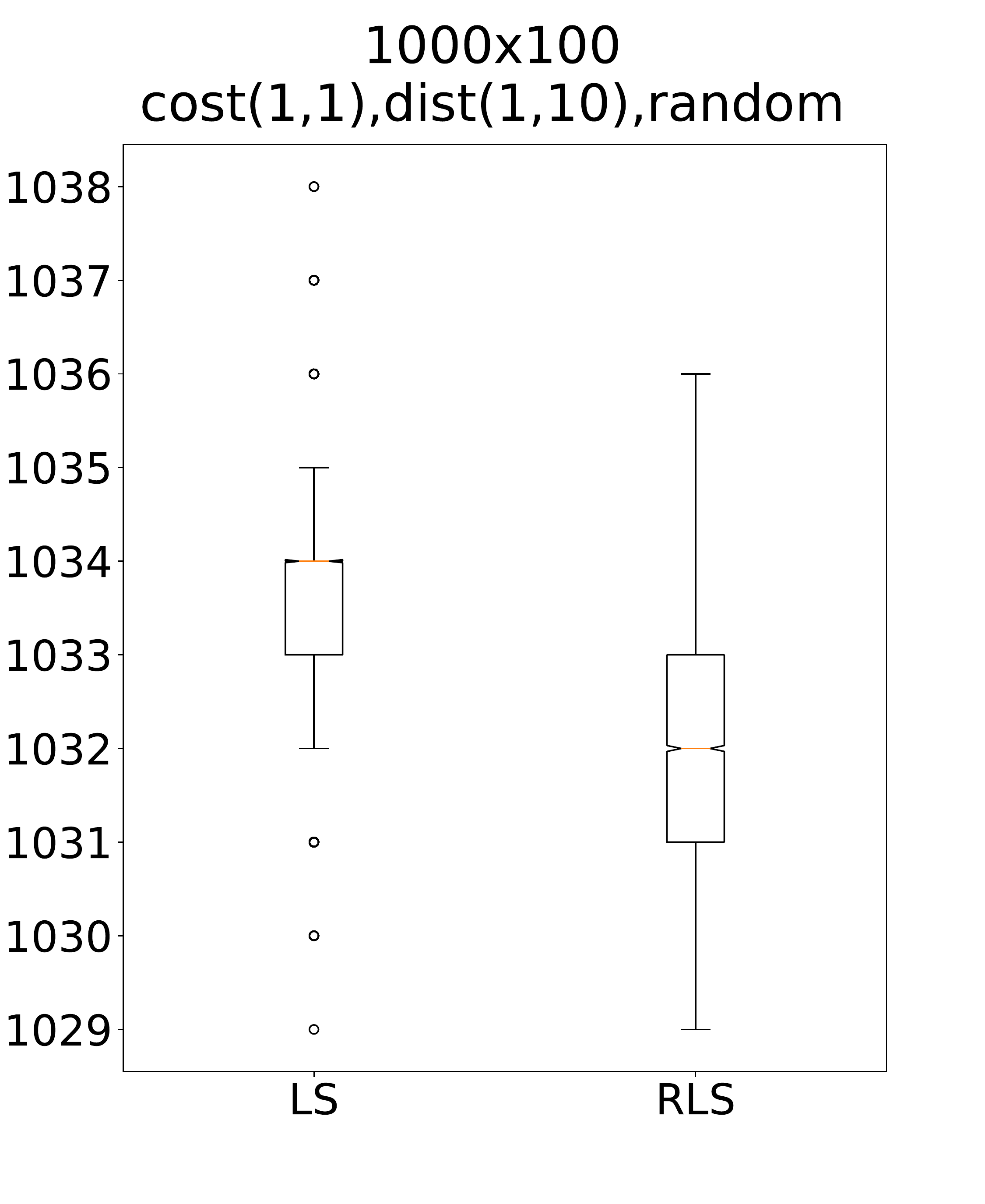}
\includegraphics[scale=0.13]{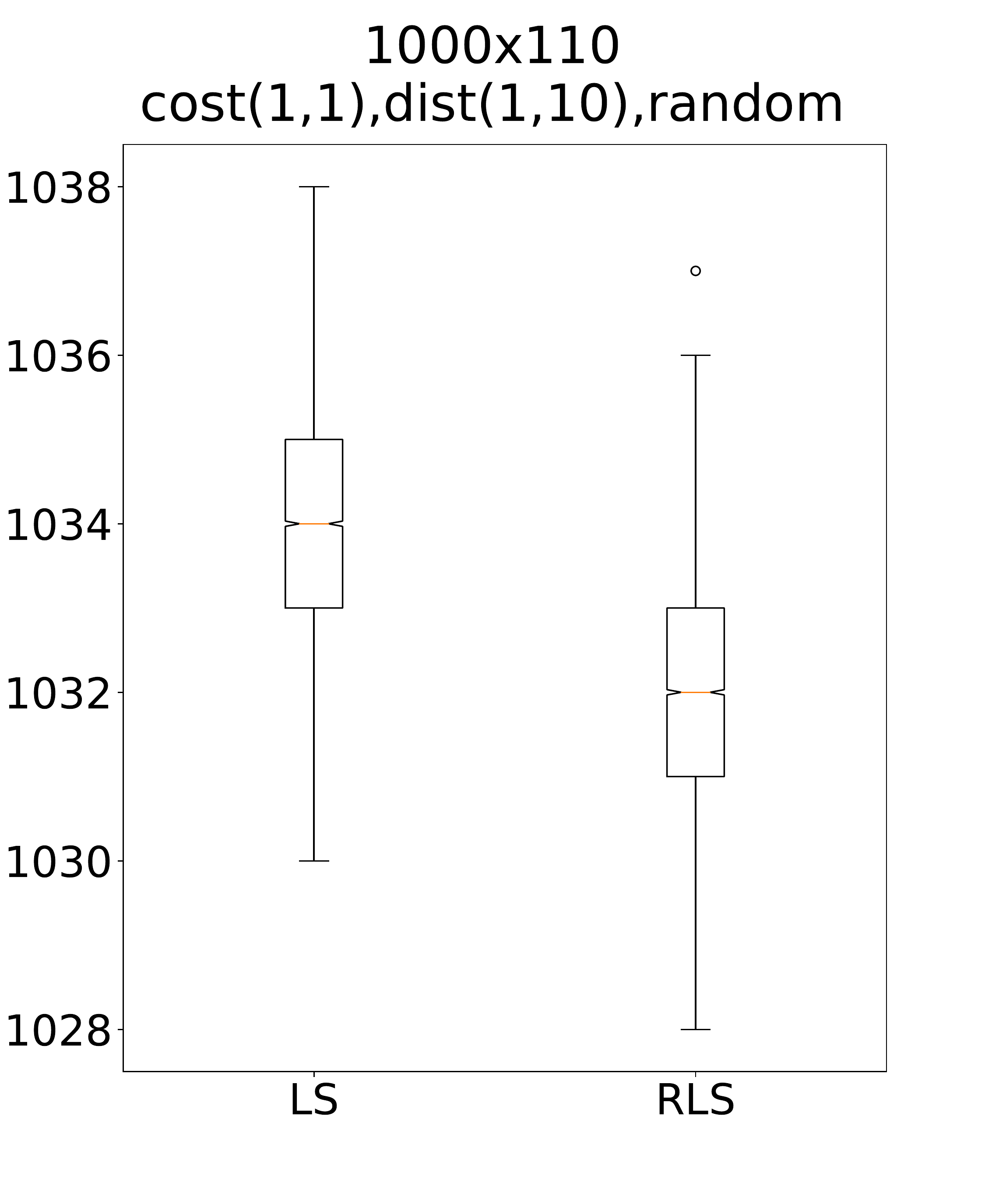}
\includegraphics[scale=0.13]{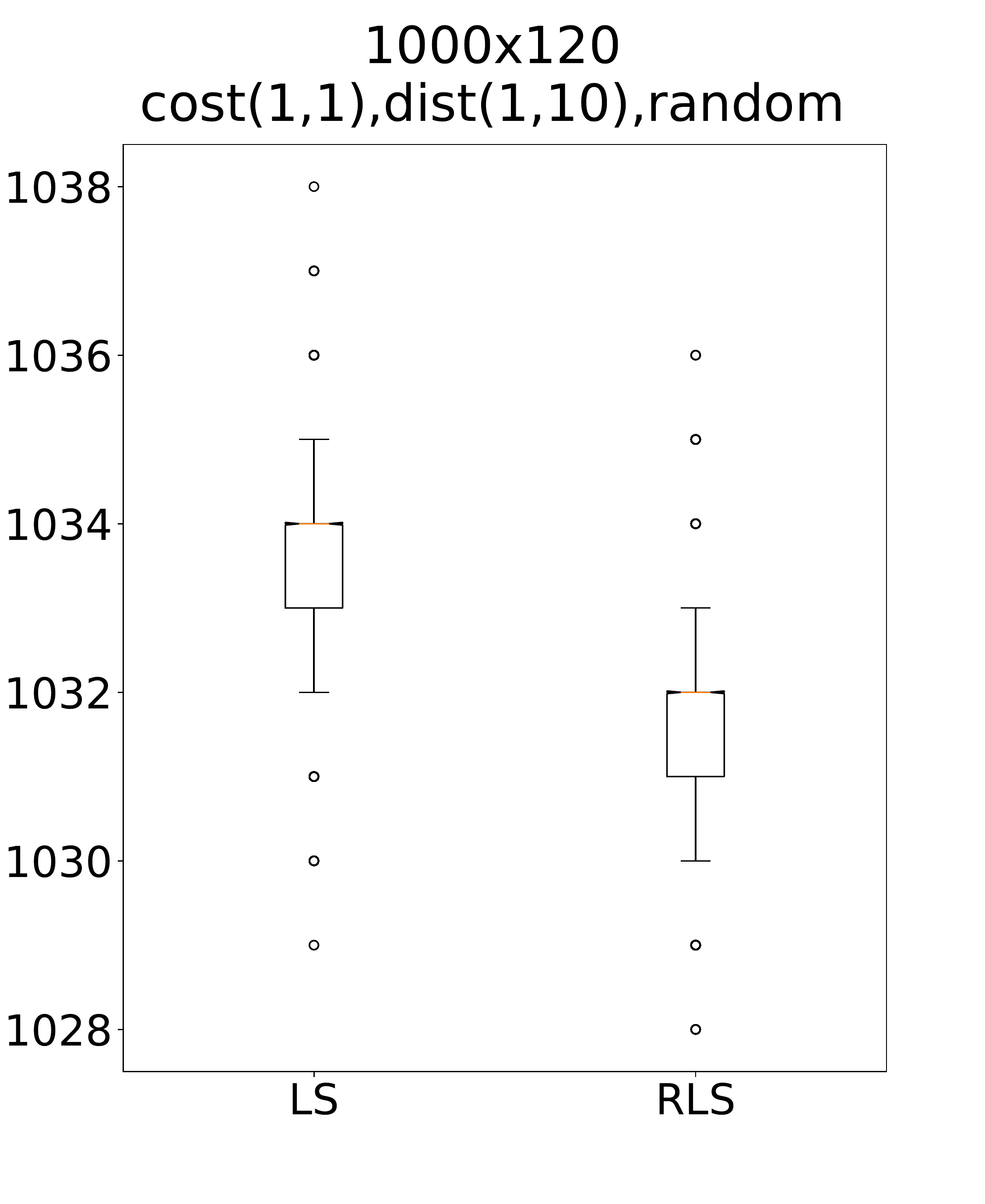}
\includegraphics[scale=0.13]{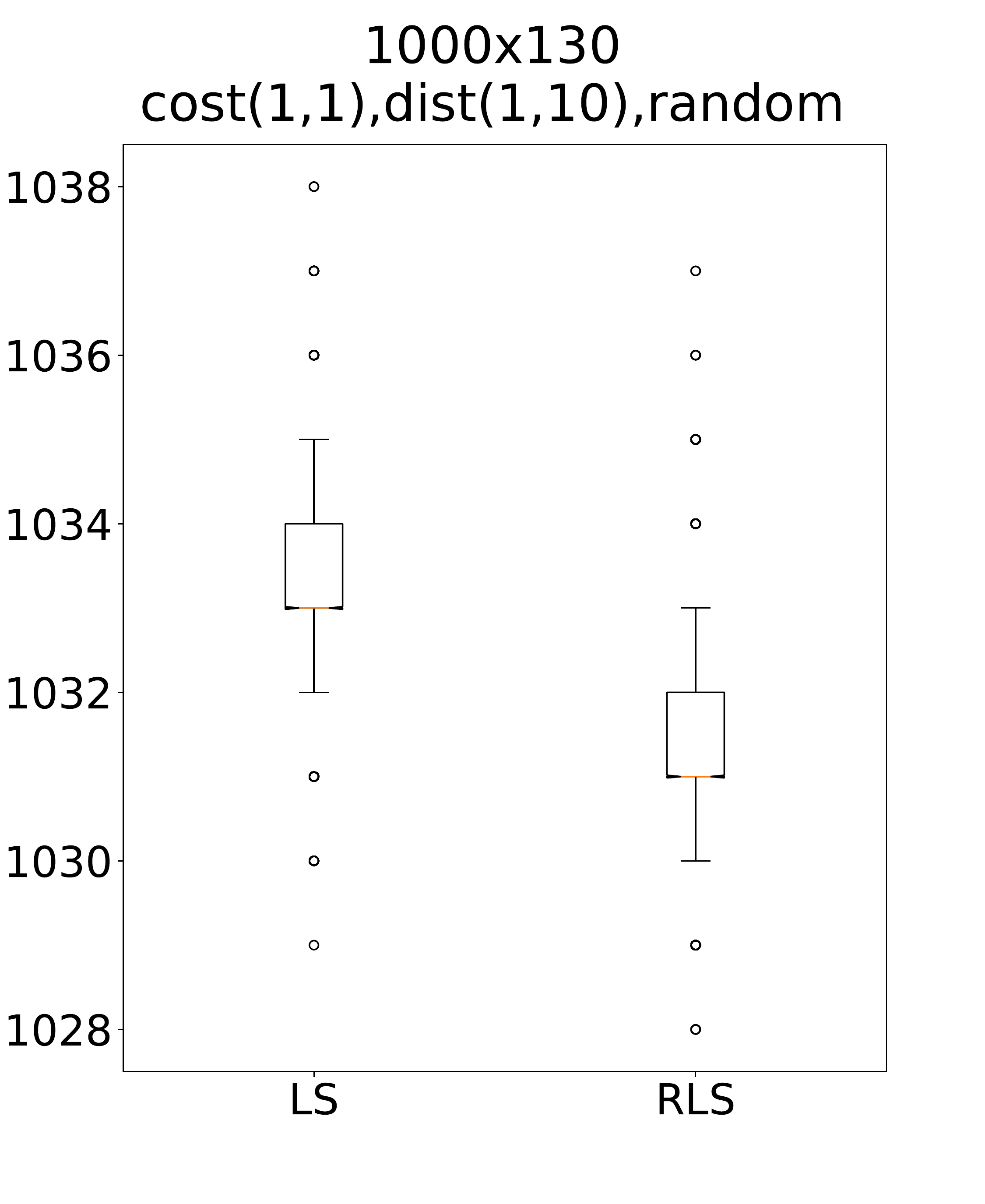}
\includegraphics[scale=0.13]{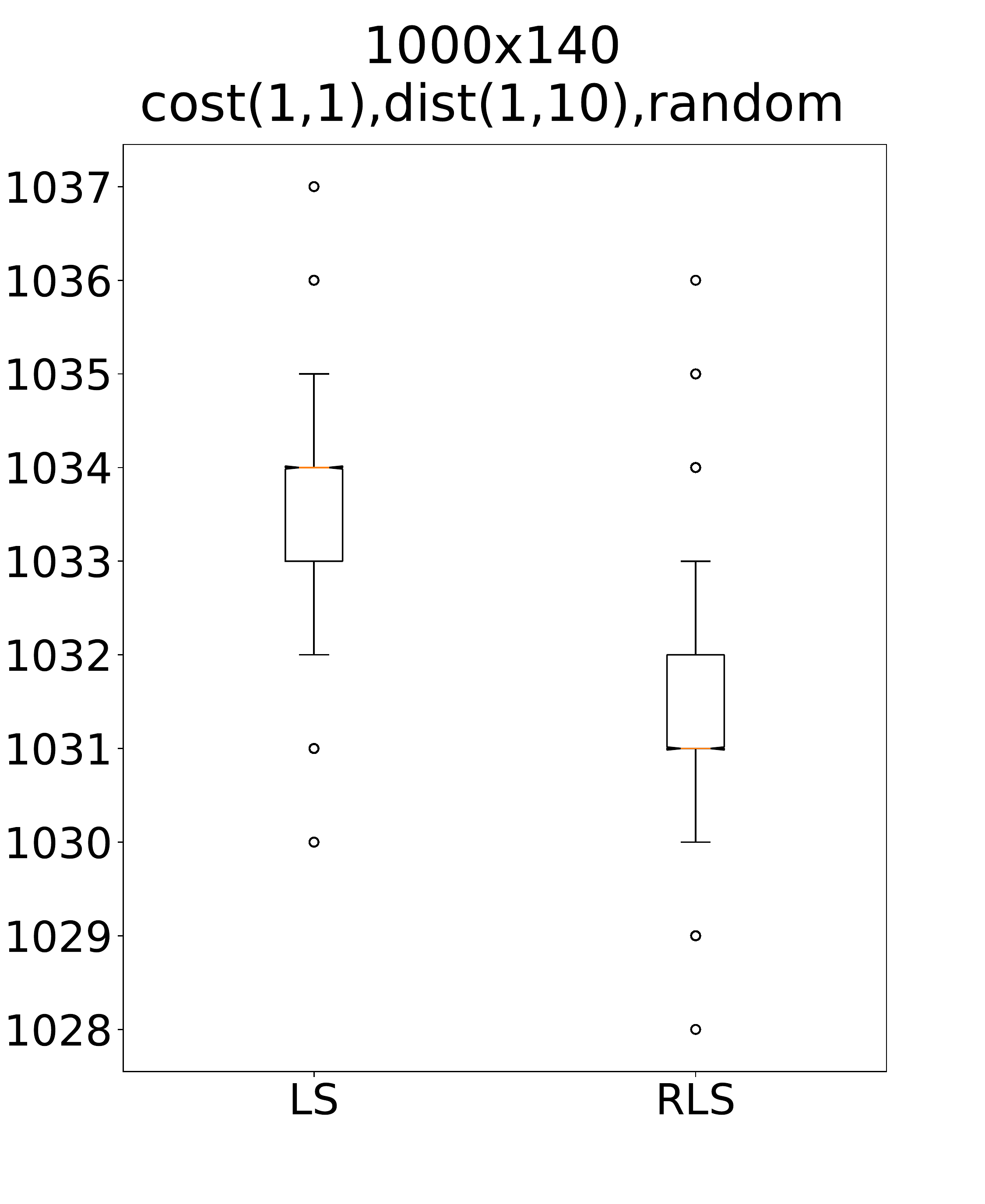}
\end{center}
\caption{Box-whisker plots obtained for $10$ problem instances generated according to Model 1. LS and RLS were used $1000$ times per instance, while the results for CBC represent optimal or near-optimal reference solutions obtained by the corresponding ILP solving procedure.}
\end{figure*}

\begin{figure*}
\begin{center}
\includegraphics[scale=0.13]{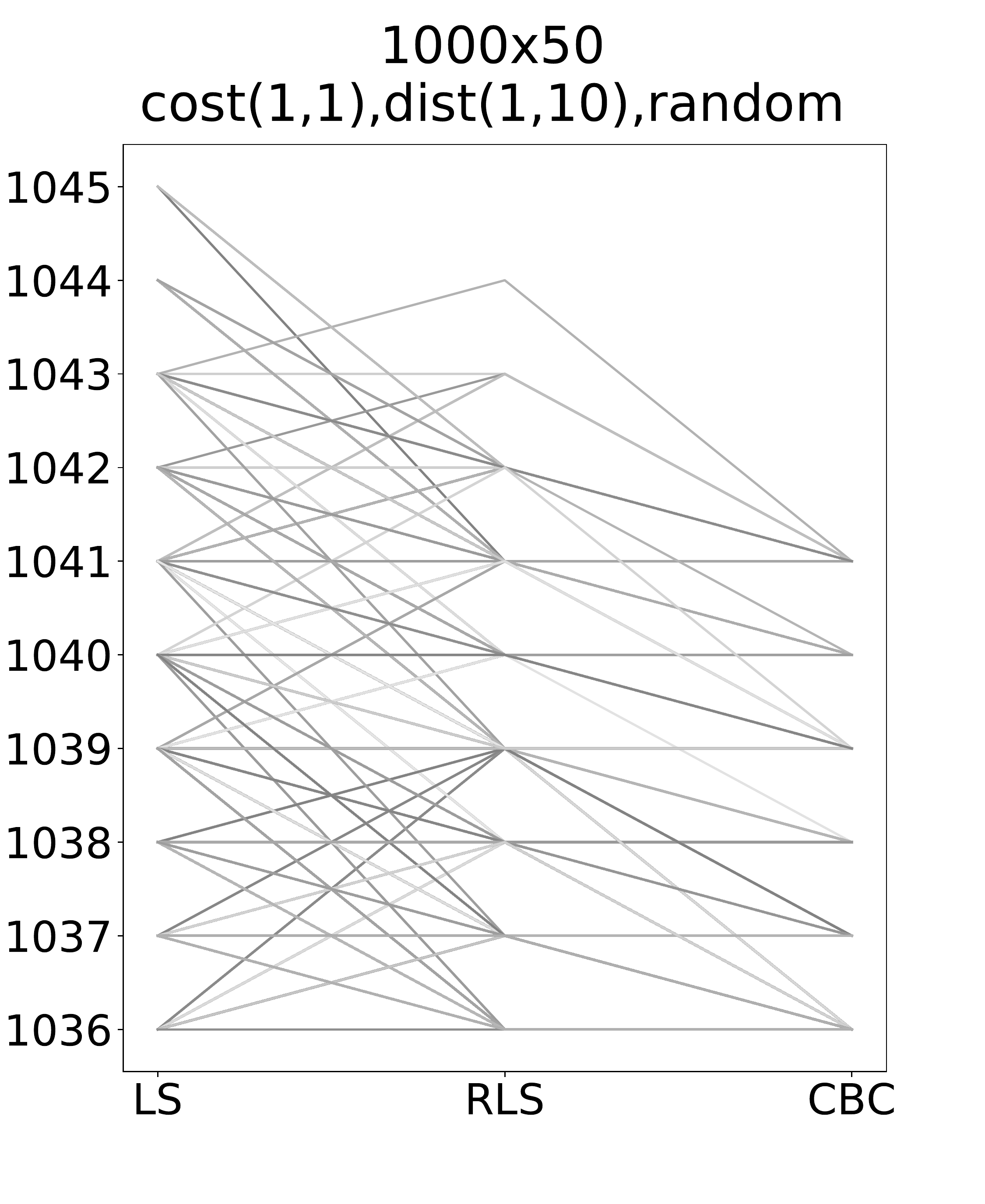}
\includegraphics[scale=0.13]{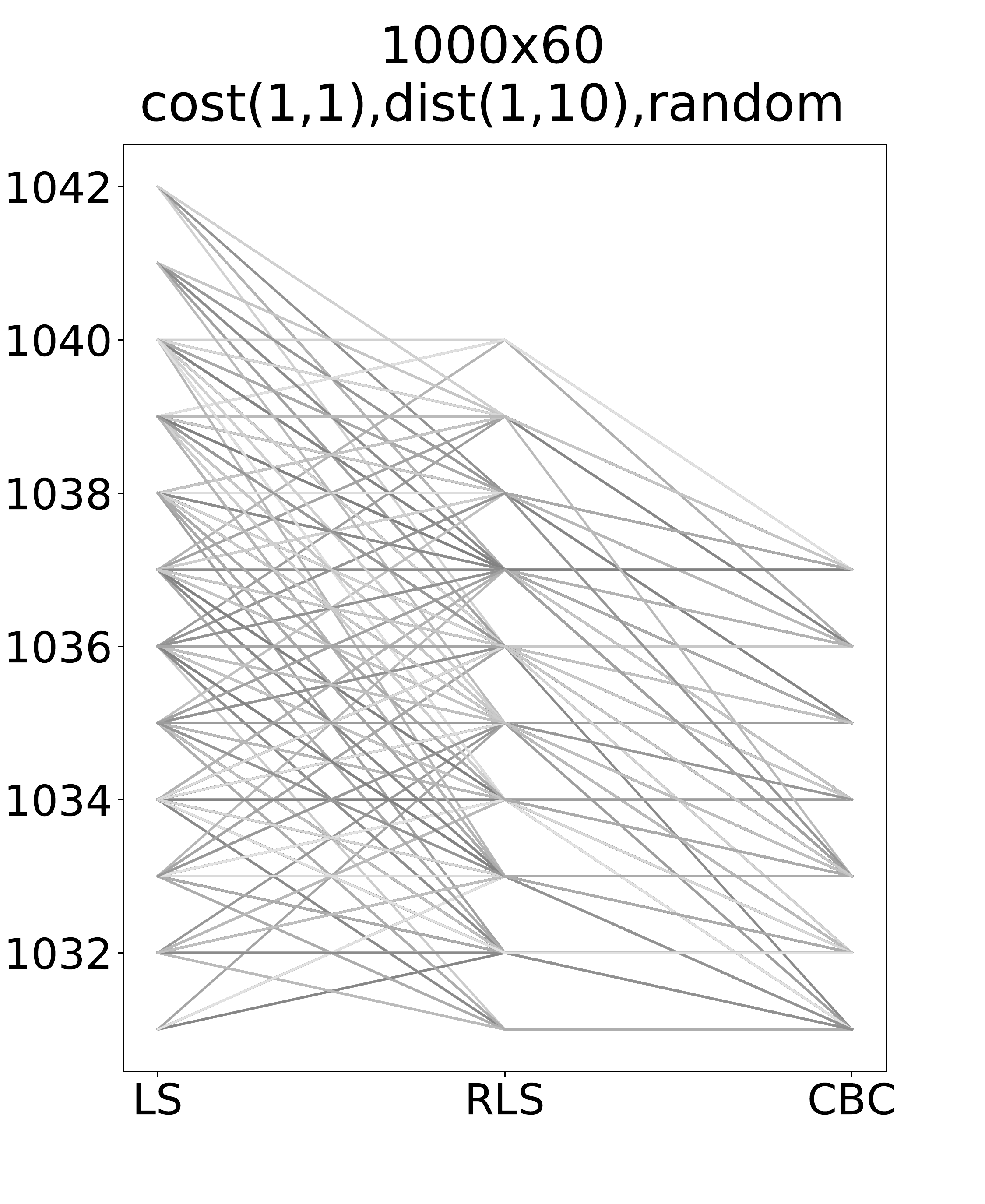}
\includegraphics[scale=0.13]{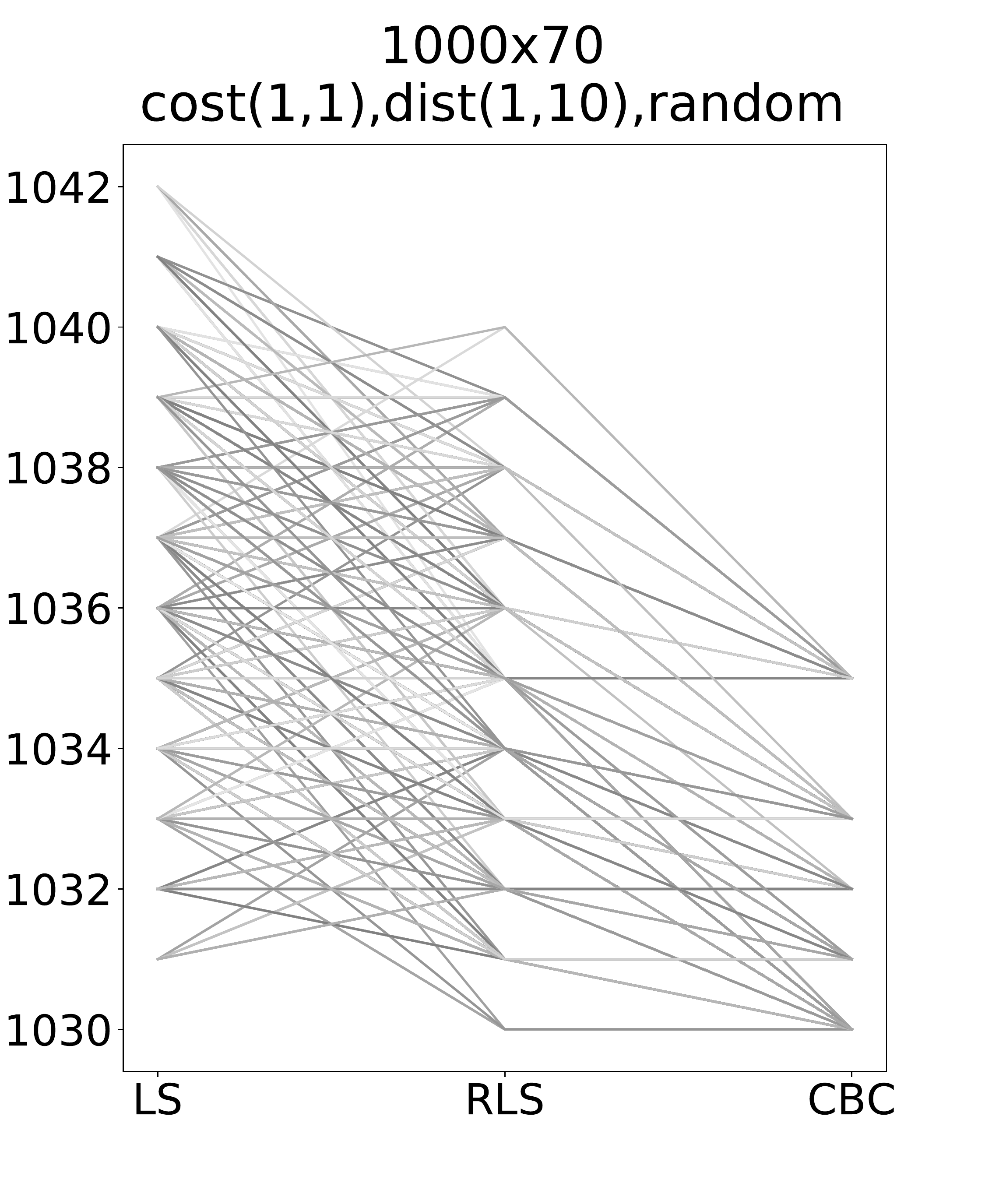}
\includegraphics[scale=0.13]{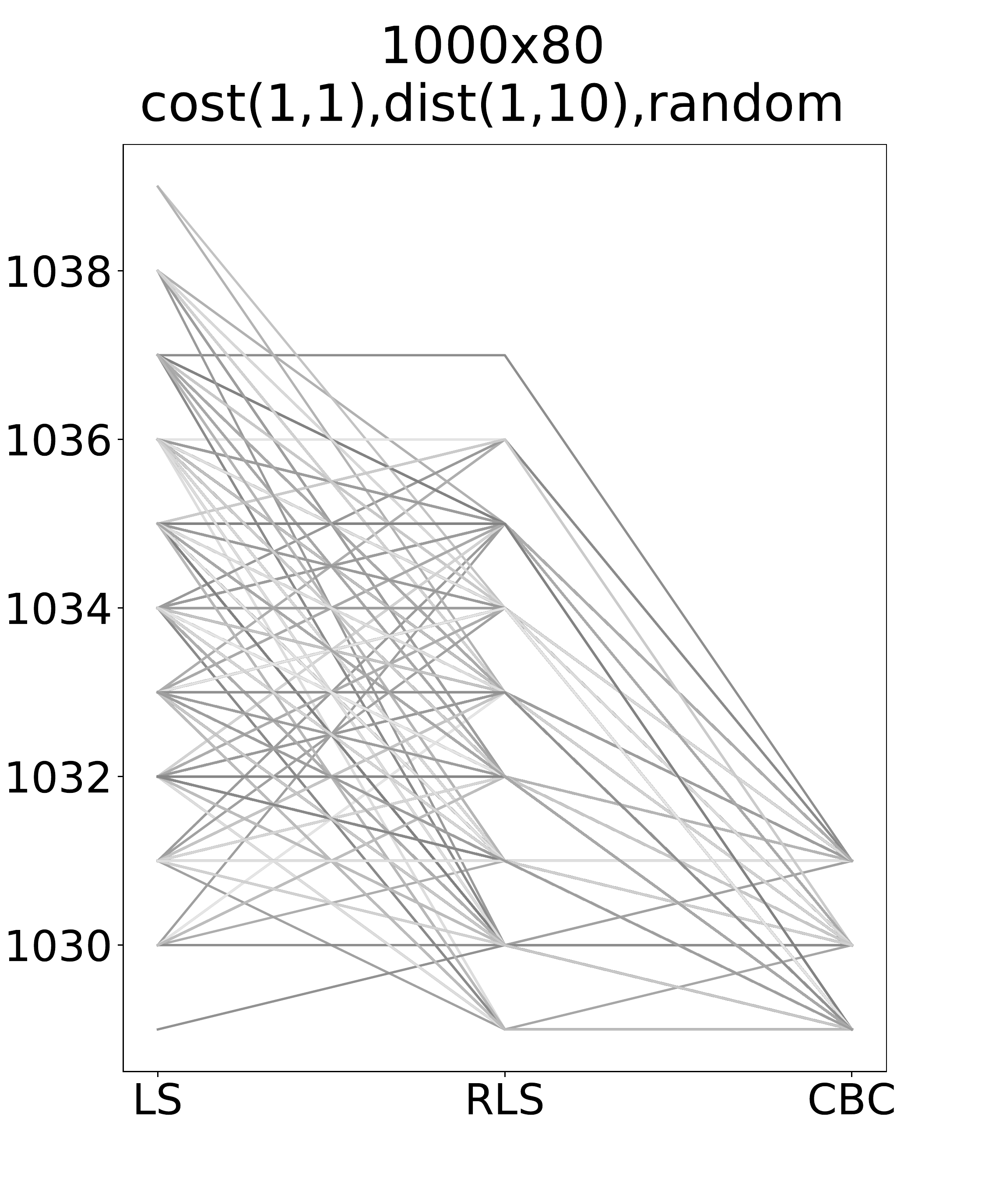}
\includegraphics[scale=0.13]{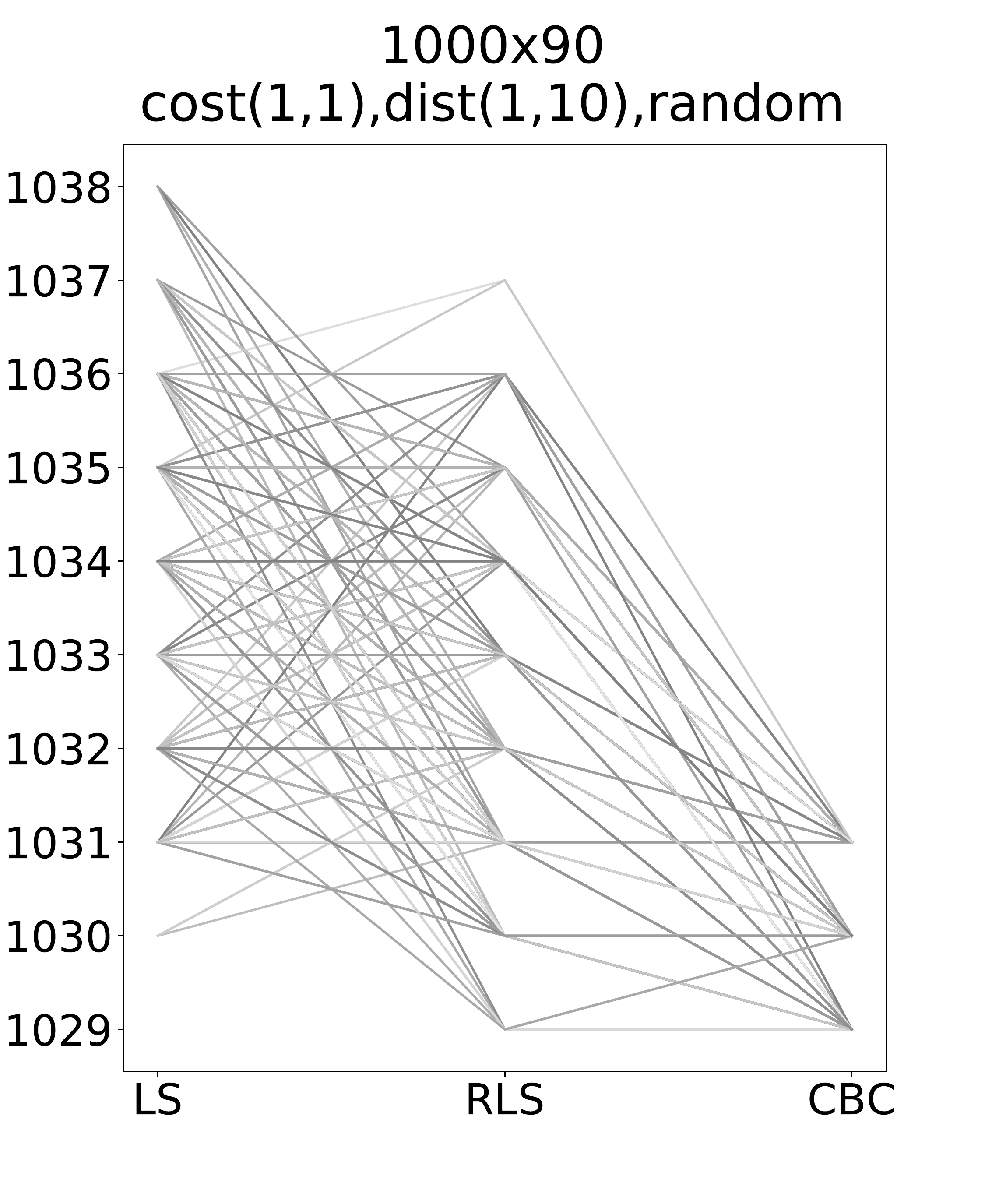}
\end{center}
\caption{A plot depicting the relation of the performance of LS, RLS and the actual optimal or near-optimal solutions found by CBC for Model 1. These results are grouped according to the problem instance. Each of the gray lines represents the objective values found by a run of LS, RLS and the objective value found by CBC.}	
\end{figure*}

\begin{figure*}
\begin{center}
\includegraphics[scale=0.13]{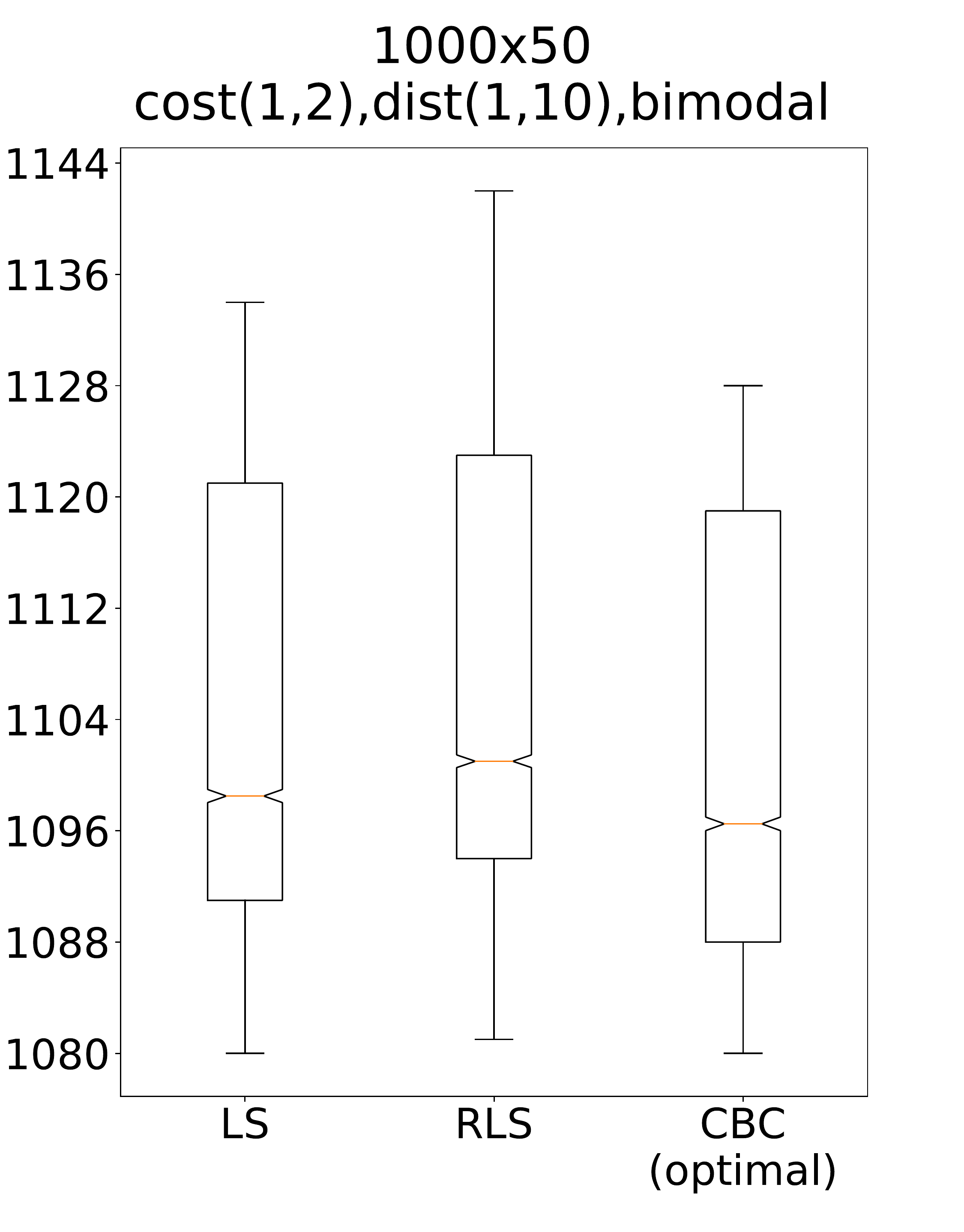}
\includegraphics[scale=0.13]{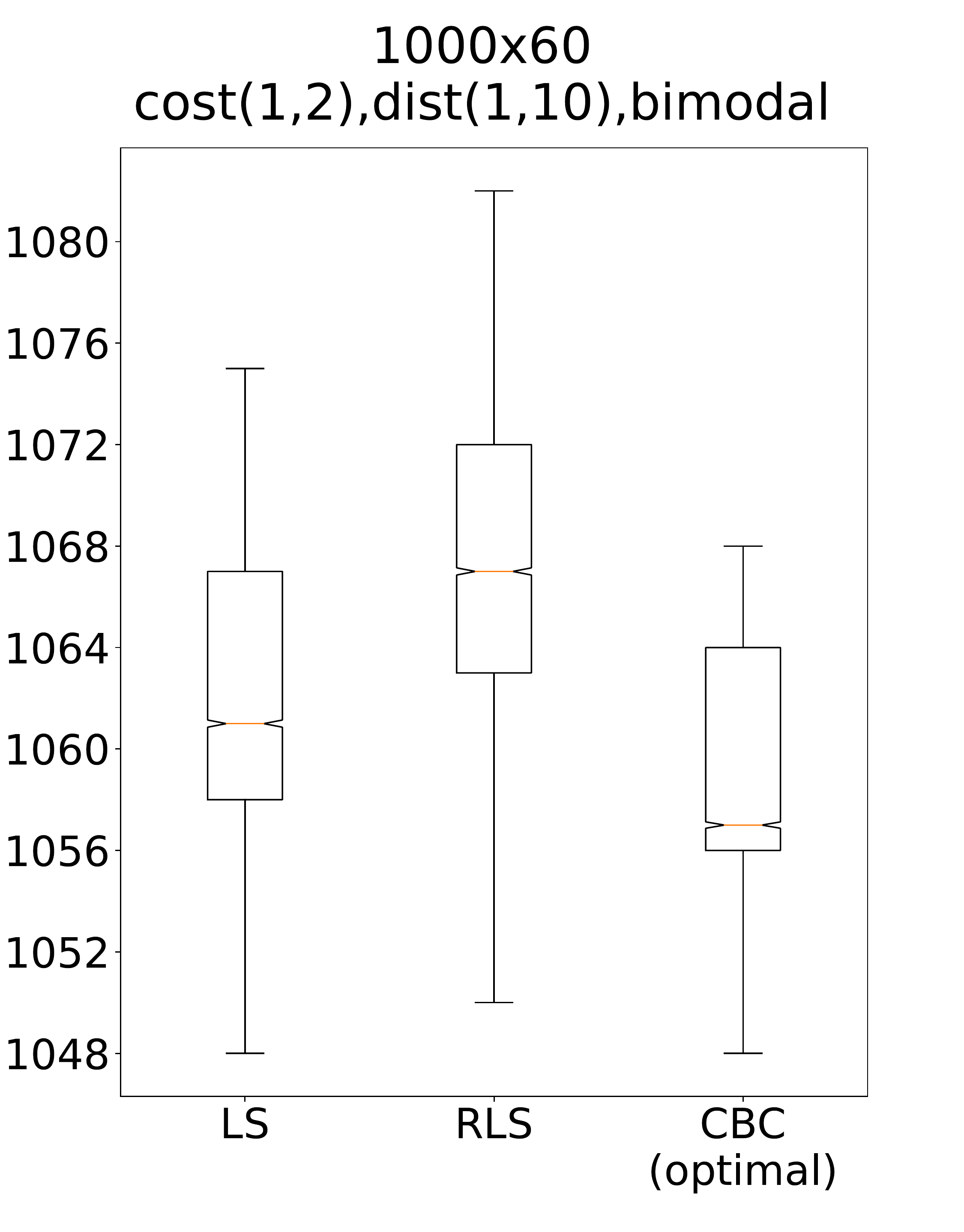}
\includegraphics[scale=0.13]{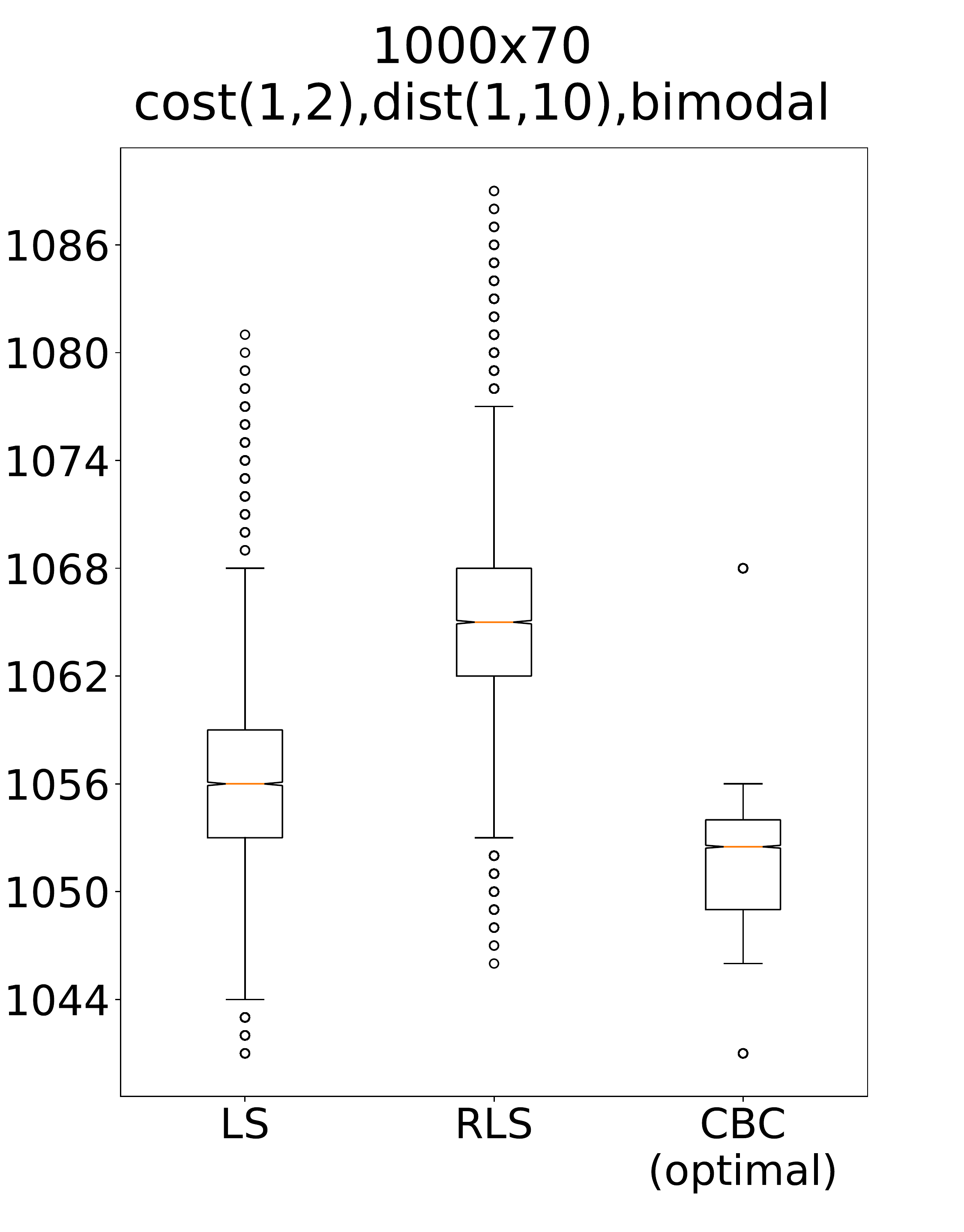}
\includegraphics[scale=0.13]{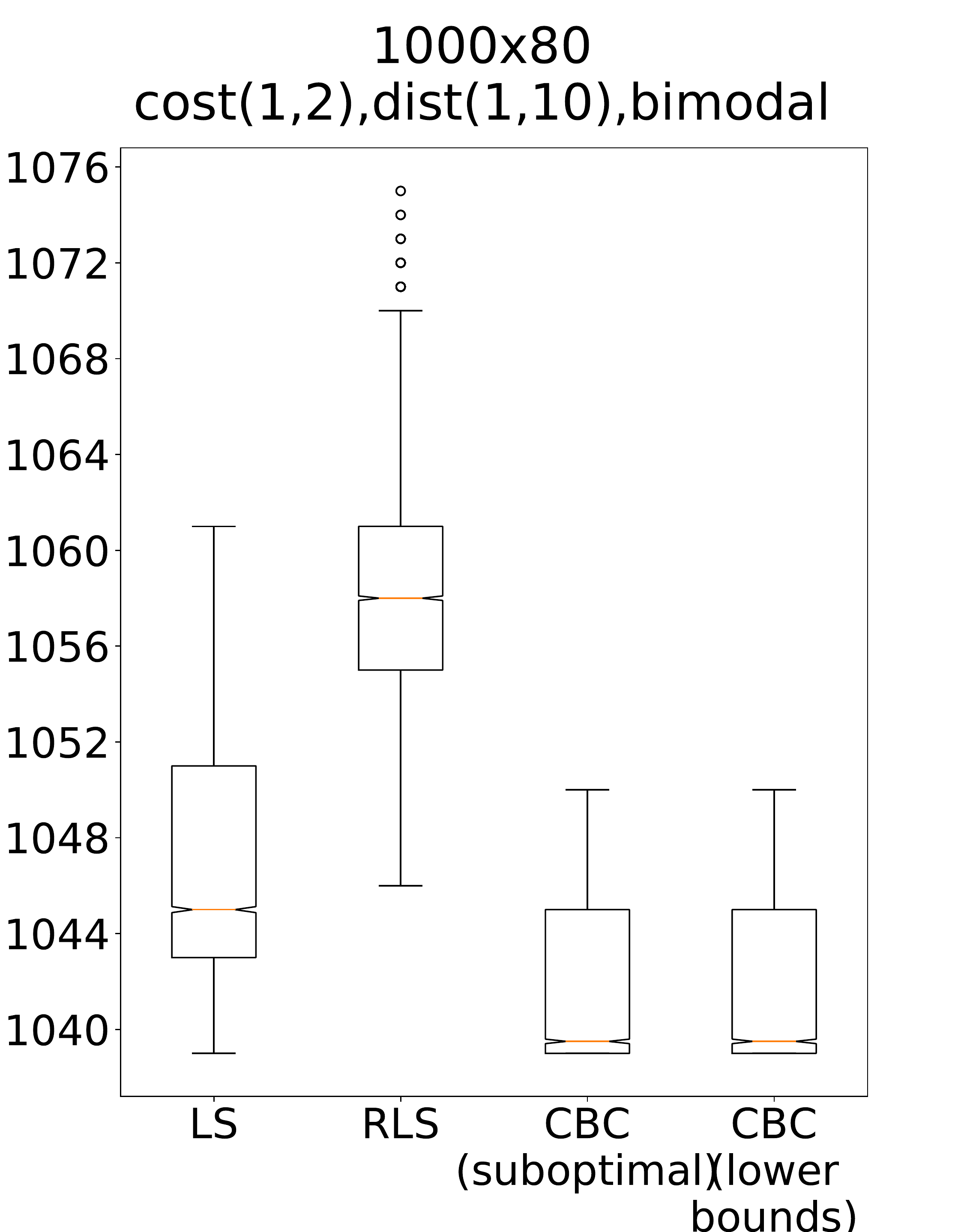}
\includegraphics[scale=0.13]{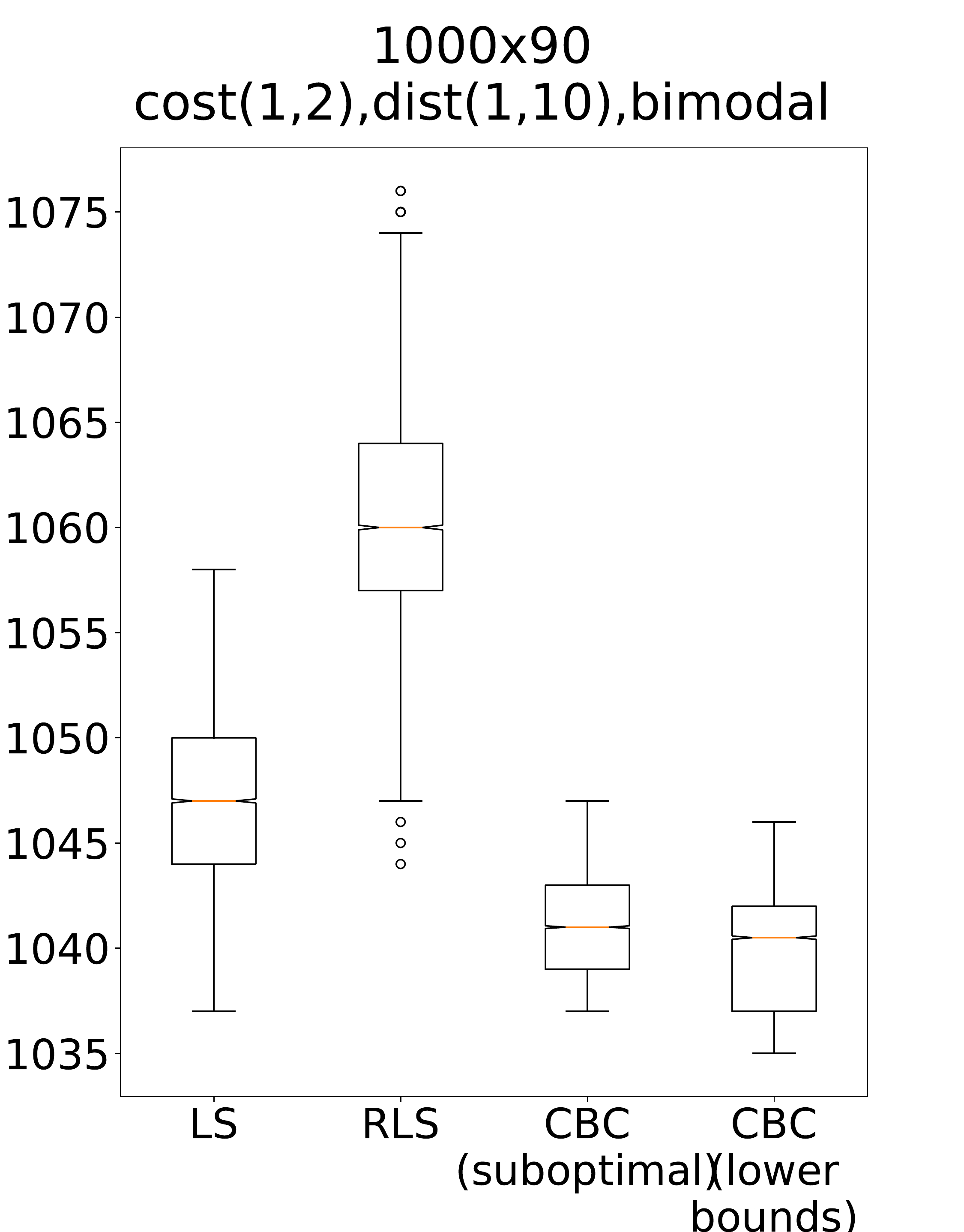}\\
\includegraphics[scale=0.13]{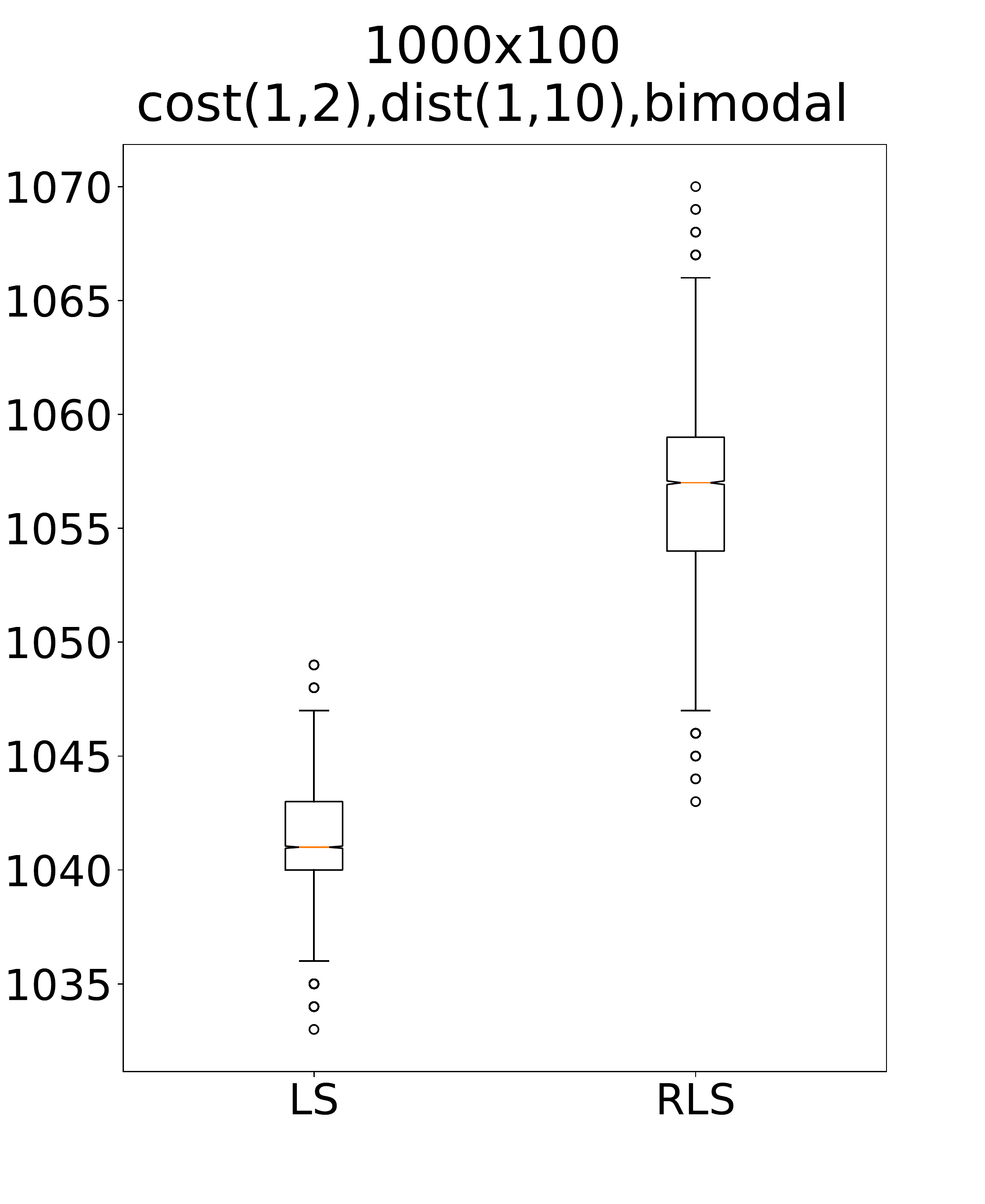}
\includegraphics[scale=0.13]{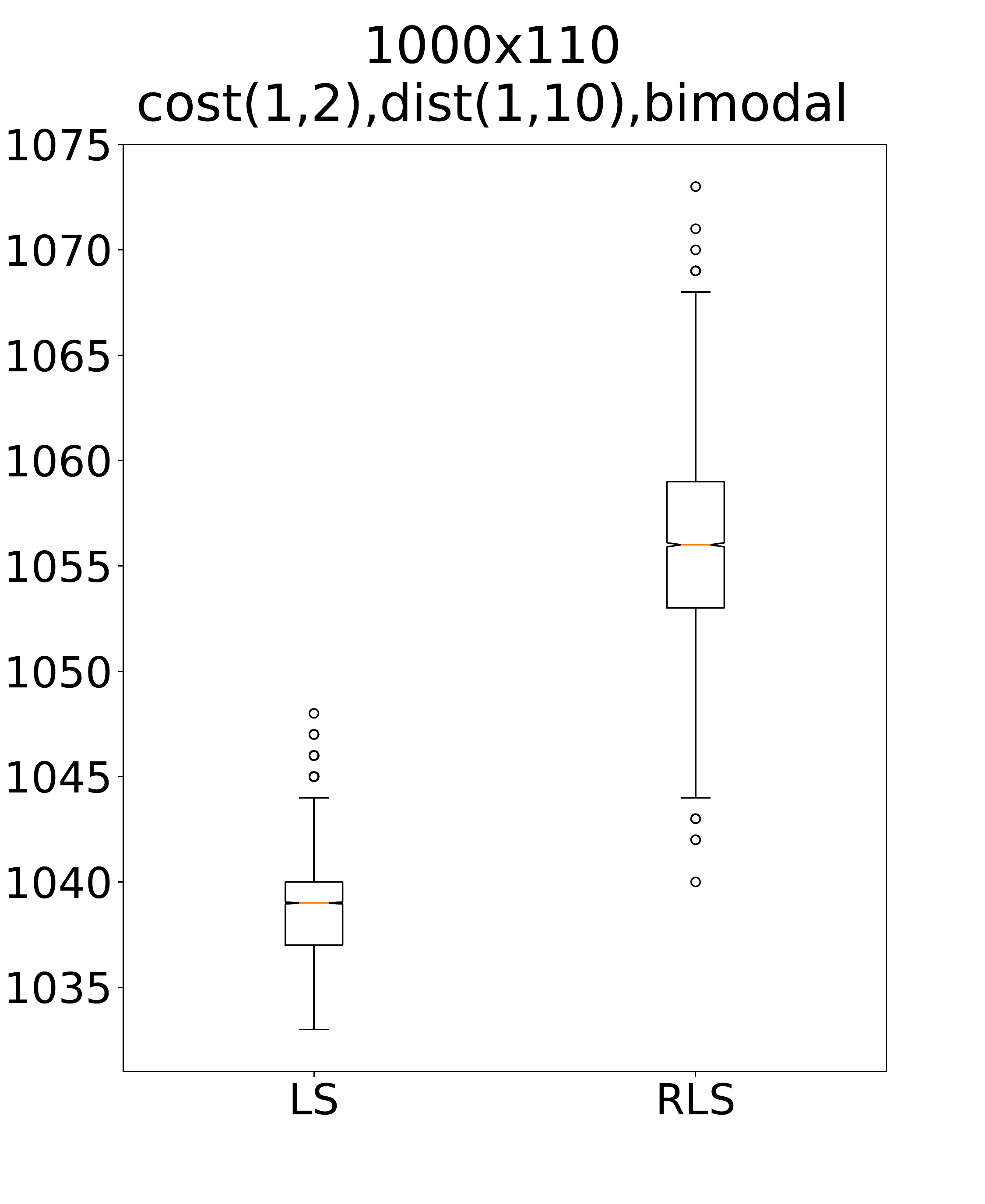}
\includegraphics[scale=0.13]{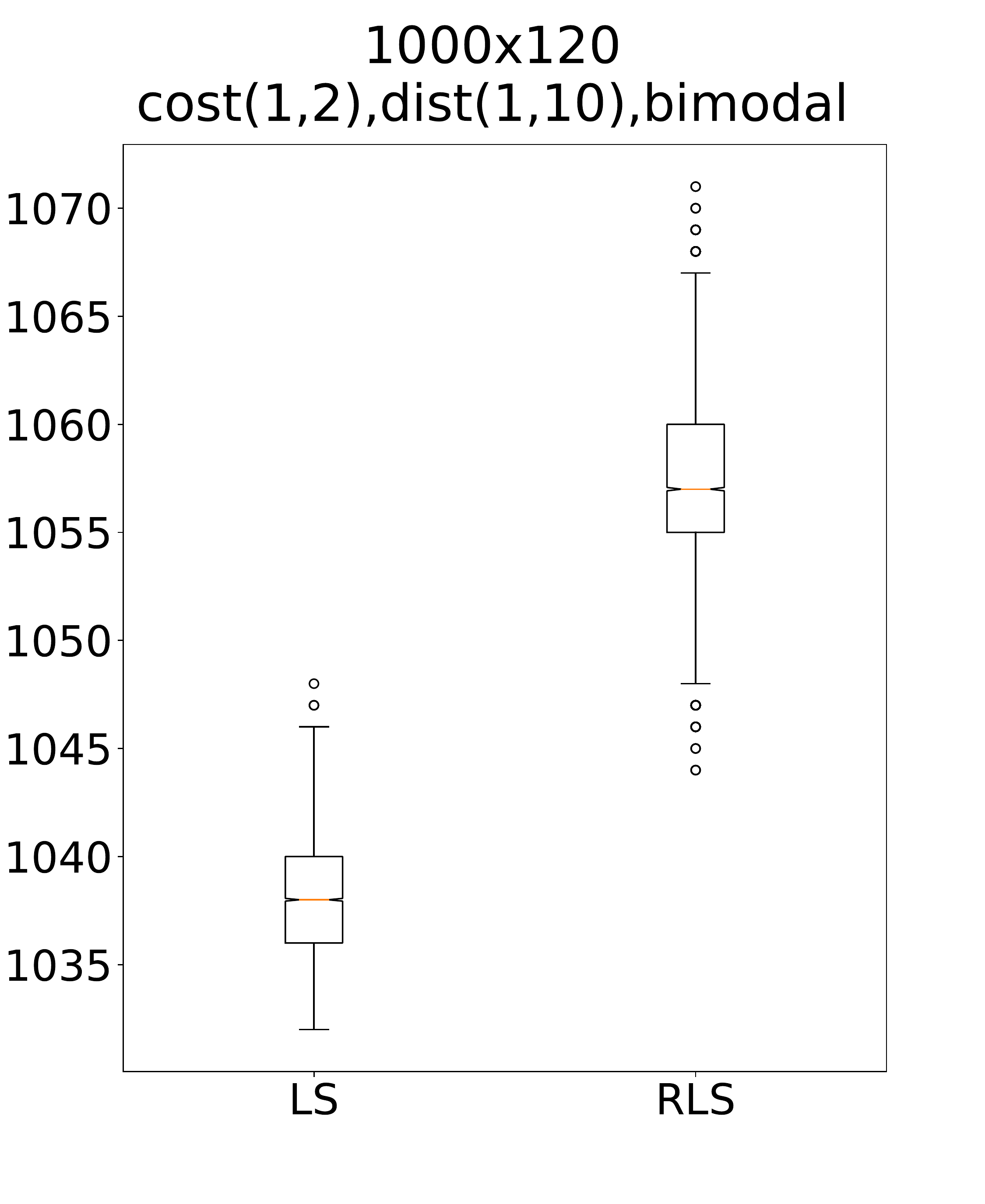}
\includegraphics[scale=0.13]{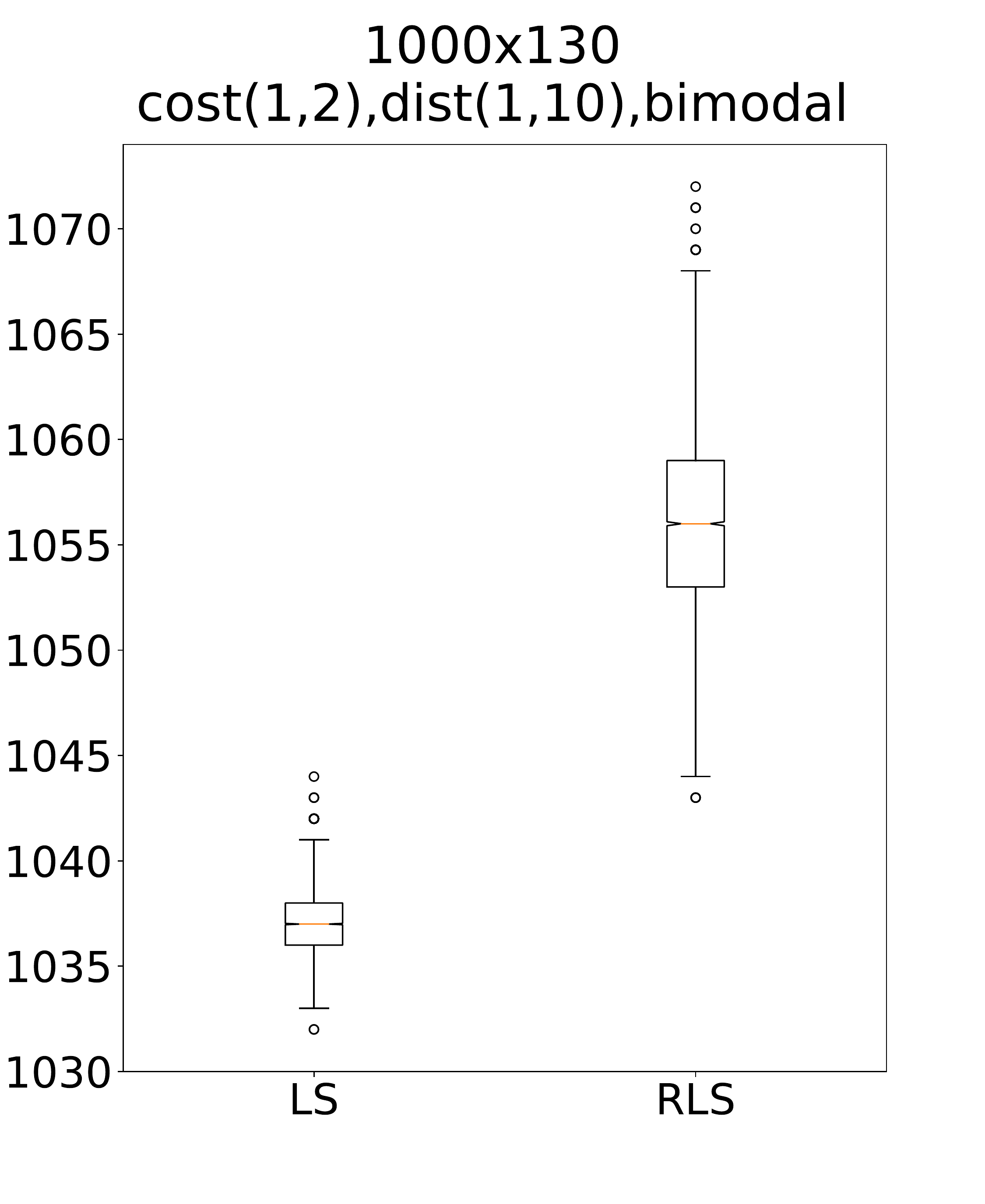}
\includegraphics[scale=0.13]{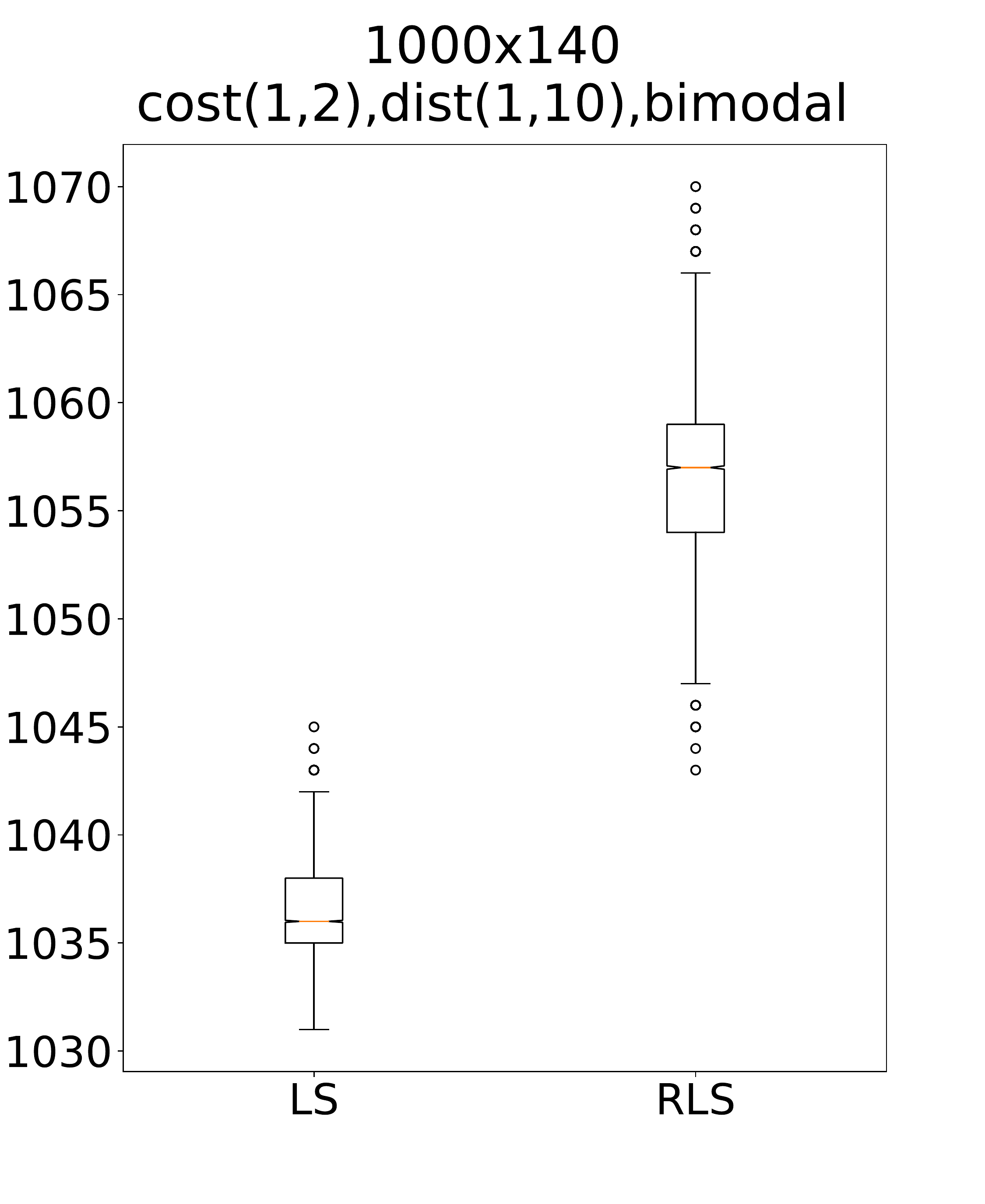}
\end{center}
\caption{Box-whisker plots obtained for $10$ problem instances generated according to Model 2. LS and RLS were used $1000$ times per instance, while the results for CBC represent optimal or near-optimal reference solutions obtained by the corresponding ILP solving procedure.}	
\end{figure*}

\begin{figure*}
\begin{center}
\includegraphics[scale=0.13]{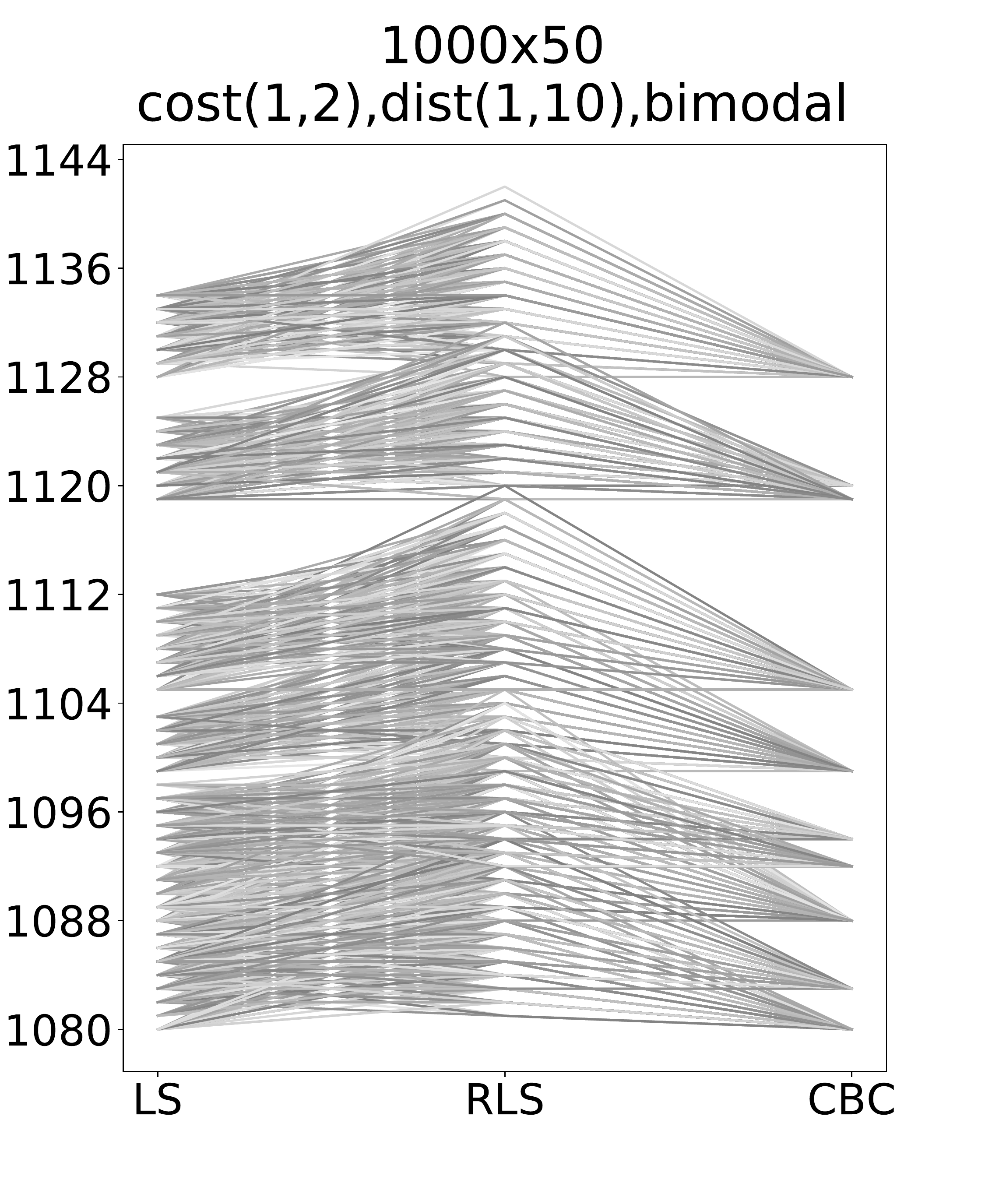}
\includegraphics[scale=0.13]{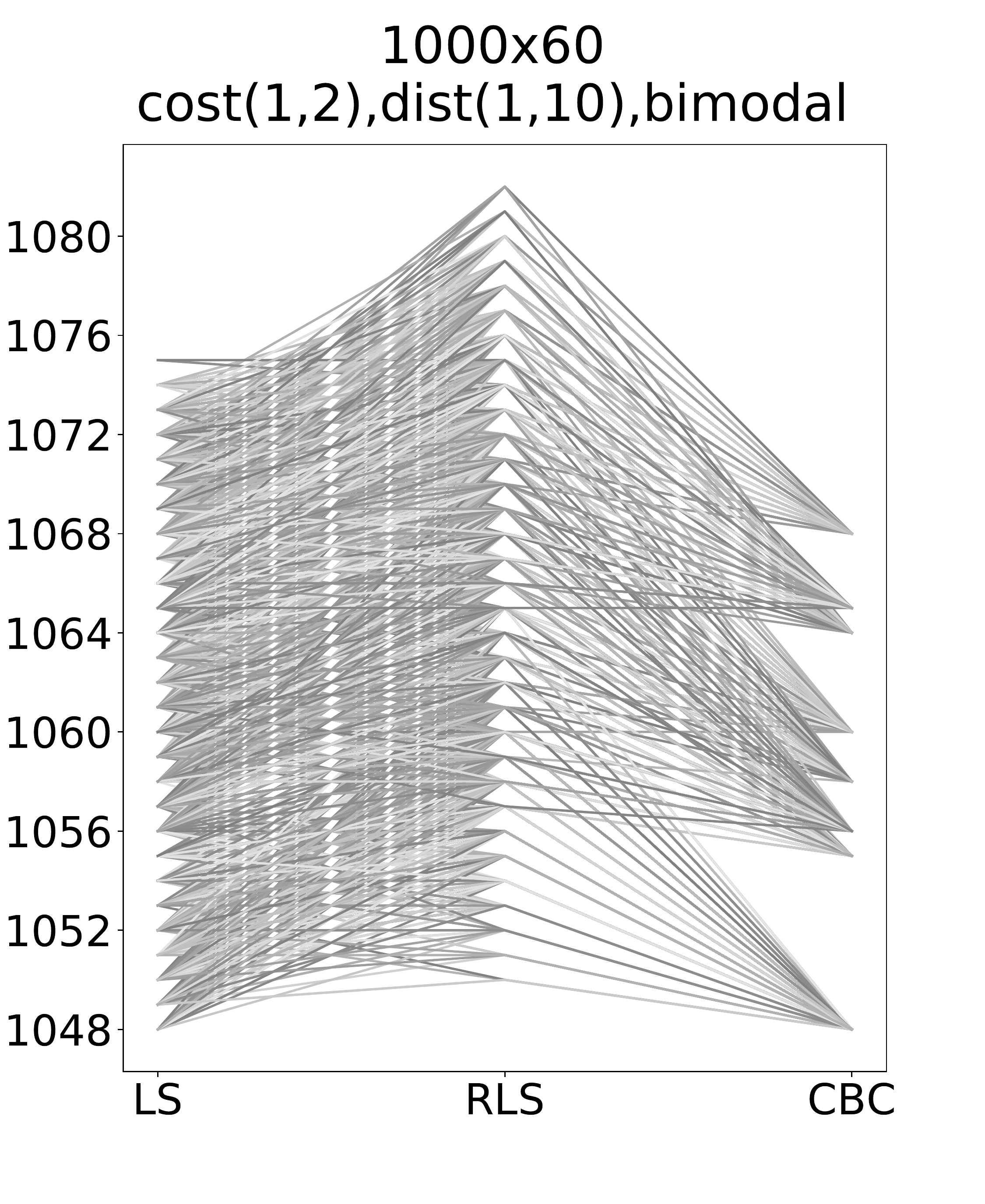}
\includegraphics[scale=0.13]{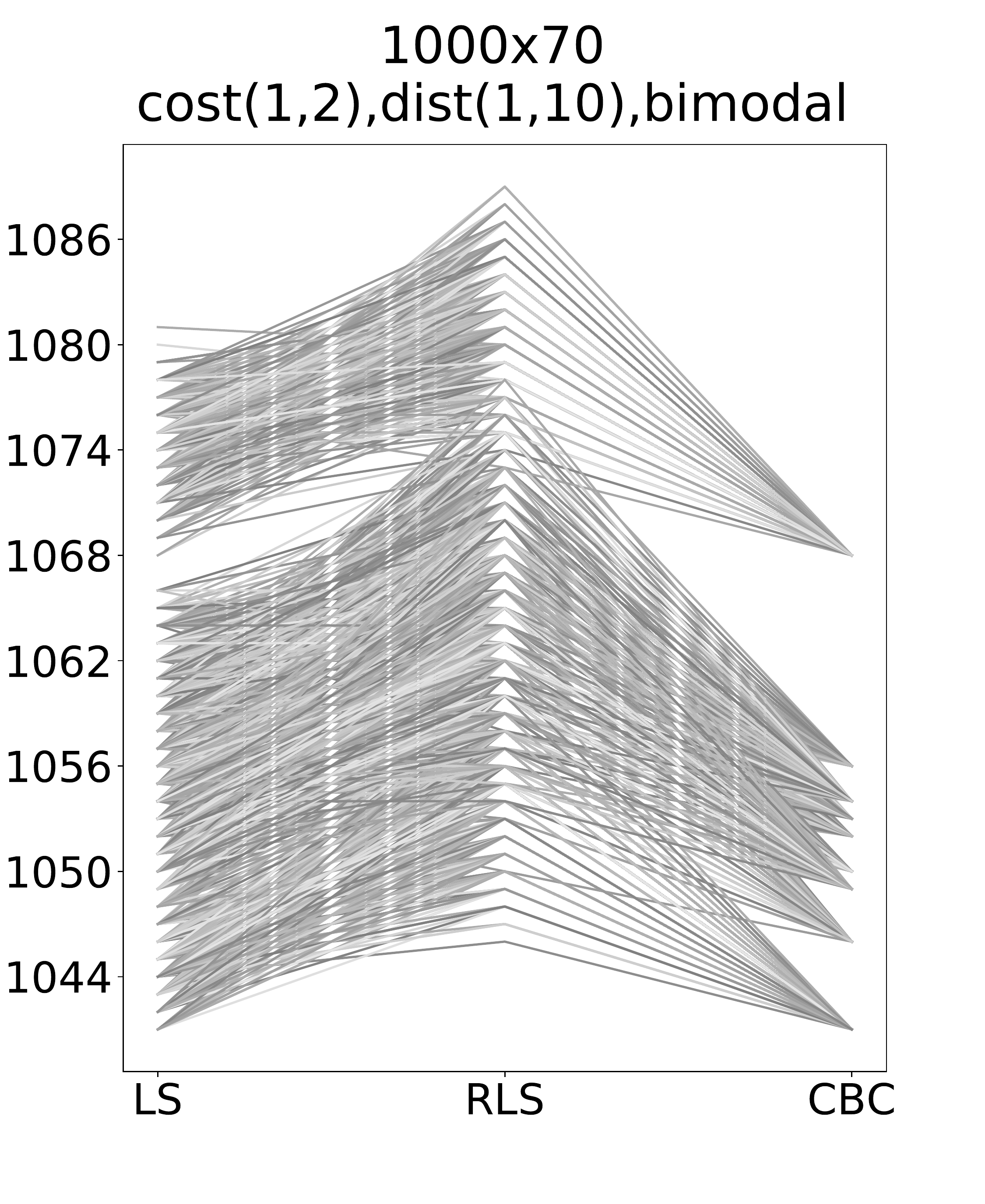}
\includegraphics[scale=0.13]{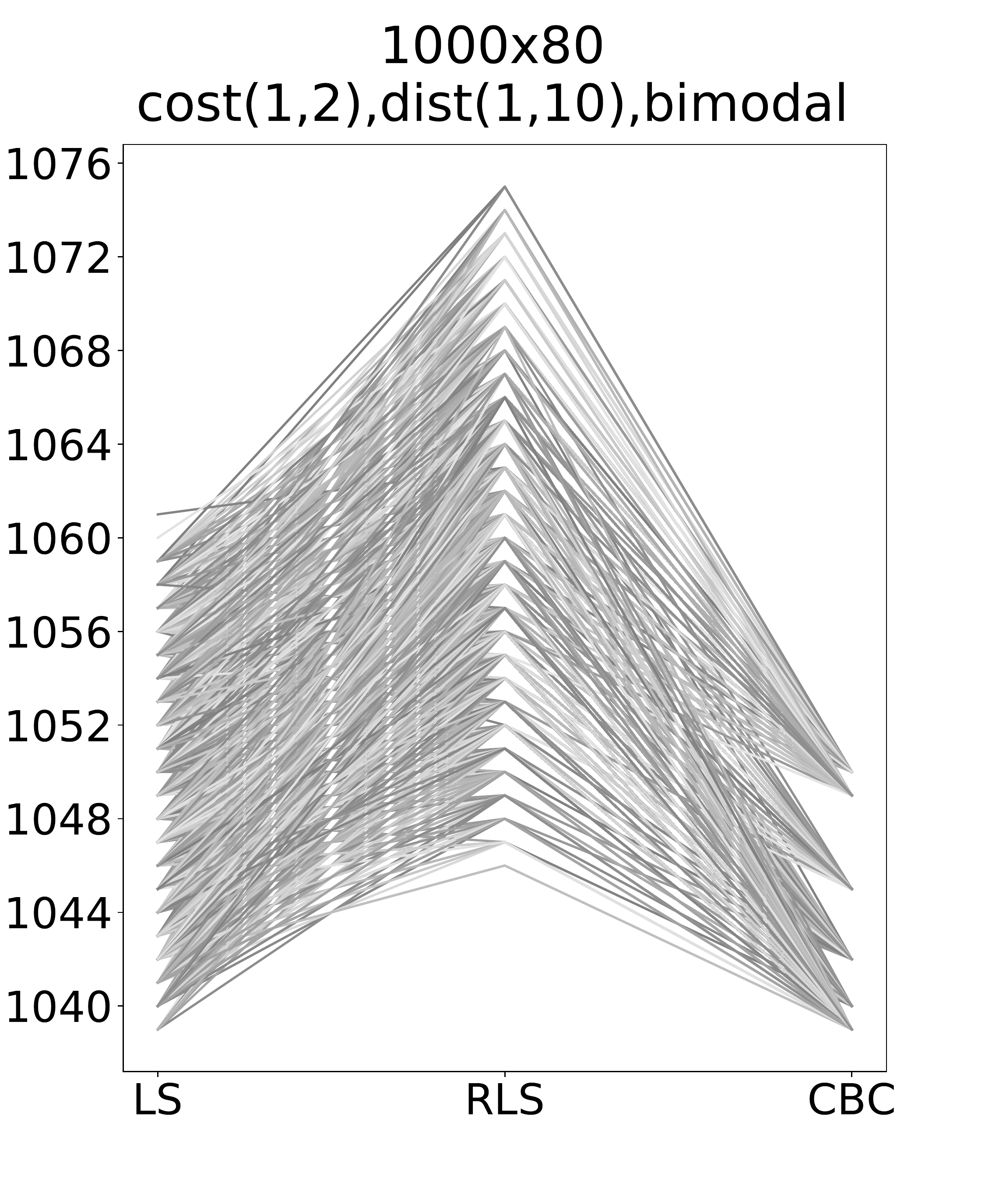}
\includegraphics[scale=0.13]{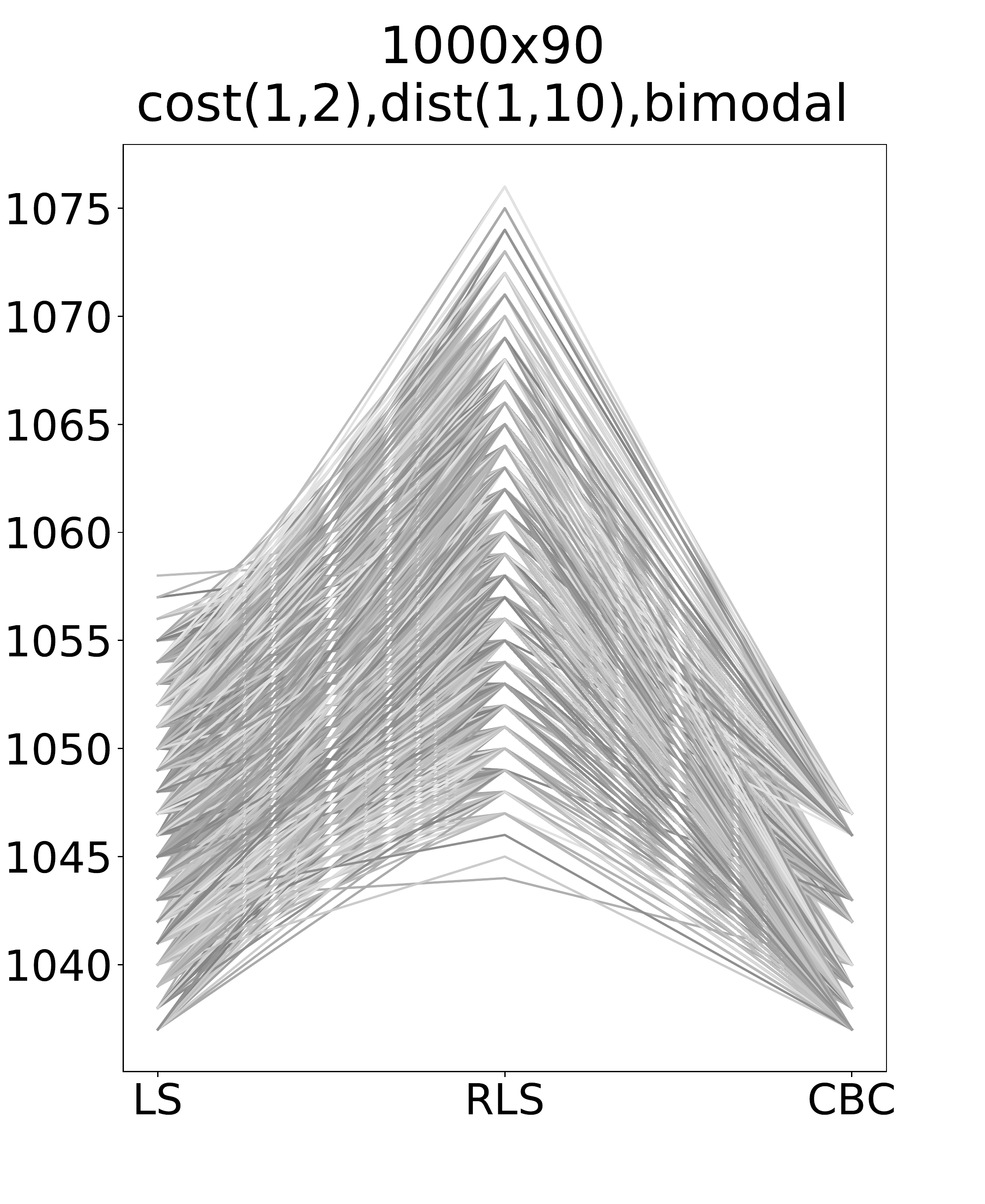}
\end{center}
\caption{A plot depicting the relation of the performance of LS, RLS and the actual optimal or near-optimal solutions found by CBC for Model 2. These results are grouped according to the problem instance. Each of the gray lines represents the objective values found by a run of LS, RLS and the objective value found by CBC.}	
\end{figure*}

Figure 4 sheds a bit more light on the relations between the performances of different techniques, grouped by instances. Each line in the plot connects results for the corresponding runs of LS, RLS and CBC, linking the solutions found by the local search algorithms to the actual optimal or near-optimal solutions. One can observe an overall ``downward'' trend from LS to RLS, confirming the better performance of RLS also on the instance level. However, one can also observe a widening gap between the individual results of RLS and CBC. This needs to be interpreted in context: the fact that the performance of an individual run of LS or RLS is inferior does not necessarily mean that a multi-start variant of the algorithm cannot succeed. This can be seen in particular for $1000 \times 80$ instances. While there is a gap between individual performances of RLS and CBC, the box-whisker plot indicates that ``lucky'' runs of RLS are still not far away from the results of CBC. The question that remains open is how many restarts would such an algorithm require to be successful and whether a hybridisation of the algorithm would be more suitable.

\subsubsection{Model 2: Binary Facility Cost, Bimodal Distribution of Distances}

Figure 5 illustrates the results obtained for Model 2 as box-wisker plots. These reveal a contrasting pattern to the one obtained for Model 1. LS outperforms RLS for these instances. At this point, it is worth noting that we have only changed the cost and distance structure. Such a change already leads to a very different result to the one observed for the previous instances. Observing the distributions of results found by LS and RLS, one can see that the gap even widens with growing number of facilities. However, one can also observe a difference between the median performance of LS and median objective value of optimal or near-optimal solutions.

However, Figure 6 does not seem to indicate a pronounced slope in the relations between objective values observed for high-quality runs of LS and the results found by CBC. It also reveals a very clear pattern of clusters, corresponding to individual problem instances. It seems that for this type of instance, a multi-start variant of the LS algorithm is a good choice.

\subsubsection{Model 3: Binary Facility Cost, Binary Distances}

Figure 7 reveals the results obtained for Model 3. These paint a more complex picture than those obtained for the previous instances. This is further supported by the results of CBC, which was able to find proven optima only for the instances with $50$ facilities. For larger instances, near-optimal solutions were found.

For these instances, LS has a very intriguing performance. One can observe that the median solution quality is better than for RLS. The distribution of solution quality for LS seems to be estimated relatively well in terms of its shape. However, one can also observe a consistent bias in this estimate. The results obtained by LS are heavily concentrated around the median, similarly to the distribution of the actual optimal or near-optimal solutions.

\begin{figure*}
\begin{center}
\includegraphics[scale=0.13]{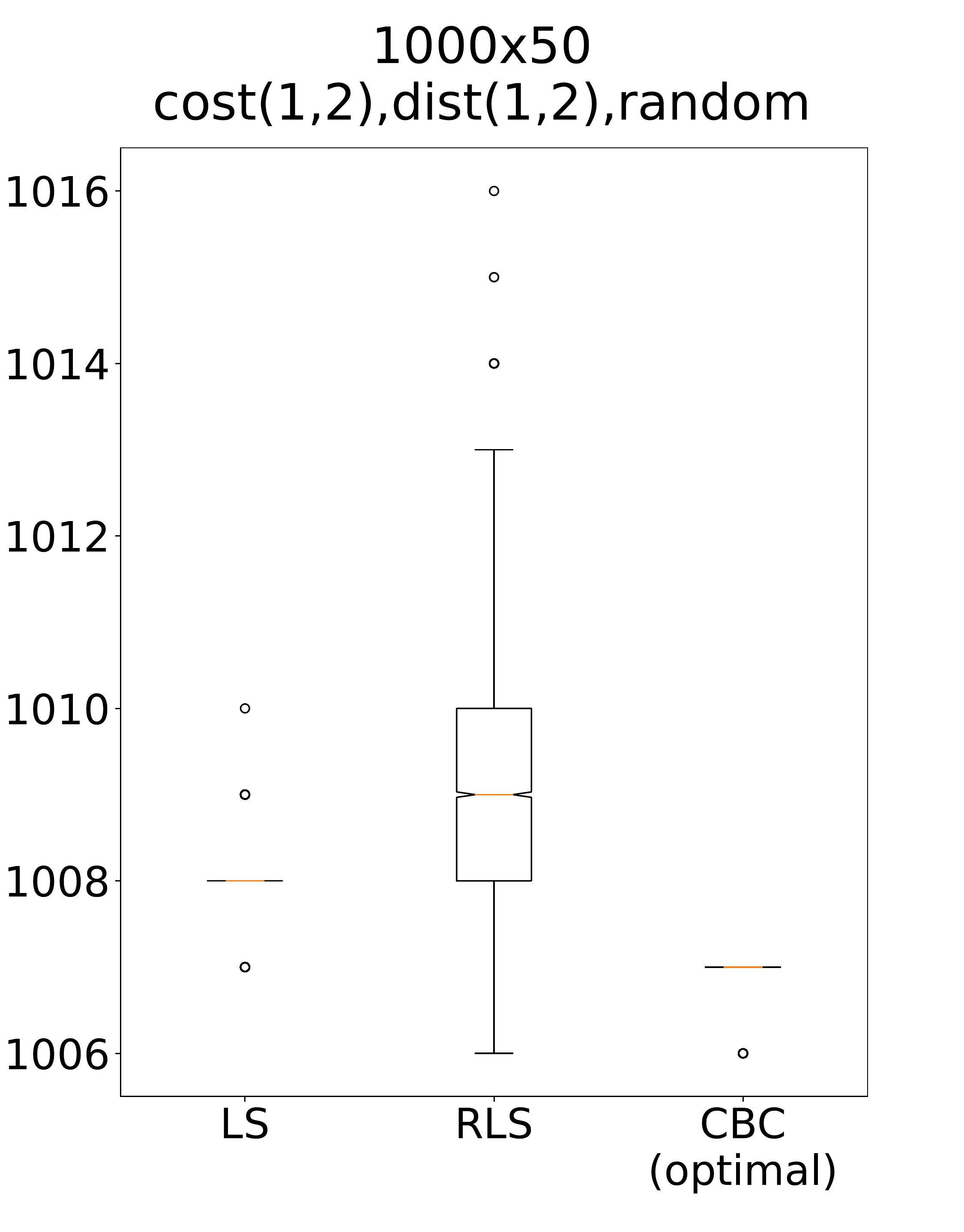}
\includegraphics[scale=0.13]{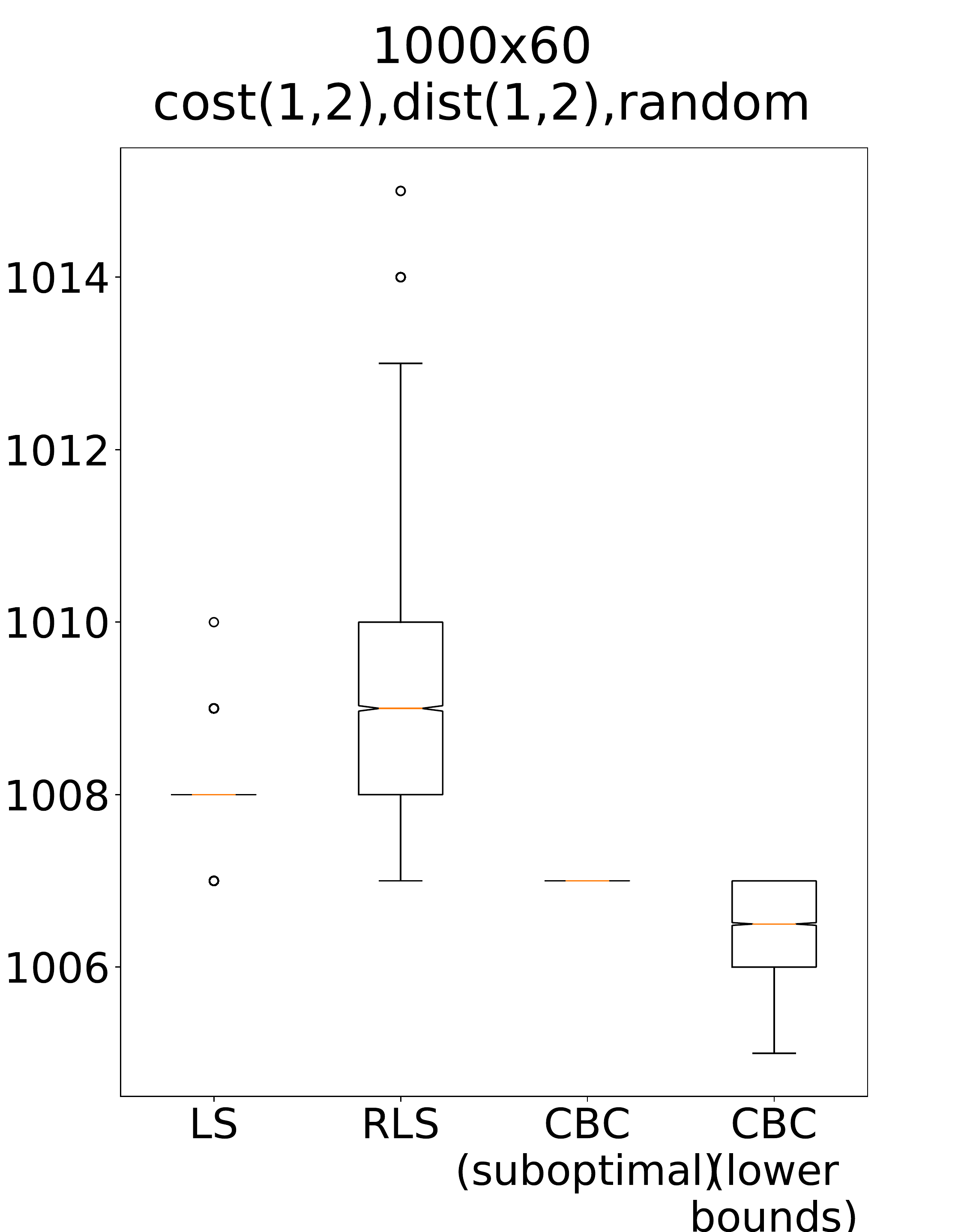}
\includegraphics[scale=0.13]{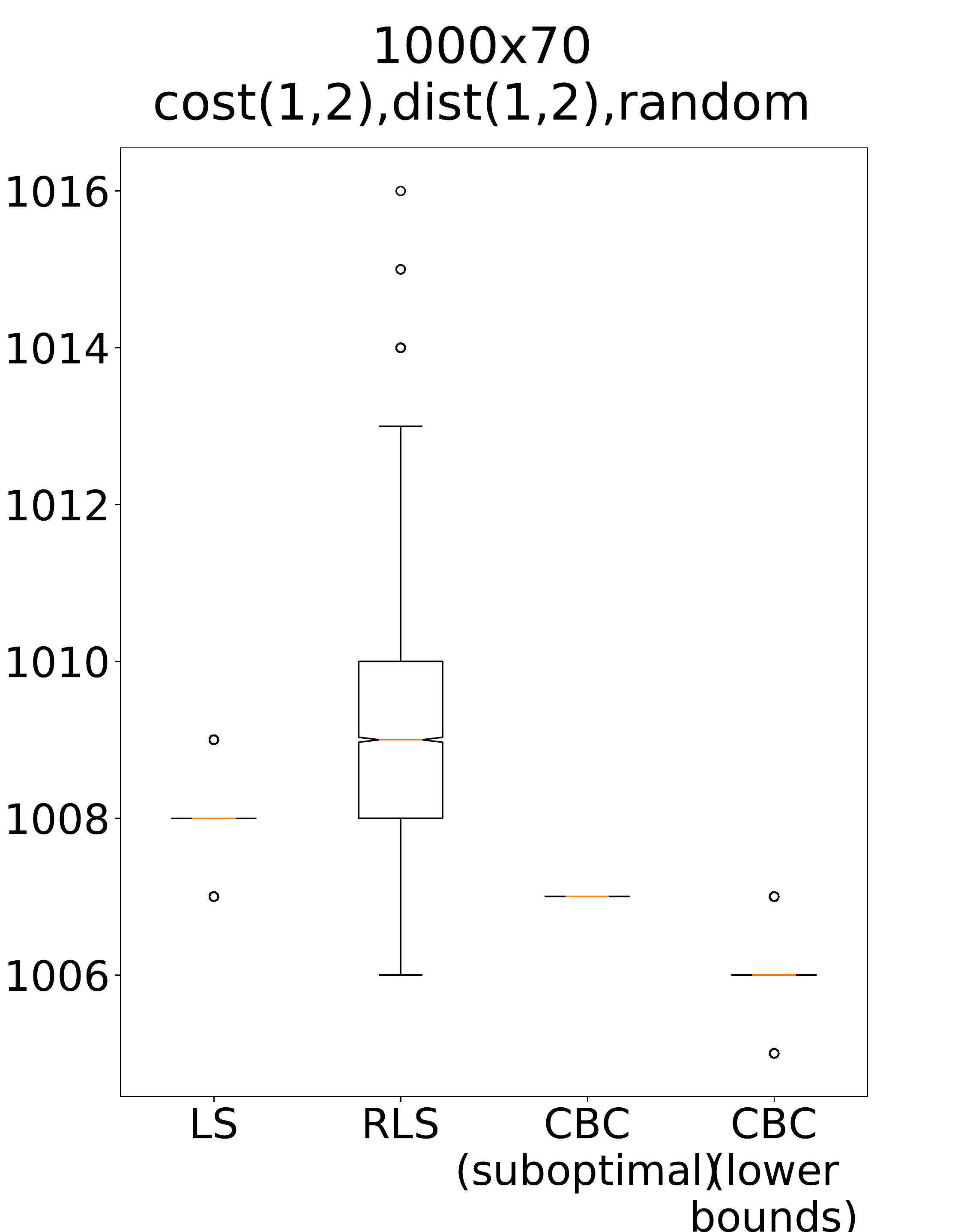}
\includegraphics[scale=0.13]{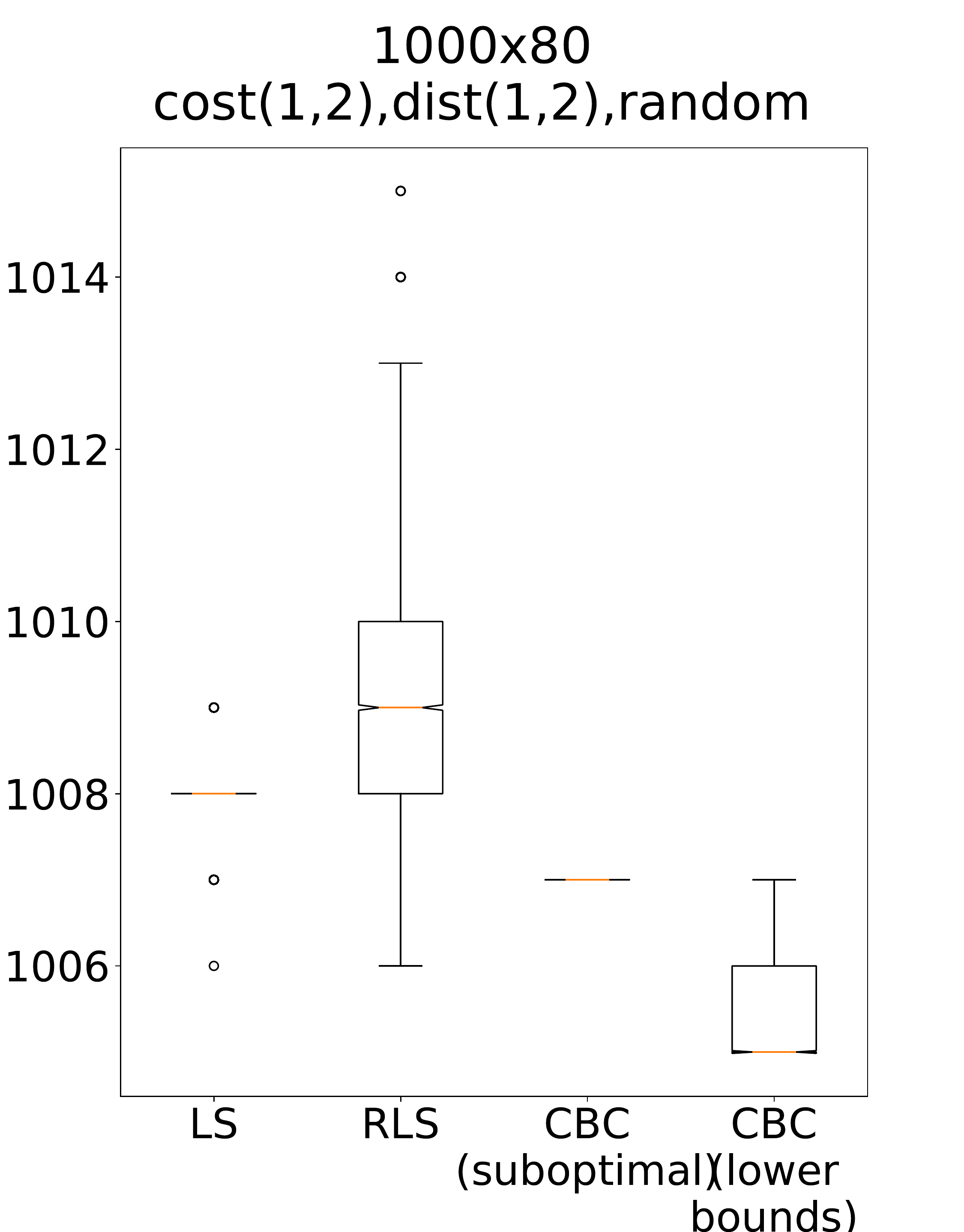}
\includegraphics[scale=0.13]{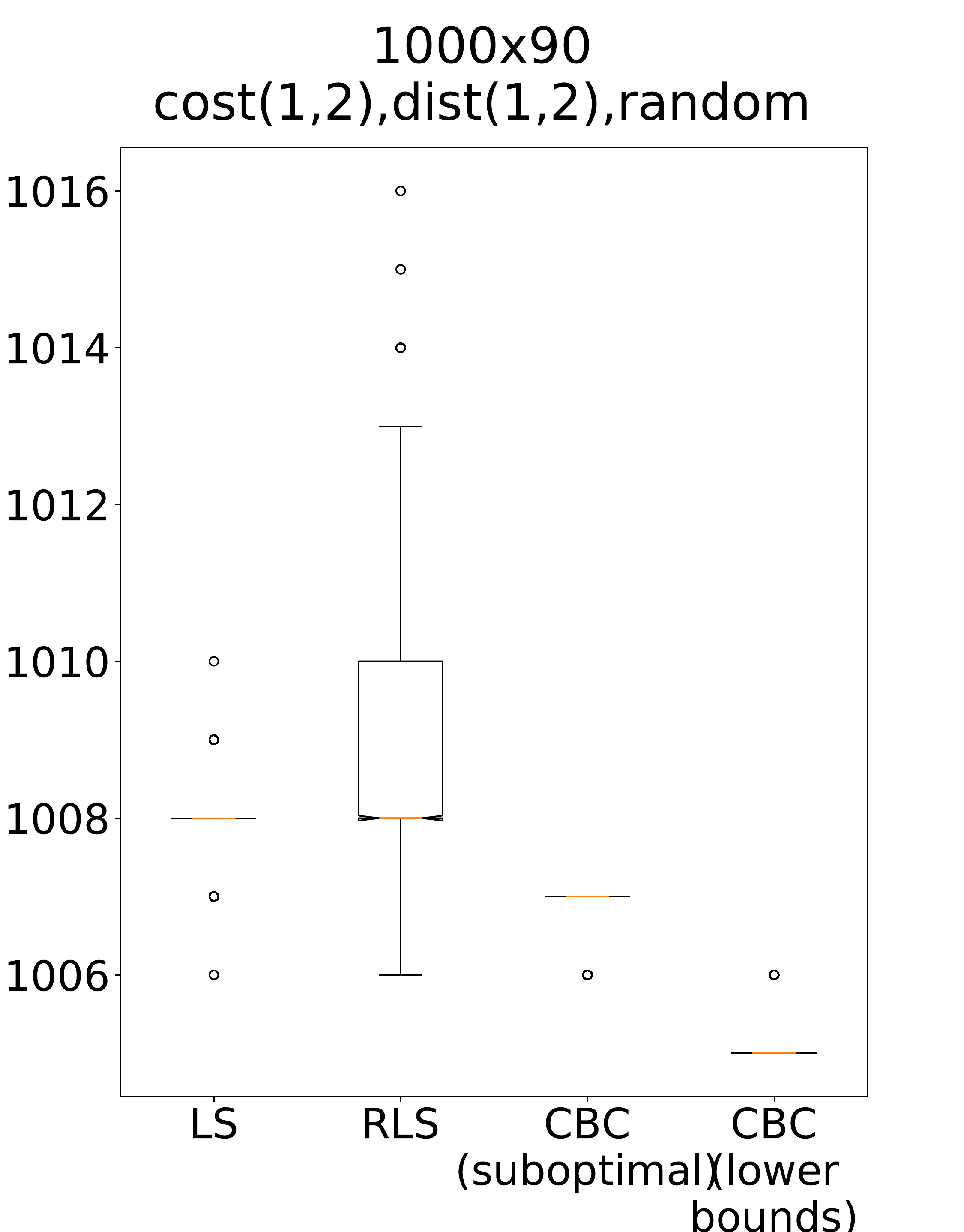}\\
\includegraphics[scale=0.13]{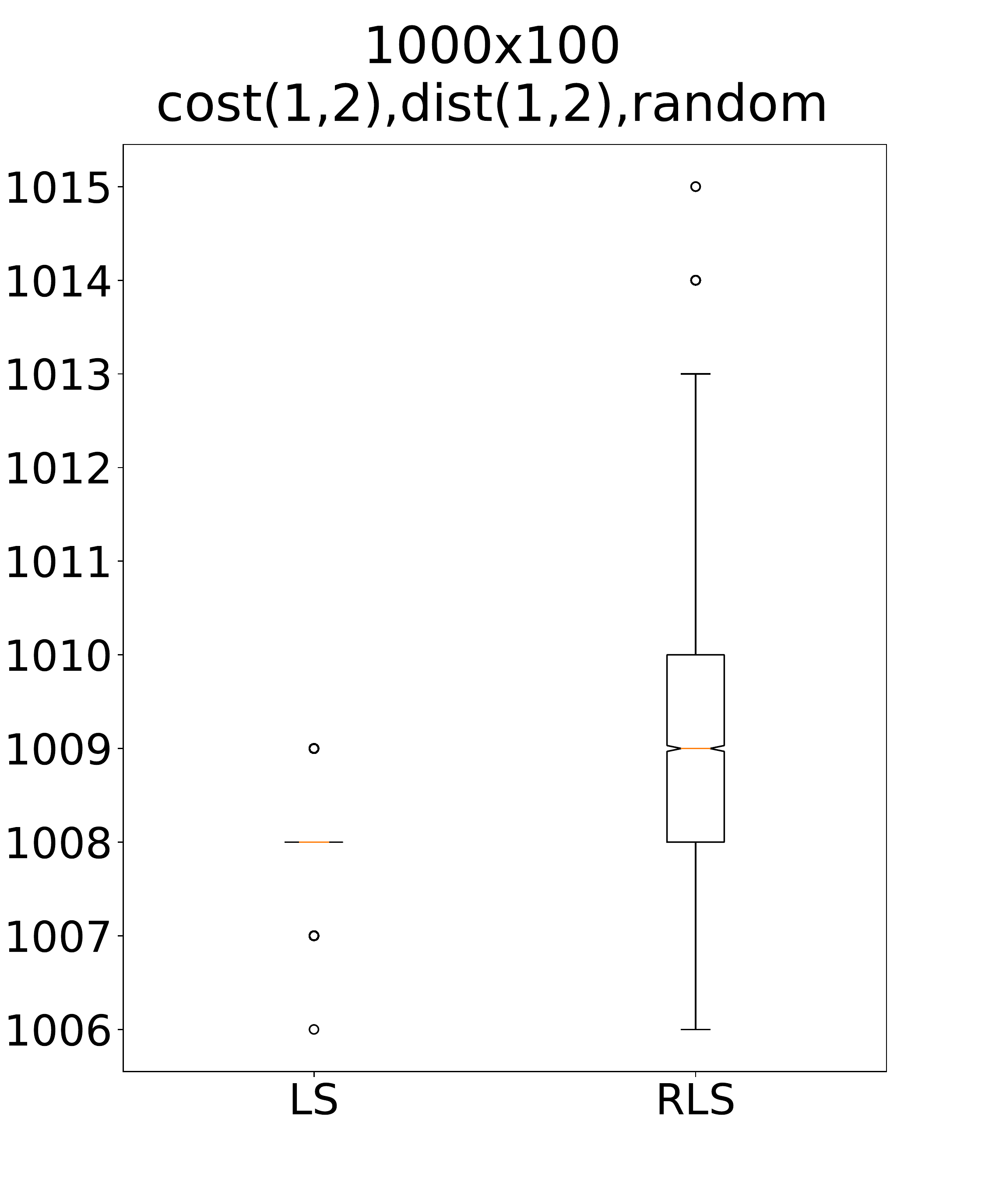}
\includegraphics[scale=0.13]{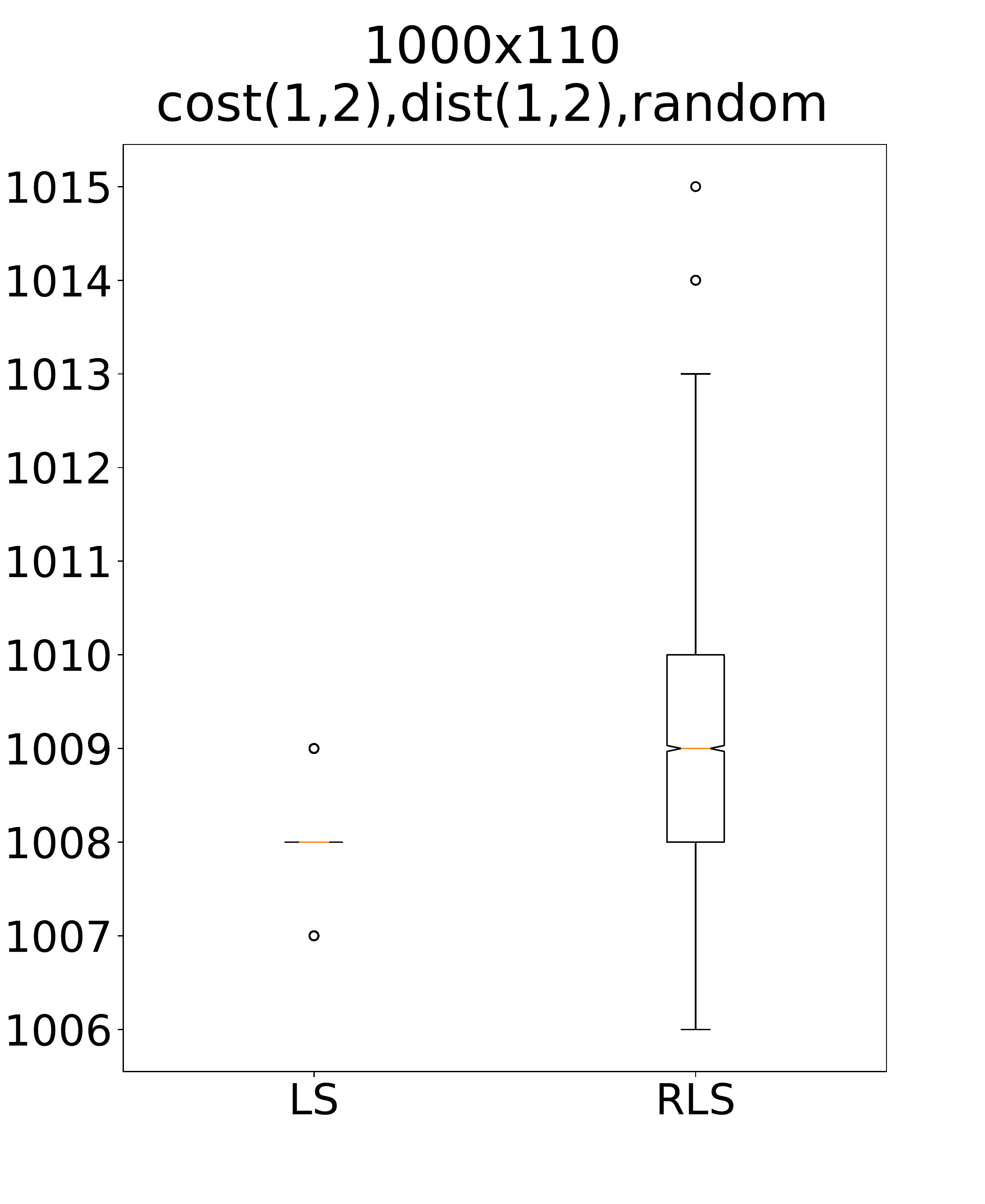}
\includegraphics[scale=0.13]{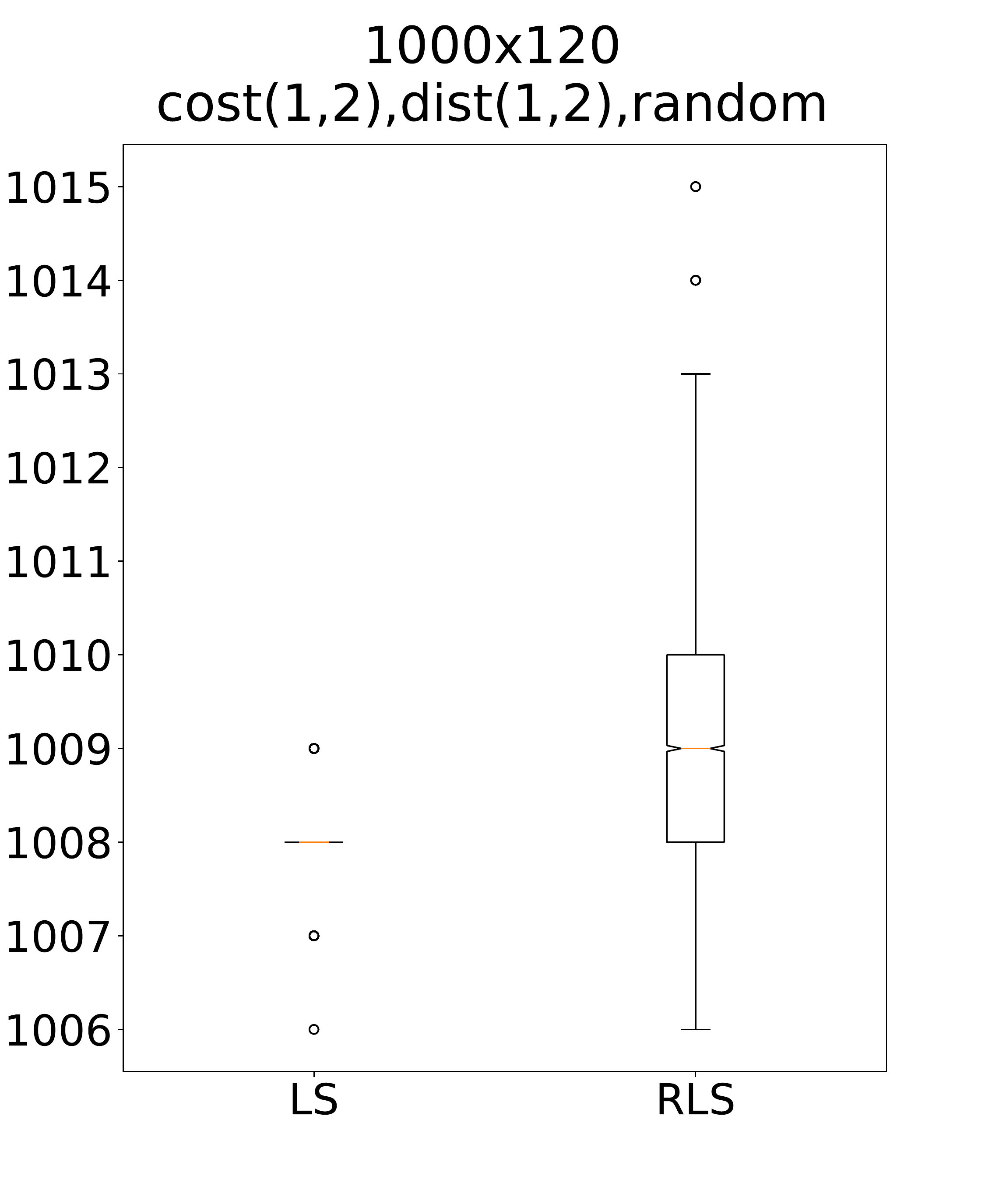}
\includegraphics[scale=0.13]{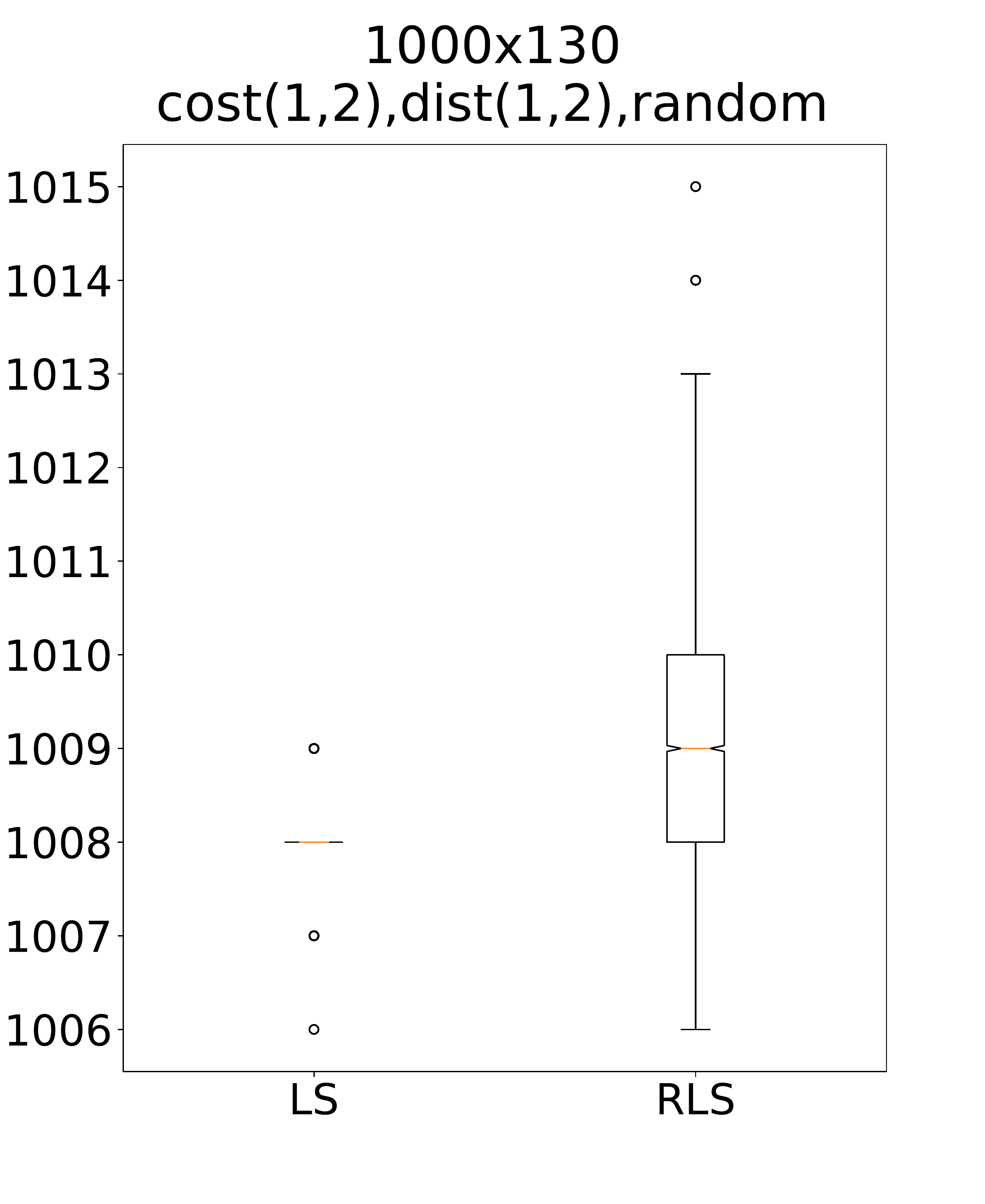}
\includegraphics[scale=0.13]{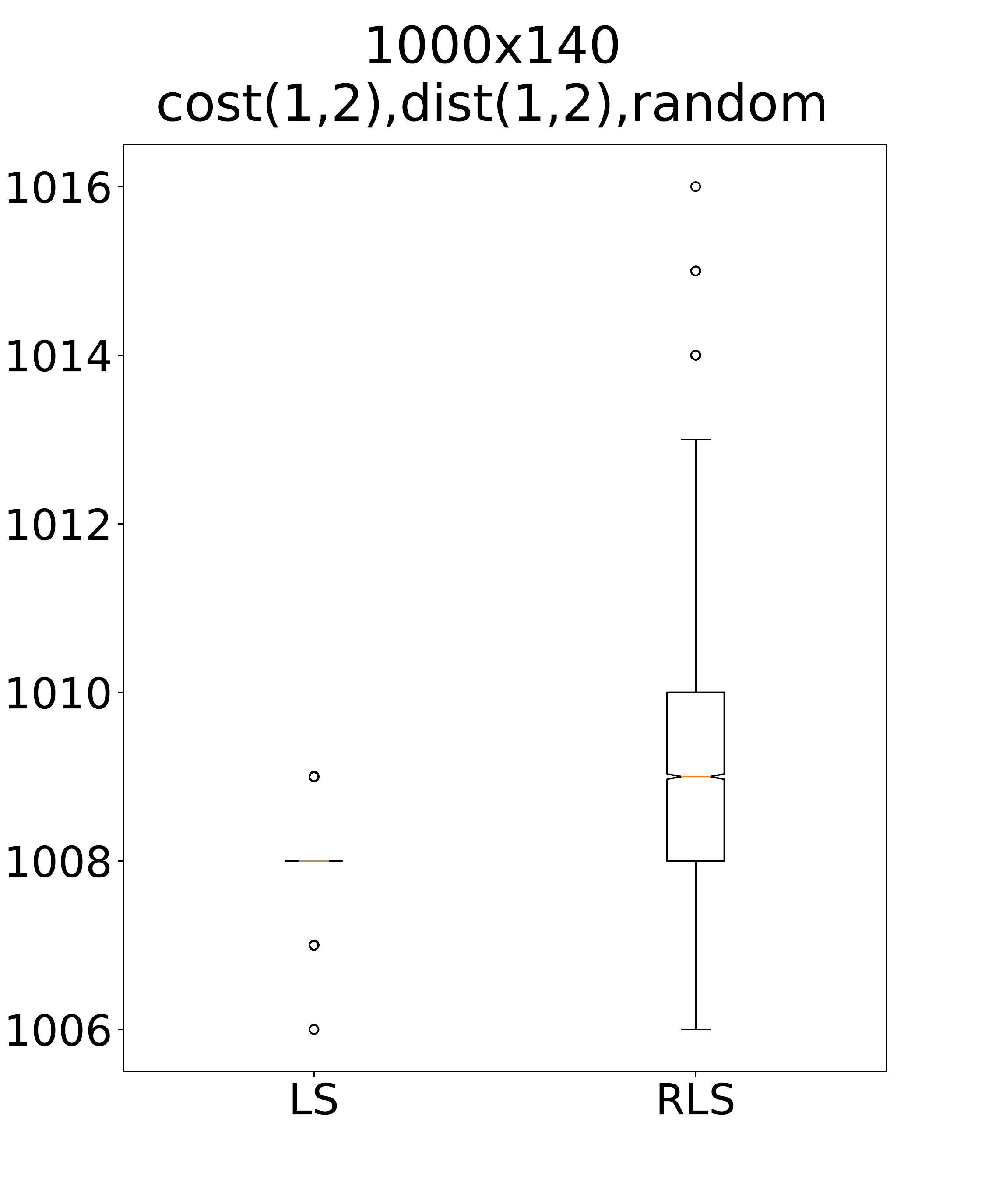}
\end{center}
\caption{Box-whisker plots obtained for $10$ problem instances generated according to Model 3. LS and RLS were used $1000$ times per instance, while the results for CBC represent optimal or near-optimal reference solutions obtained by the corresponding ILP solving procedure.}
\end{figure*}

\begin{figure*}
\begin{center}
\includegraphics[scale=0.13]{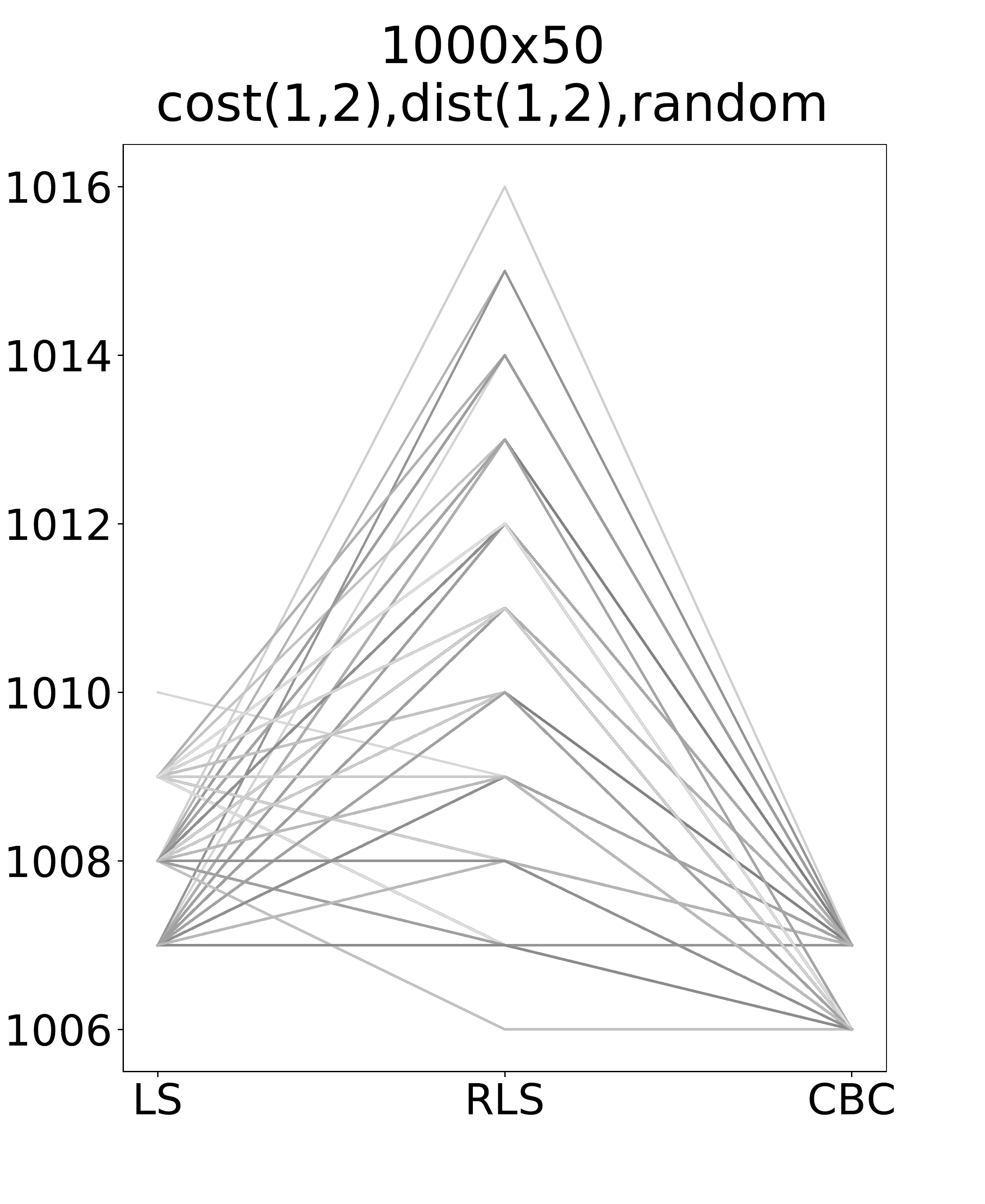}
\includegraphics[scale=0.13]{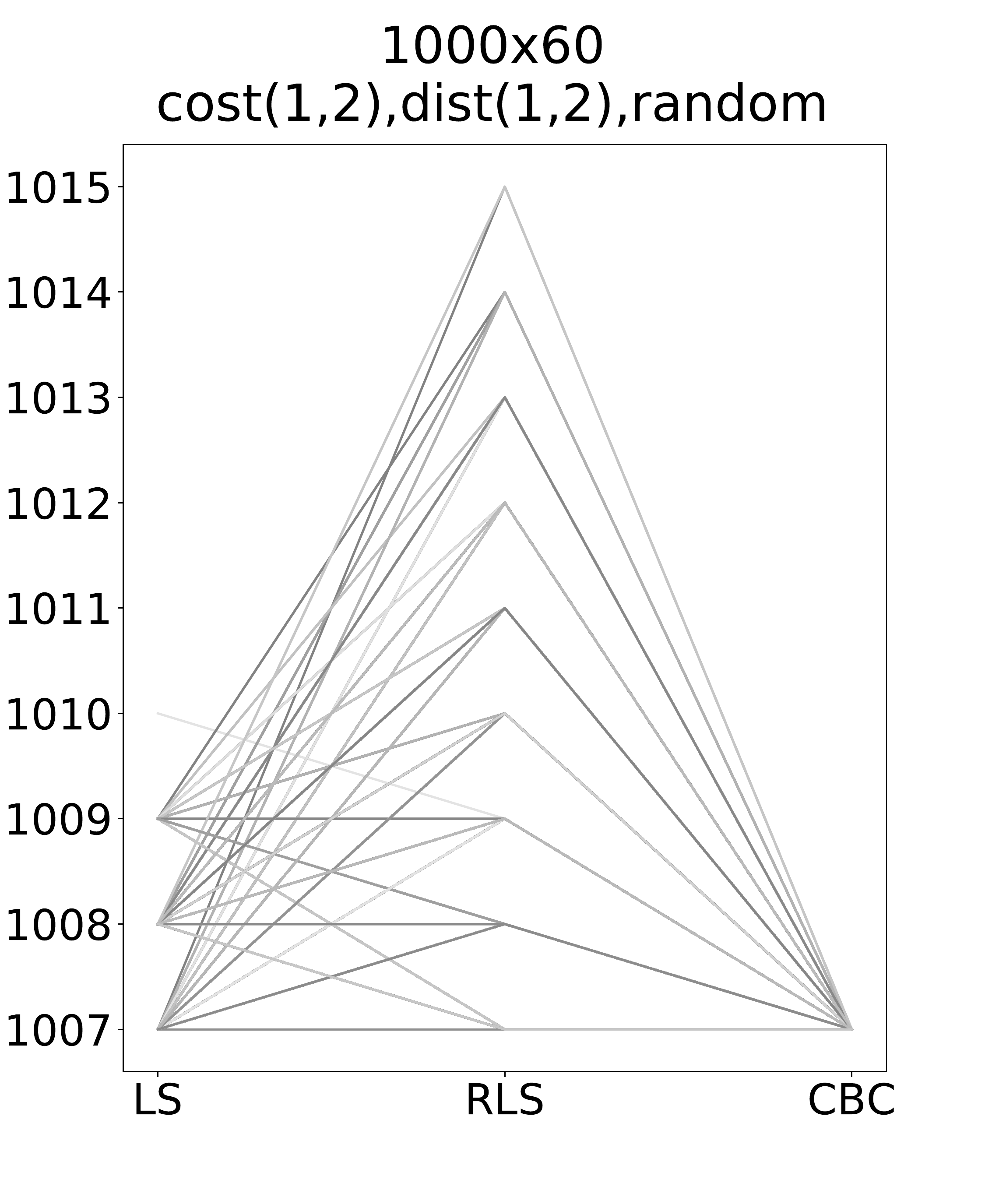}
\includegraphics[scale=0.13]{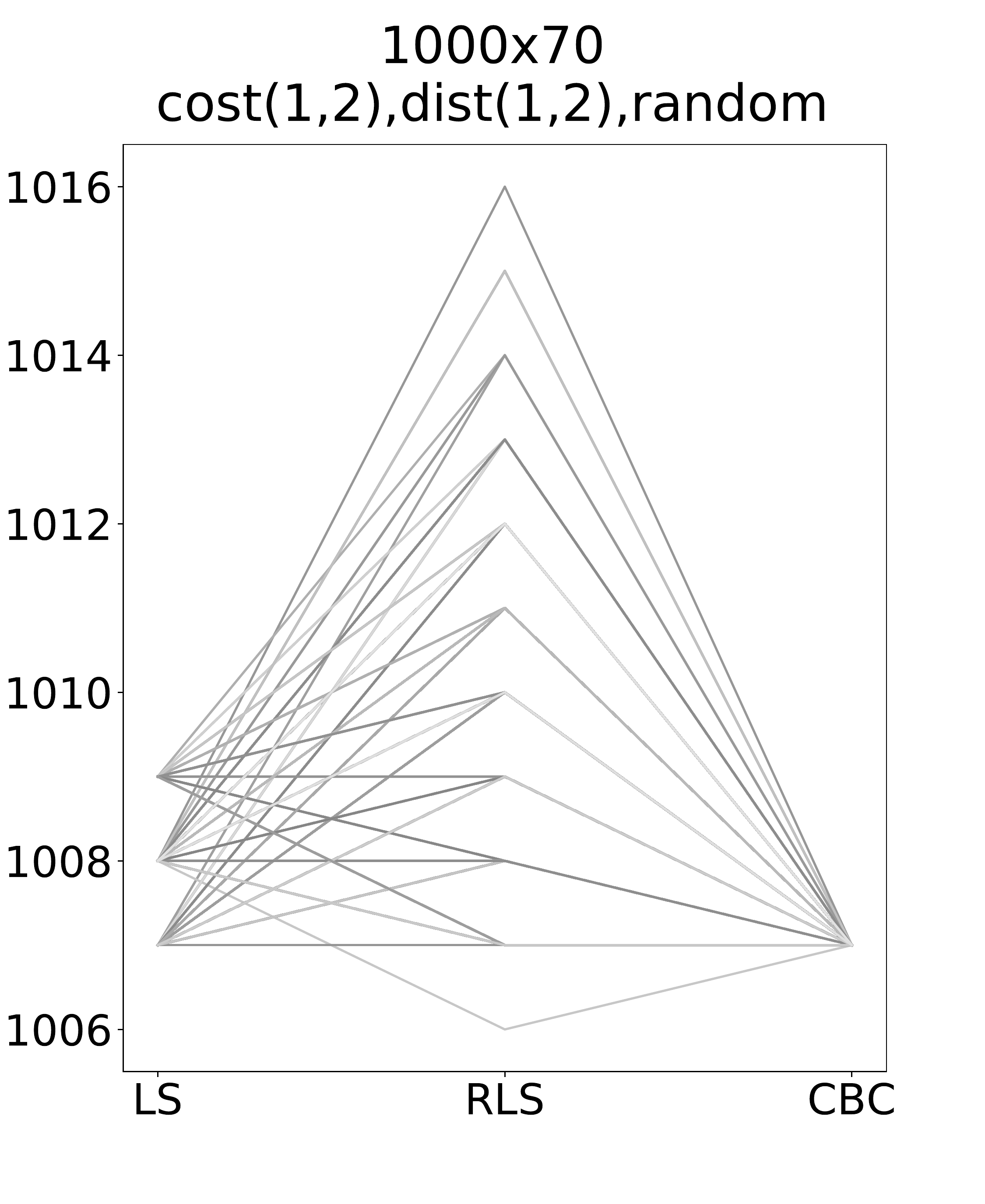}
\includegraphics[scale=0.13]{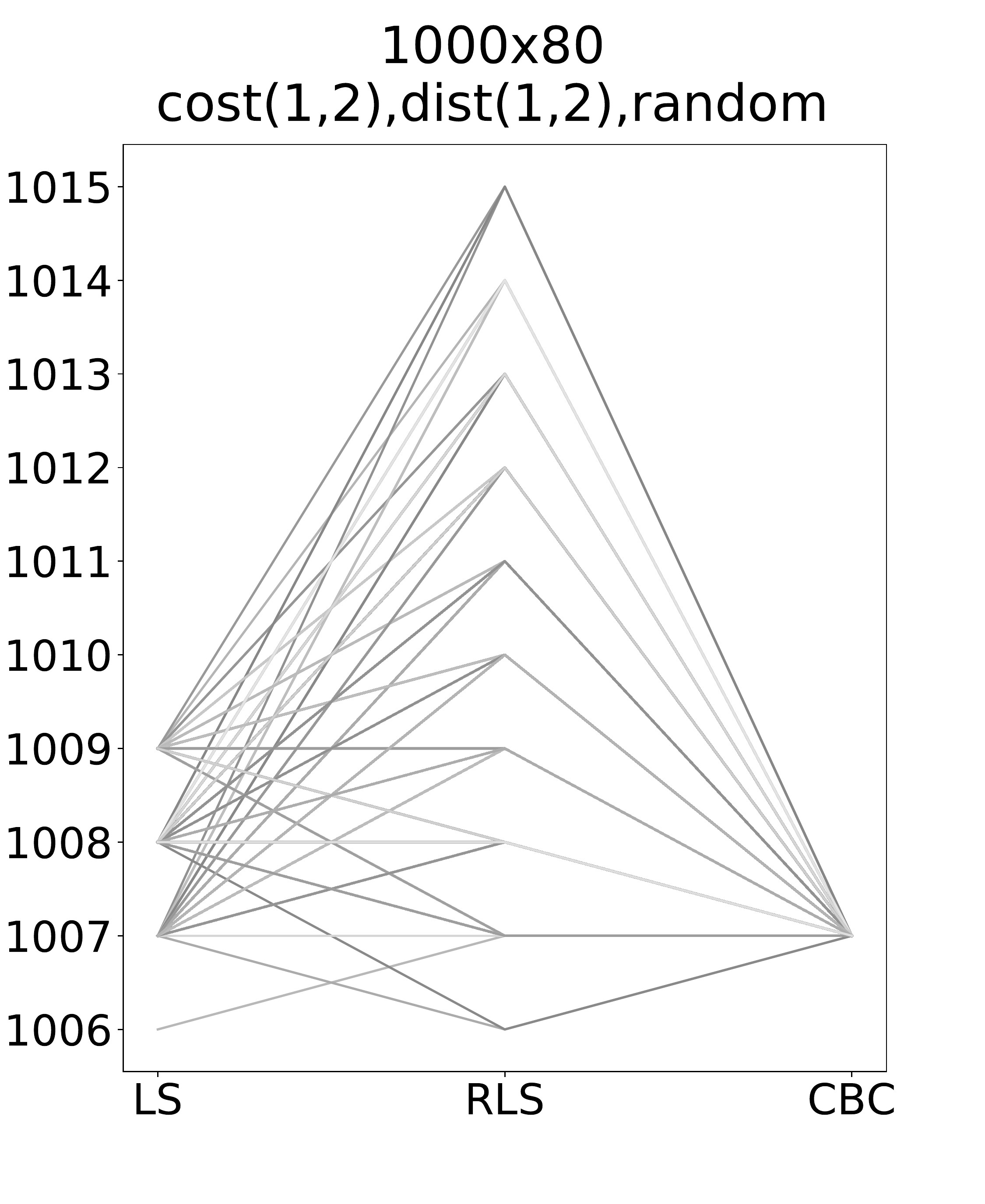}
\includegraphics[scale=0.13]{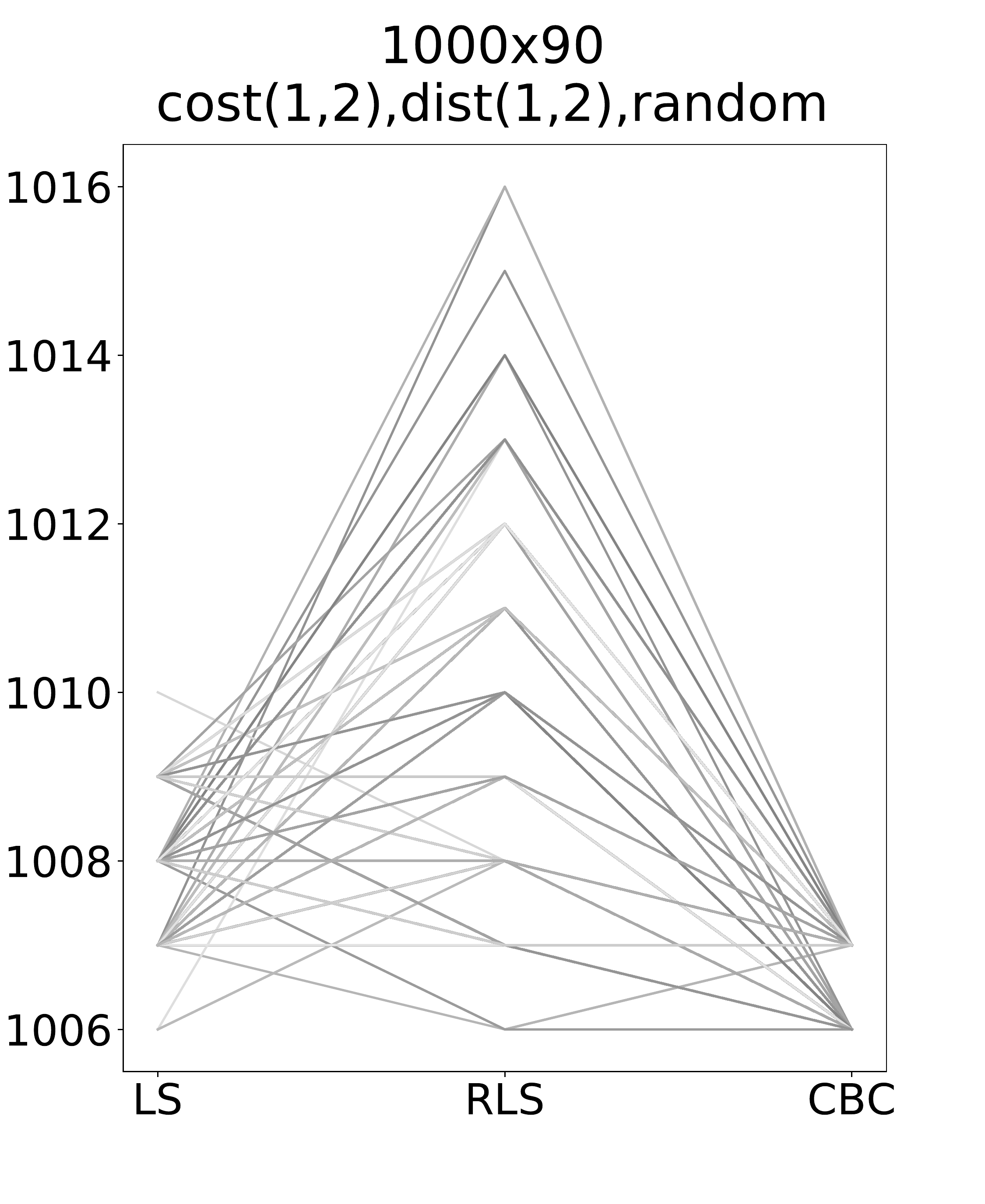}
\end{center}
\caption{A plot depicting the relation of the performance of LS, RLS and the actual optimal or near-optimal solutions found by CBC for Model 3. These results are grouped according to the problem instance. Each of the gray lines represents the objective values found by a run of LS, RLS and the objective value found by CBC.}	
\end{figure*}

\begin{figure*}
\begin{center}
\includegraphics[scale=0.13]{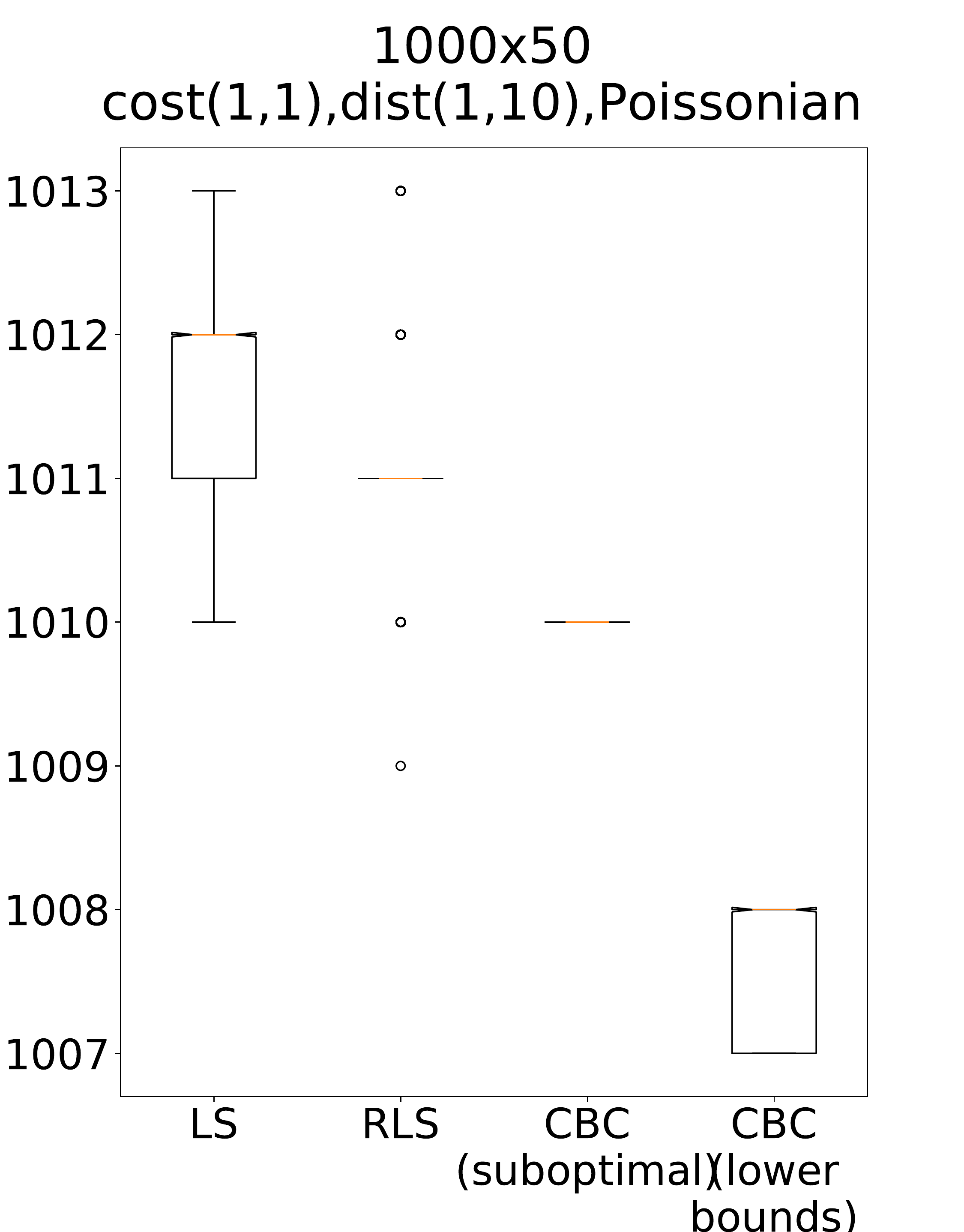}
\includegraphics[scale=0.13]{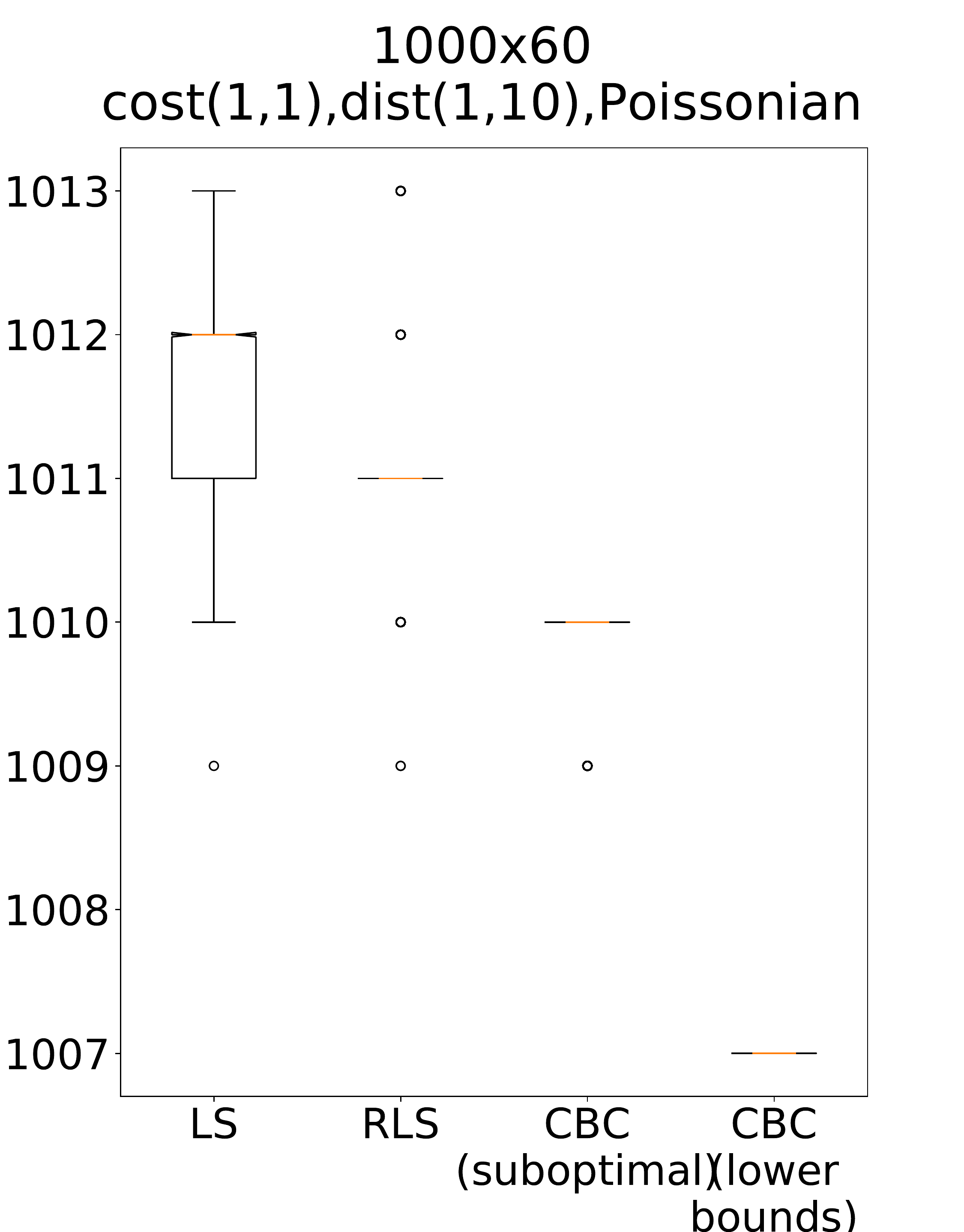}
\includegraphics[scale=0.13]{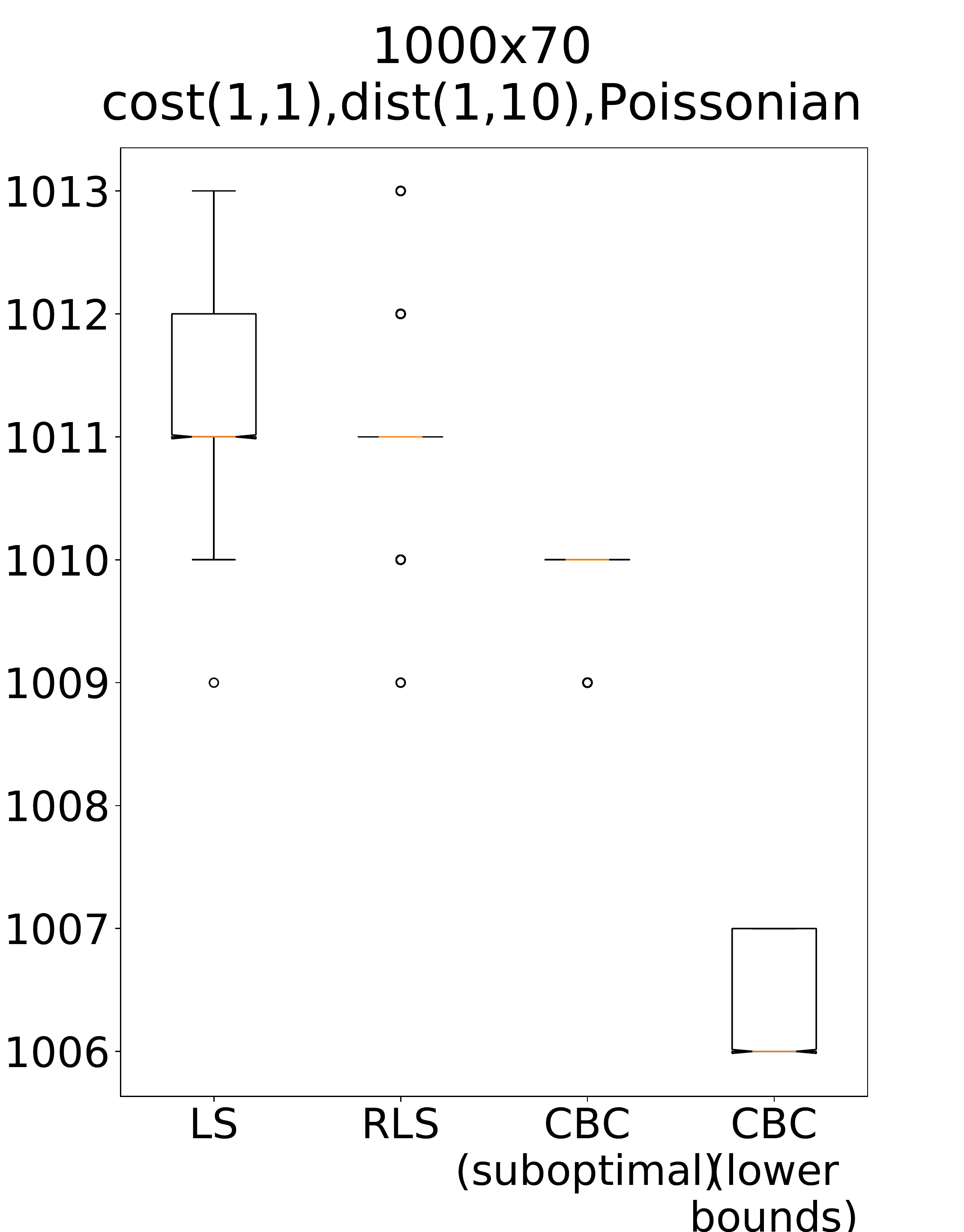}
\includegraphics[scale=0.13]{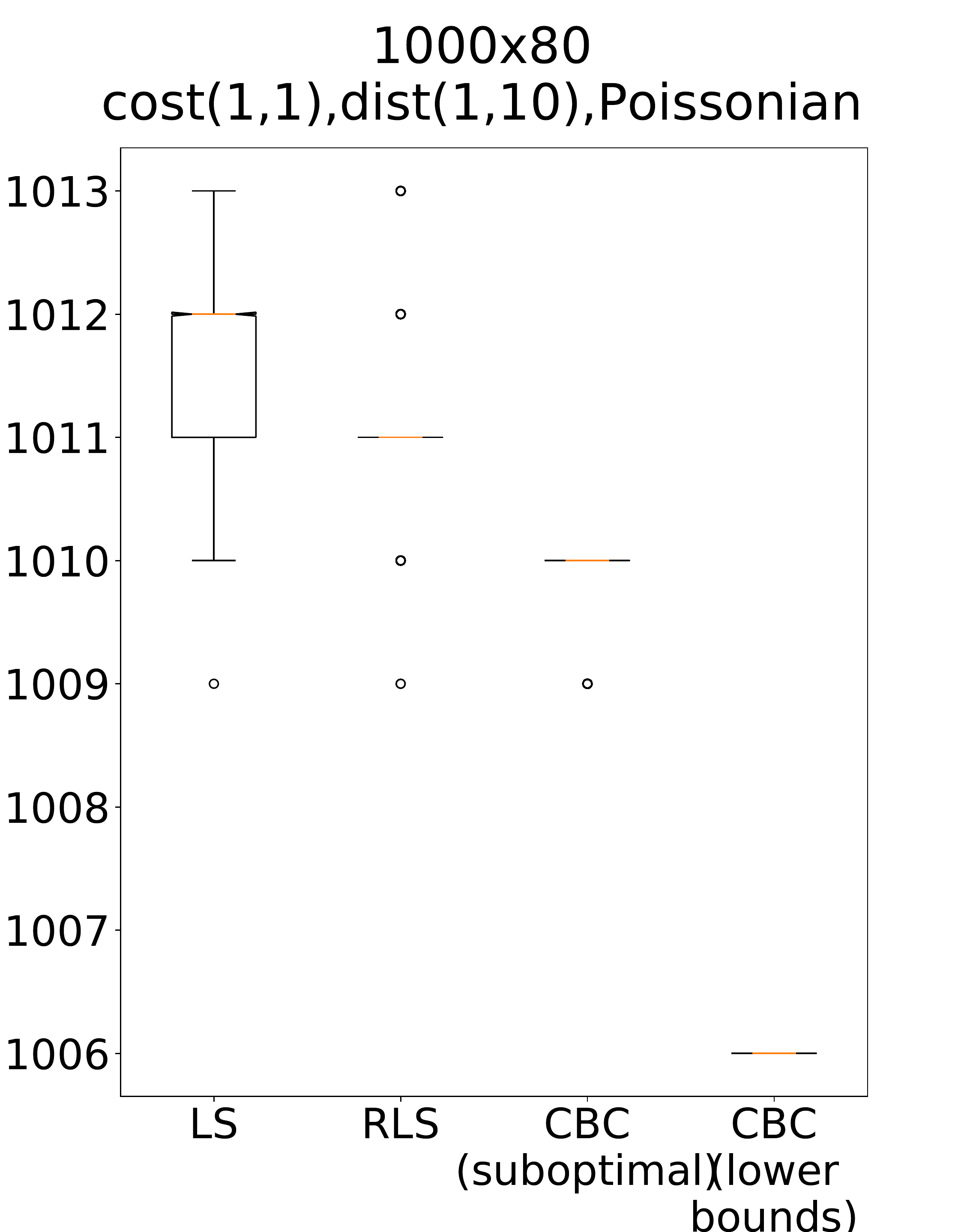}
\includegraphics[scale=0.13]{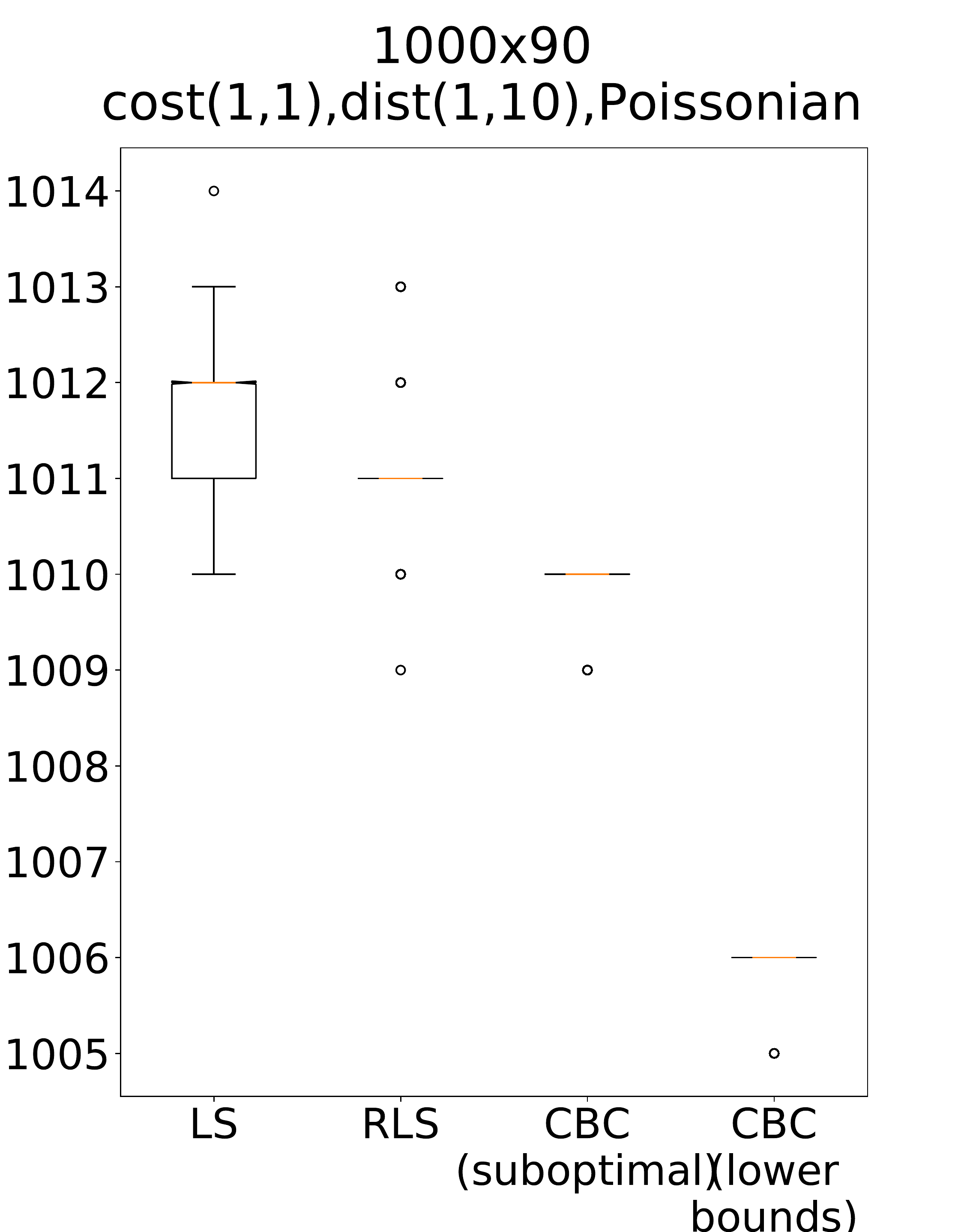}\\
\includegraphics[scale=0.13]{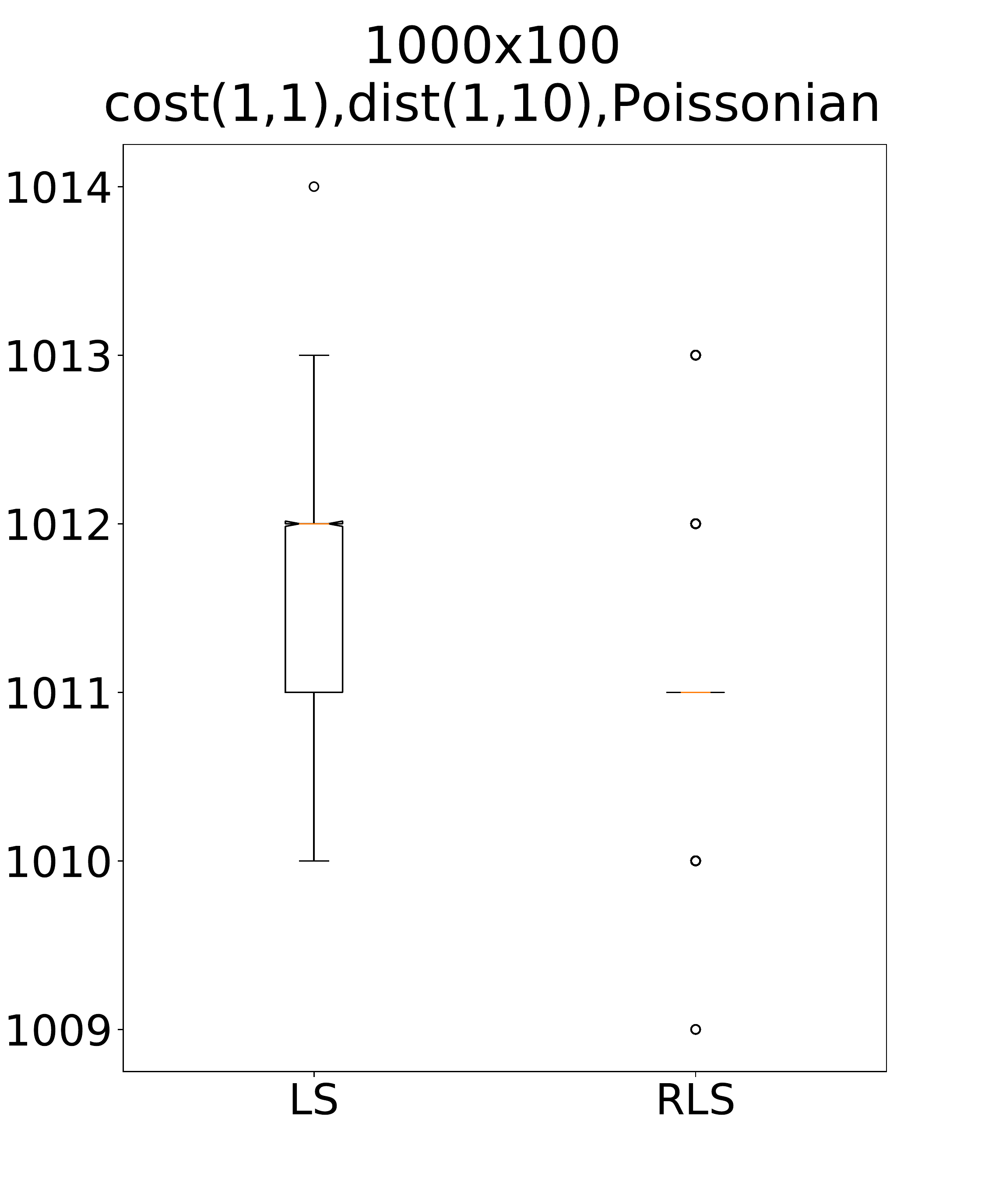}
\includegraphics[scale=0.13]{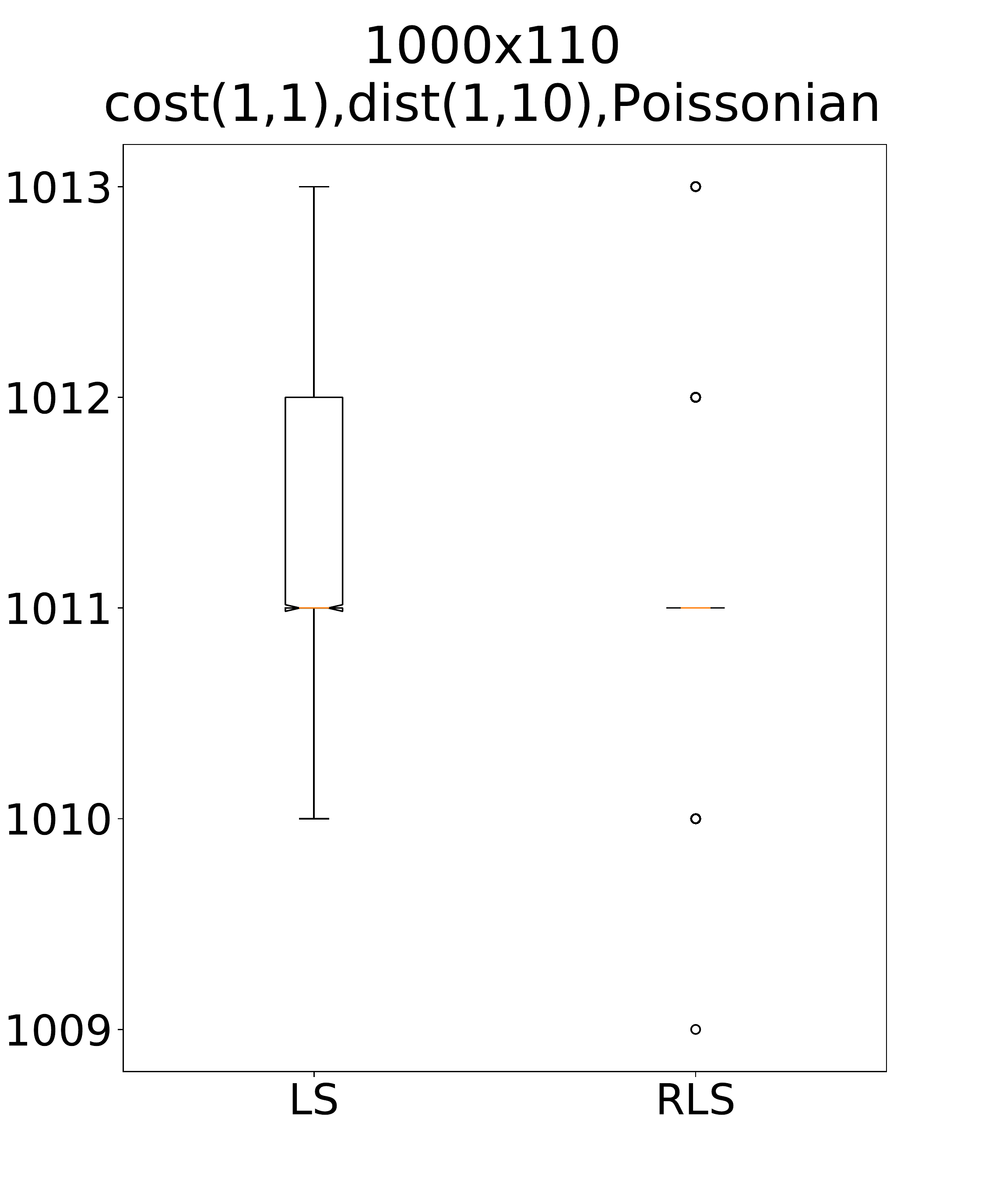}
\includegraphics[scale=0.13]{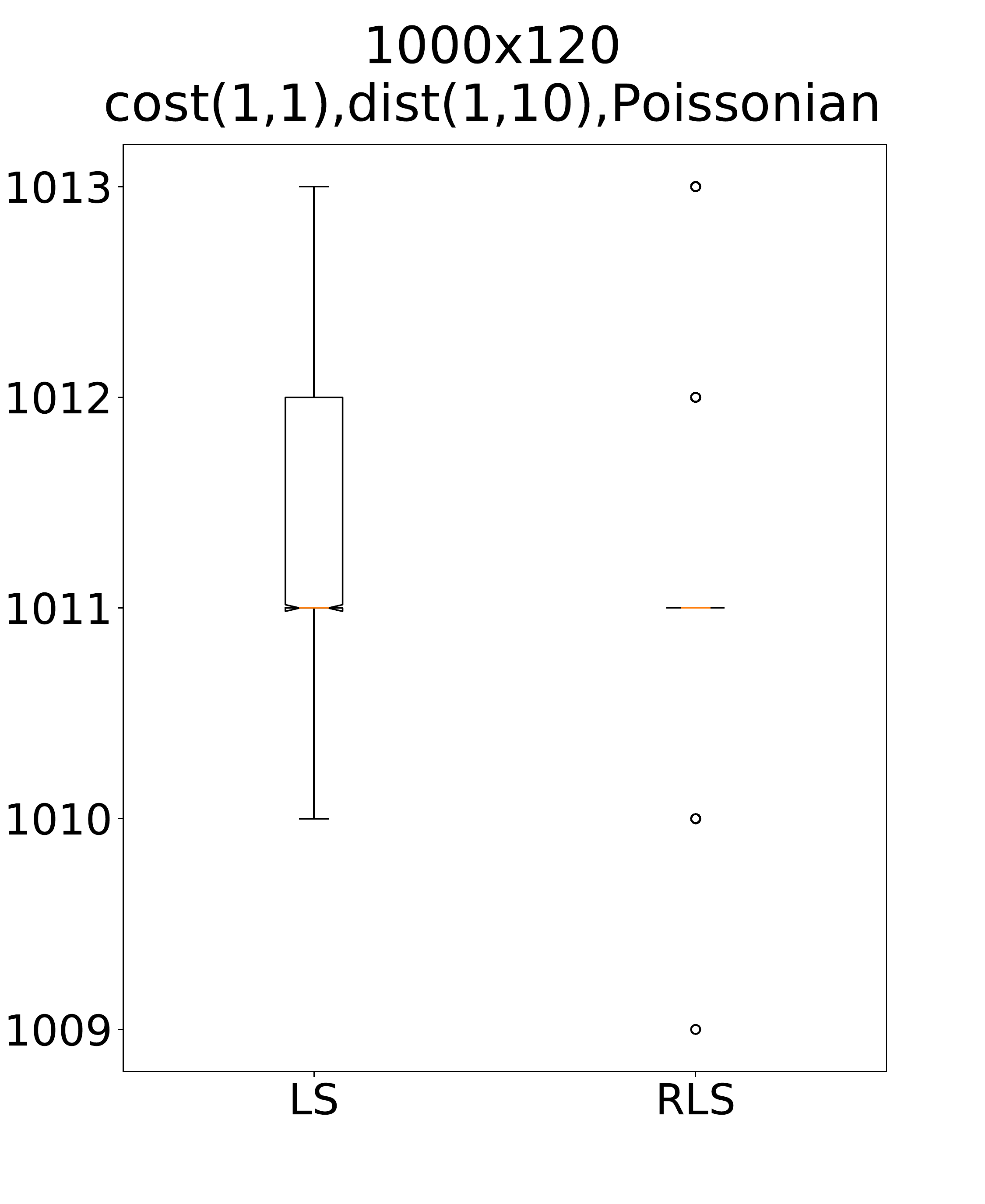}
\includegraphics[scale=0.13]{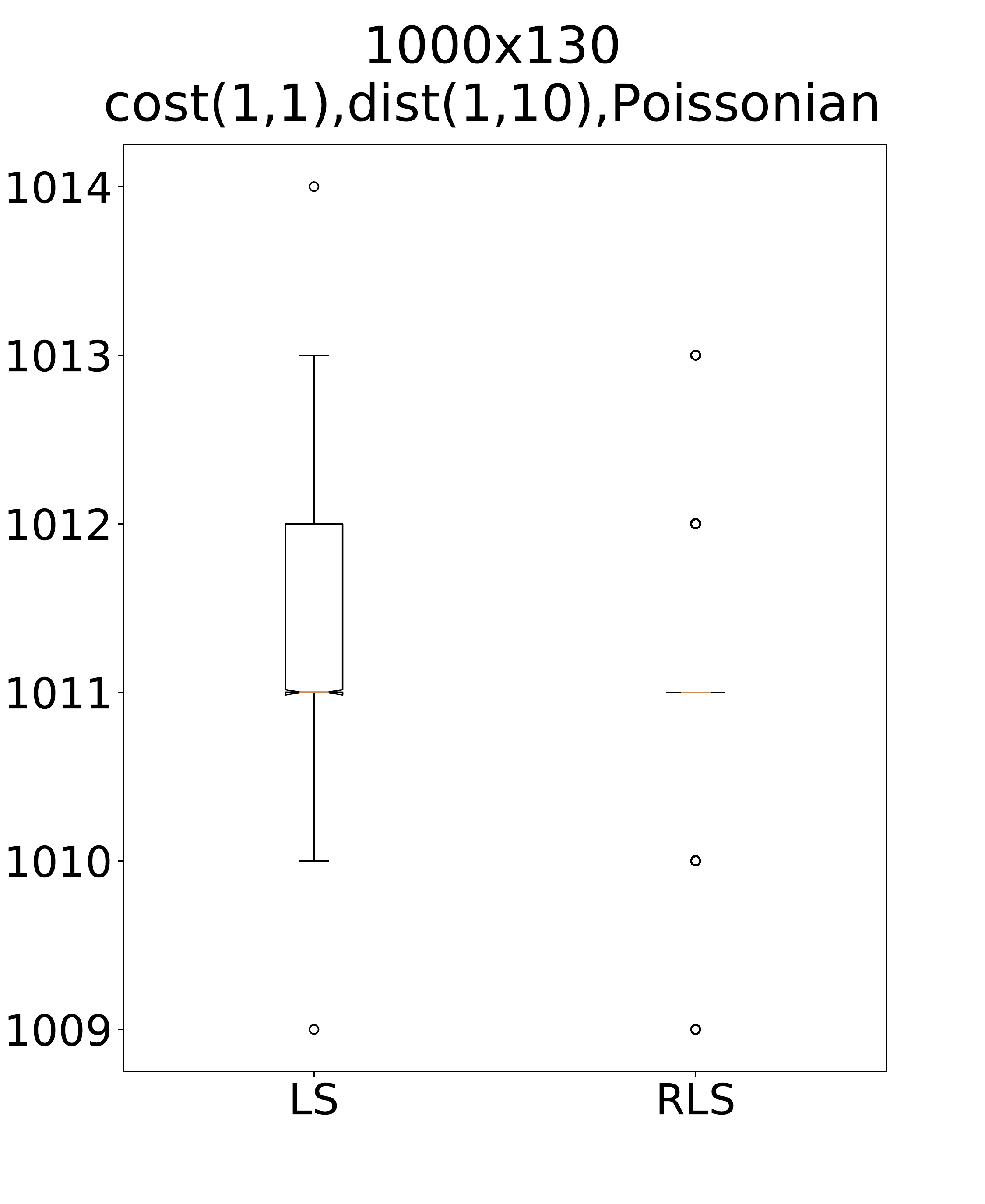}
\includegraphics[scale=0.13]{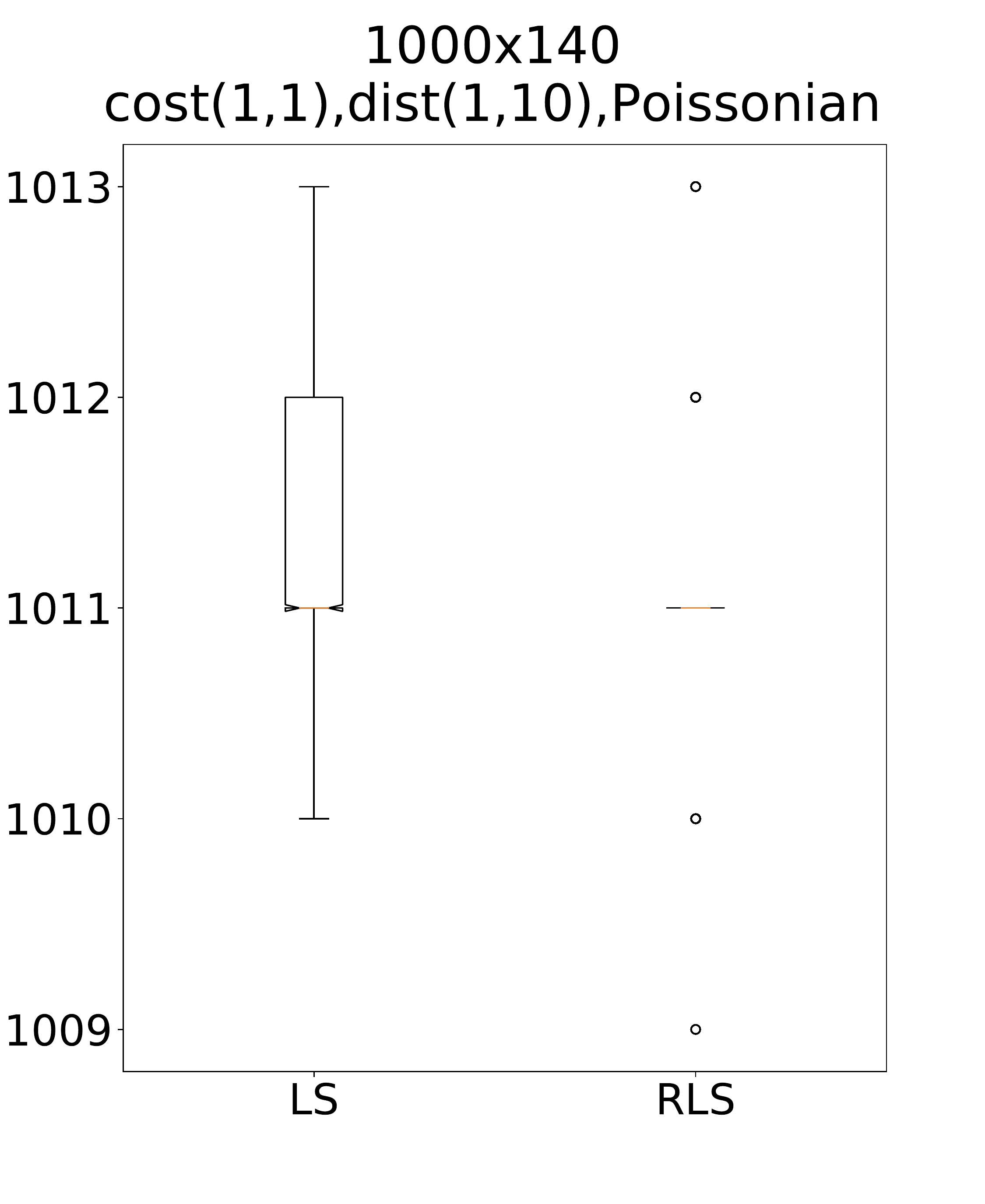}
\end{center}
\caption{Box-whisker plots obtained for $10$ problem instances generated according to Model 4. LS and RLS were used $1000$ times per instance, while the results for CBC represent optimal or near-optimal reference solutions obtained by the corresponding ILP solving procedure.}	
\end{figure*}

\begin{figure*}
\begin{center}
\includegraphics[scale=0.13]{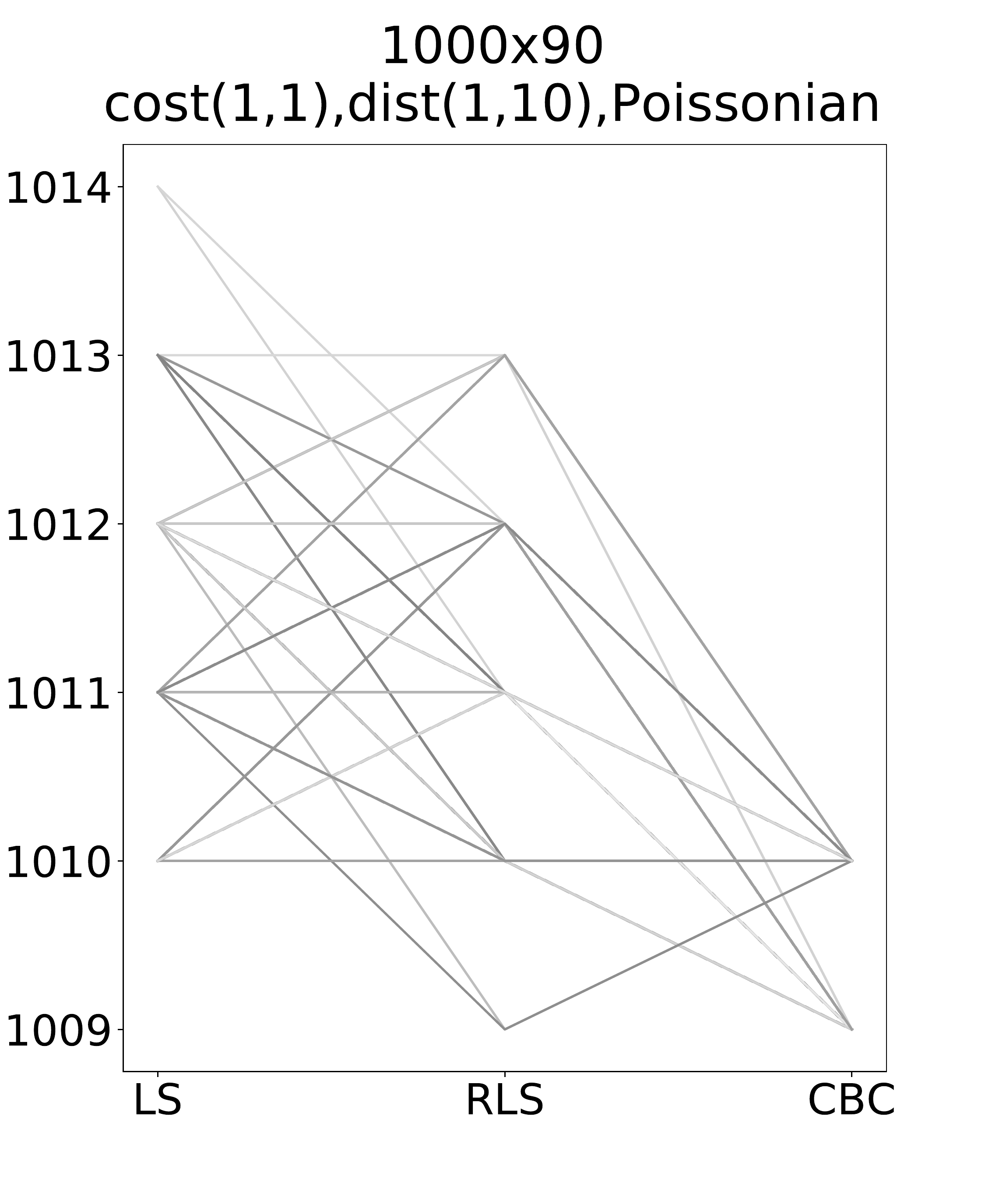}
\includegraphics[scale=0.13]{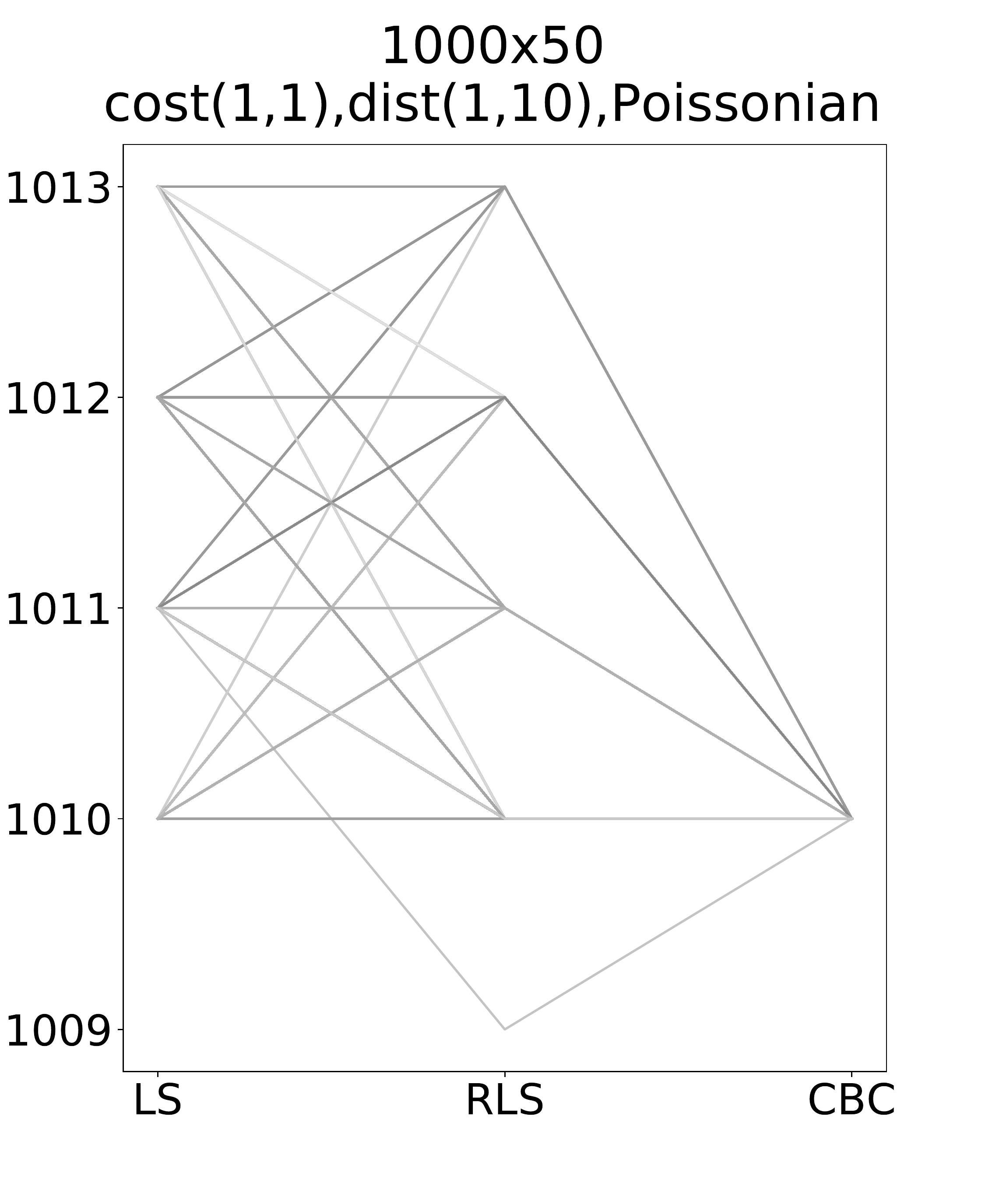}
\includegraphics[scale=0.13]{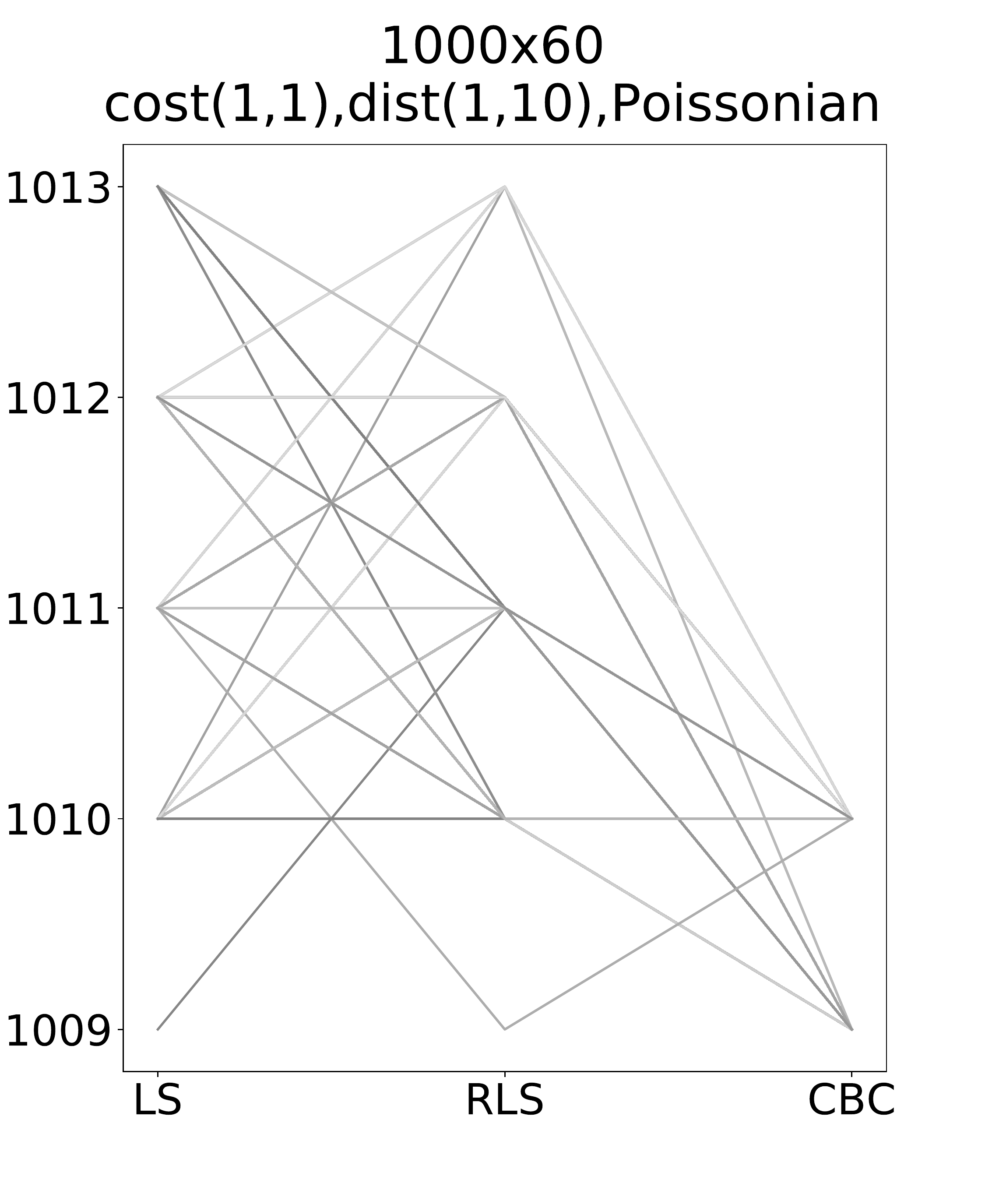}
\includegraphics[scale=0.13]{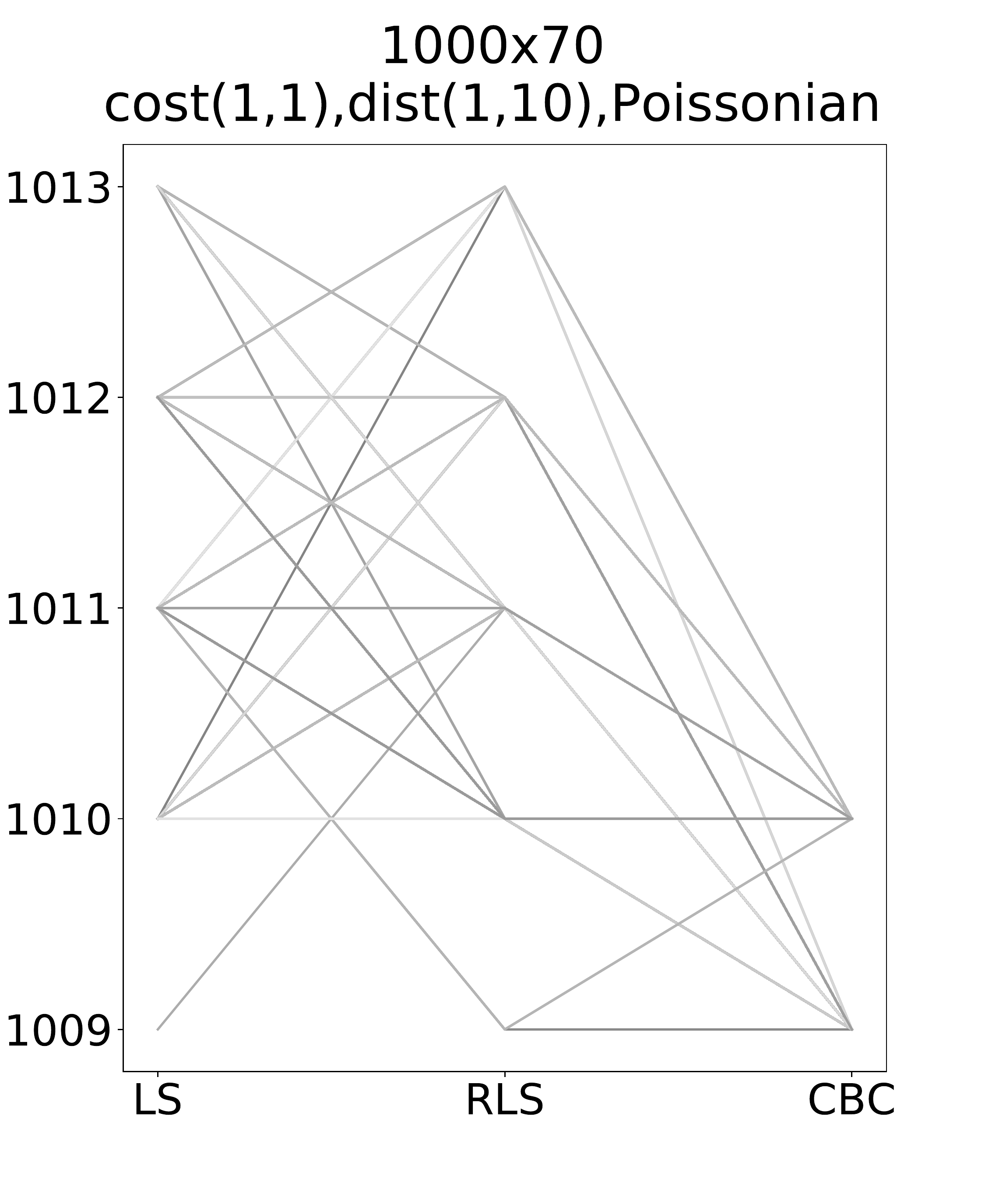}
\includegraphics[scale=0.13]{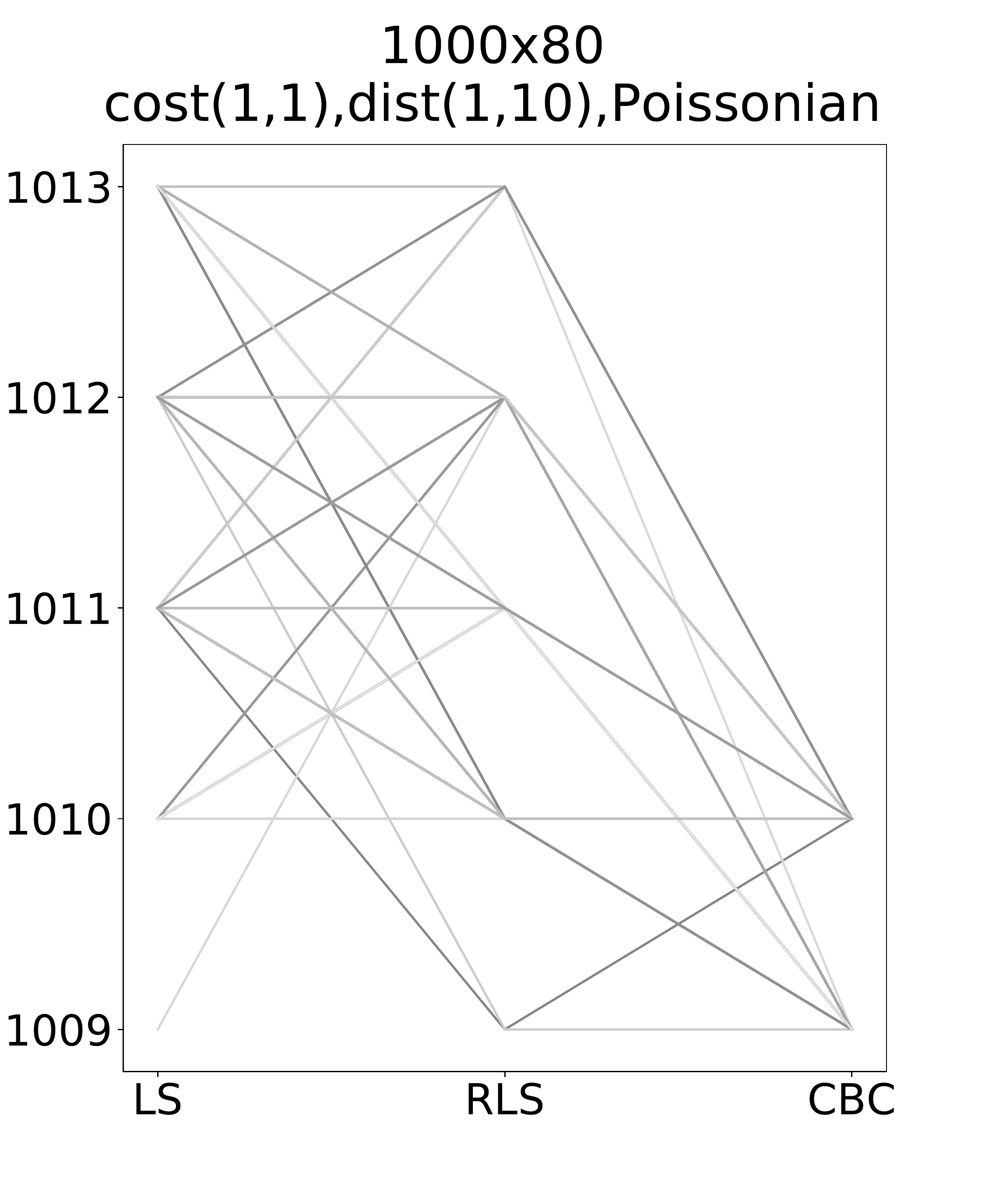}
\end{center}
\caption{A plot depicting the relation of the performance of LS, RLS and the actual optimal or near-optimal solutions found by CBC for Model 4. These results are grouped according to the problem instance. Each of the gray lines represents the objective values found by a run of LS, RLS and the objective value found by CBC.}	
\end{figure*}

A surprising observation is that while RLS often has inferior performance, it can actually outperform LS at times due to its randomised nature. This can be observed in Figure 8. The triangular shape of the structures indicates that RLS is often far from high-quality solutions. However, one can also observe occasional high-quality runs, such as the solution with the optimal objective value $1006$ for one of the $1000 \times 50$ instances, that LS failed to produce.

Model 3 represents an instance type, for which LS is a good choice to rapidly obtain a good but not necessarily optimal solution. LS estimates the distribution of optima better but with a consistent bias, possibly guaranteeing a good approximation but getting frequently stuck in local optima. If one needs to obtain a near-optimal solution, then RLS might be a better choice of an algorithm due to its randomised nature and better robustness. In summary, this seems to be a model, for which a suitable hybrid of both LS and RLS may work quite well.

\subsubsection{Model 4: Flat Facility Cost, Poissonian Distribution of Distances}

Figure 9 presents the results for the last model in the form of box-whisker plots. This model seems to be the most intriguing one out of the four models investigated. In contrast to Model 3, it is RLS, for which the results are concentrated around the median. However, RLS still seems to perform better than LS in most cases, even though the results indicate that it also has a tendency to get stuck in local optima. It is also worth noting that CBC was only able to find near-optimal solutions for these instances for the sizes we investigated, including the smallest instances with $50$ facilities. The near-optimal solutions sampled have a very flat distribution. In this context, RLS seems to be better at estimating these distributions, albeit with a bias.

In Figure 10, we present a more detailed picture of the performances of the algorithms. The patterns reveal that in most cases, LS and RLS produce solutions of comparable quality. RLS is therefore likely to sample solutions with better objective values more frequently, thus, obtaining a better median performance. However, for the $1000 \times 50$ instances, one can also observe that while CBC consistently produced solutions with objective value $1010$, RLS was able to provide a better solution with objective value $1009$, standing out from the corresponding plot. This is also in line with the relatively distant lower bound distribution sampled. This reveals that for Model 4, hybrid algorithms incorporating RLS may be a good choice of problem solving techniques.

\subsection{Discussion}

The previous results have indeed painted a very complex picture of problem difficulty landscape and algorithm performance. On a more fine-grained level, we have obtained four partially or heavily contrasting instance models and algorithm behaviours. For simplicity, we summarise our findings in the following:

\begin{itemize}
	\item Model 1. We obtained that a hybrid or a multi-start algorithm using RLS would be a good technique for this type of instances. The use of a solution polishing feature of an out-of-the-box ILP solver could also be a good choice, given the results obtained by CBC. This idea was previously successfully used to solve the minimum weight dominating set problem \cite{mwdshybrid}.
	\item Model 2. For this model, we obtained that a multi-start variant of LS would likely be the best choice of an algorithm.
	\item Model 3. This model showed much more complex patterns of landscape and algorithm behaviour. However, it seems that a multi-start variant of RLS and possibly its hybrid could work well for these instances.
	\item Model 4. For the Poissonian distance model, we obtained a very flat picture of algorithm behaviours. However, we observed a phenomenon of ``very lucky'' runs of RLS, that may provide even better solutions than CBC in hours of systematic search. For such an instance type, hybrid algorithms based on RLS seem to be well-suited.
\end{itemize}

\begin{table*}
\caption{Detailed results obtained for each of the instance models by CBC with $24$ hour time limit. The results presented represent averages over $10$ instances per each configuration.}
{\small
\begin{center}
\begin{tabular}{l l l l l l}
\toprule
instance						&	$\mathcal{C} \times \mathcal{F}$			& average						& average						& result *		& average \\
model								&																				& upper bound				& lower bound				& 						& CPU time \\
\midrule
Model 1							& $1000 \times 50$											&	1038.4						&	1038.4						&	OPT					& 101 s \\
										& $1000 \times 60$											&	1033.8						&	1033.8						&	OPT					& 1437 s \\
										& $1000 \times 70$											&	1032.4						&	1032.2						&	TLE (1)			& 27557 s \\
										& $1000 \times 80$											&	1030							&	1027.3						&	TLE (9)			& 82238 s \\
										& $1000 \times 90$											&	1030.1						&	1025							&	TLE (10)		& \\
										\midrule
Model 2							& $1000 \times 50$											&	1100.8						&	1100.8						&	OPT					& 45 s \\
										& $1000 \times 60$											&	1058.6						&	1058.6						&	OPT					& 254 s\\
										& $1000 \times 70$											&	1052.3						&	1052.3						&	OPT					& 1175 s \\
										& $1000 \times 80$											&	1042.1						&	1041.9						&	TLE (1)			& 24506 s \\
										& $1000 \times 90$											&	1041.4						&	1040							&	TLE (4)			& 55092 s \\
\midrule
Model 3							& $1000 \times 50$											&	1006.9						&	1006.9						&	OPT 				& 20002 s \\
										& $1000 \times 60$											&	1007							&	1006.4						&	TLE (5)			& 65511 s \\
										& $1000 \times 70$											&	1007							&	1006							&	TLE (8)			& 81083 s \\
										& $1000 \times 80$											&	1007							&	1005.6						&	TLE (8)			& 81369 s \\
										& $1000 \times 90$											&	1006.9						&	1005.1						&	TLE (10)		& \\
\midrule
Model 4							& $1000 \times 50$											&	1010							&	1007.6						&	TLE (10)		& \\
										& $1000 \times 60$											&	1009.9						&	1007							&	TLE (10)		& \\
										& $1000 \times 70$											&	1009.9						&	1006.4						&	TLE (10)		& \\
										& $1000 \times 80$											&	1009.9						&	1006							&	TLE (10)		& \\
										& $1000 \times 90$											&	1009.9						&	1005.9						&	TLE (10)		& \\
\bottomrule
\end{tabular}
\end{center}
* Result values represent the following:\\
OPT - optimal solution found for all $10$ instances generated,\\
TLE - time limit exceeded and near-optimal solution found for at least one instance (the value in parentheses represents the number of instances with TLE result).
}
\end{table*}

\noindent
It is worth pointing out that these results are not particularly surprising in the context of the \textit{No Free Lunch} theorem \cite{culberson1998futility,ho2002simple,wolpert1997no}. However, their implications for the design of efficient algorithms are profound for solving related problems in real-world scenarios. Such a case becomes even more pressing if generalisation and adaptability to instances with unknown numerical properties is desired. This includes popular problems such as job shop scheduling \cite{jobshopscheduling}, flow shop scheduling problems \cite{flowshopscheduling}, variants of knapsack problems \cite{garcia2014tabu} or container pick-up problems \cite{Do2016285}.

The way out of this vicious circle of metaheuristic algorithm design is two-fold. Firstly, one can argue that a reliable metaheuristic for a real-world problem can only be designed if specific structural and numerical properties of the instance to encounter are known in advance. This is the case for several real-world applications, for which the problem-specific knowledge may be used to describe the expected numerical structure of the instance. Alternatively, if a solution of only a certain quality is expected, then it is desirable to use as simple algorithm as possible. This has several advantages, including scalability and minimised need for parameter tuning. However, the results we obtained for LS and RLS also suggest combining several of such algorithms may be a good way forward.

It has previously been verified that tabu search is the most universally successful technique for solving the facility location problem, providing better results than genetic algorithms and simulated annealing in most cases \cite{arostegui2006empirical}. However, as tabu search can be based on both systematic and randomised local search strategies, an interesting open problem is an extension of our investigations to tabu search algorithms based on LS and RLS. In addition, it is known the tabu tenure, i.e. the tabu list size, is crucially important for the performance of metaheuristics in combinatiorial optimisation. It may even be decisive on whether the algorithm converges or will get stuck with an overwhelming probability \cite{tabucolanalysis}. We therefore believe this study is an important step towards a different approach to performance investigation for metaheuristics. This also has an impact on their design, either in its numerically focused or exploratory form.

\section{Conclusions}

We proposed four cost and distance models for the uncapacitated facility location problem instances. These models enabled us to uncover a complex relation between the numerical properties of a problem instance and efficiency of the optimisation algorithms used to solve the problem. Since facility location is a simple example of a scalable choice-and-assignment optimisation problem, these phenomena are likely to observed also in more complex optimisation problems in science and engineering.

An investigation of the efficiency of two metaheuristic algorithms was presented, namely systematic local search (LS) and randomised local search (RLS). An out-of-the-box mixed-integer linear programming solver has also been used to obtain reference results and determine how far LS and RLS are from the optimal or near-optimal solutions. In the experimental results, one can observe a rich variety of behaviours that is preserved even though the algorithm performance was investigated for instances with the same size. RLS outperformed LS for Model 1, while this was reversed in Model 2. Models 3 and 4 show even more intricate landscape properties and behaviours of LS, RLS, as well as the exact solver.

Based on the results obtained by LS and RLS, one can conclude that the choice of the right local search strategy for a particular model is closely related to its internal numerical properties. As a consequence, efficient state-of-the-art algorithms for each model would look somewhat different in design. This strengthens the case for design of hybrid metaheuristics for this type of problems, but also highlights the intriguing need for the numerical properties of a particular instance to be taken into account in such a design. In some real-world applications, these properties may be well-known intuitively. However, for applications where these properties are unknown, this shows a pressing need for simple and scalable algorithms with a compact set of parameters.

Our results can likely be generalised to other related assignment problems and combinatorial problems in general, such as job shop and flow shop scheduling, knapsack problems or pallet stacking. We believe that these results may motivate a somewhat new bottom-up approach to metaheuristic design and investigation, especially in relation to hybrid metaheuristics.

\bibliography{common}{}
\bibliographystyle{plain}

\end{document}